\documentclass[twocolumn,aps,prx,groupedaddress,longbibliography]{revtex4-1}
\usepackage{graphicx}
\usepackage{color}
\usepackage{amsmath}
\usepackage{marvosym}
\usepackage{bm}

\newcommand{\be}{\begin{equation}}
\newcommand{\ee}{\end{equation}}
\newcommand{\bea}{\begin{eqnarray}}
\newcommand{\eea}{\end{eqnarray}}

\newcommand{\la}{\langle}
\newcommand{\ra}{\rangle}

\newcommand{\bk}{{\bf k}}
\newcommand{\Tr}{{\rm \, Tr\,}}

\newcommand{\ket}[1]{| #1\ra}
\newcommand{\Bra}[1]{\la #1|}
\newcommand{\Ket}[1]{| #1\ra}
\newcommand{\MatEl}[3]{\langle #1|#2|#3\rangle}

\renewcommand{\vec}[1]{{\bf #1}}
\renewcommand{\epsilon}{\varepsilon}

\renewcommand{\thesubsubsection}{\thesubsection.\arabic{subsubsection}}
\begin{document}

\title{Floquet topological insulators: from band structure engineering to novel non-equilibrium quantum phenomena} 
\author{Mark S. Rudner$^1$ and Netanel H. Lindner$^2$}
\affiliation{$^{1}$Niels Bohr International Academy and the Center for Quantum Devices, Niels Bohr
Institute, University of Copenhagen, 2100 Copenhagen, Denmark} \affiliation{$^{2}$Physics
Department, Technion, 320003, Haifa, Israel}

\begin{abstract}
 We review  methods for using time-periodic fields (e.g., laser or microwave fields) to induce non-equilibrium topological phenomena in quantum many-body systems. 
 We discuss how such fields can be used to change the topological properties of the
   single particle spectrum, and key experimental demonstrations in solid state, cold atomic, and
   photonic systems. The single particle Floquet band structure provides a stage on which the
   system's dynamics play out; the crucial question is then how to obtain robust topological behaviour in the many-particle setting.
   In the regime of mesoscopic transport, we discuss manifestations of topological edge states induced in the Floquet spectrum.
   Outside the context of mesoscopic transport, the main challenge of
   inducing stable topological phases in many-body Floquet systems is their tendency to absorb
   energy from the drive and thereby to heat up.  We discuss three routes to overcoming this
   challenge: long-lived transient dynamics and prethermalization, disorder-induced many-body localization, and engineered couplings to
   external baths. We discuss the types
   of phenomena that can be explored in each of these regimes, and their experimental
   realizations.

\end{abstract}

\maketitle

%%%%%%%%%%%%%%%%%%%%%%%%%%%%%%%%%%%%%%%%%%%%%%%%%%%%%%%%%%%%%%%%%%%%%%%%%%%%%%%%%%
\section{Introduction}
\label{sec:intro}

Recently developed experimental tools for probing and controlling quantum systems provide access to
quantum many-body dynamics on time and length scales, and with levels of precision, that were
nearly unimaginable just a few decades ago. In addition to enabling deep new insights into the
properties of quantum materials, these tools allow us to explore fundamentally new regimes of
quantum many-body dynamics. On a practical level, laser and microwave driving fields also open the
possibility of dynamically controlling and modifying quantum material properties ``on
demand''~\cite{Basov2017}.

As briefly outlined below, Floquet theory~\cite{Floquet1883} provides a powerful framework for
analyzing periodically-driven quantum systems; for this reason, the names ``periodically-driven
systems'' and ``Floquet systems'' are used interchangeably. Of particular interest for this review,
periodic driving provides means to manipulate the dispersion and geometry of (Floquet)-Bloch bands
in atomic or electronic systems with periodic lattice potentials~\cite{Yao2007}.
Inspired by the recent discovery of topological insulators (TIs)~\cite{HasanTI_RMP}, a series of
early works developed the notion of a ``Floquet topological insulator'' (FTI): by appropriately
choosing the drive frequency, amplitude, and symmetry, periodic driving can be used to change the
{\it topological} features of a system's Bloch bands~\cite{Oka2009, Kitagawa2010, Lindner2011}.

Crucially, while a periodic drive may be used to induce a topologically nontrivial Floquet band structure for electrons or cold atoms, this is not enough to ensure that the system displays physical properties that one might expect or hope to produce by analogy to equilibrium
TIs. The Floquet bands provide the stage upon which the system's dynamics play out; the physical
properties of the system are determined by the many-body state that is obtained, and the extent to
which it can be described by an insulator-like filling in the Floquet basis. Finding the conditions
when a periodically driven system can host a stable stationary (or nearly stationary) state with
non-trivial topological characteristics is therefore one of the central challenges in the
field.

Importantly, in the absence of coupling to an external environment, it is widely believed that a generic, {\it closed}, periodically driven quantum system containing many interacting particles will absorb energy from the driving field and increase its local entropy density, %and at long times
tending toward a featureless state at long times in which all local correlations are fully random
(as in an ``infinite temperature'' state)~\cite{DAlessio2014,Lazarides2014}. In such a state, no
topological phenomena can persist. Therefore, in order to stabilize nontrivial topological
phenomena in Floquet systems, it is necessary to develop strategies to avoid such heating~\footnote{Exceptions to
the runaway heating scenario, e.g., based on localization in energy space~\cite{Prosen1998,
Kukuljan2015, Citro2016, Chandran2016, Haldar2018} or finite-size effects~\cite{Seetharam2018}, are
also being explored.}.

A variety of approaches %strategies
for stabilizing FTIs and other Floquet phases are being investigated, leveraging regimes of high or
low frequency driving, disorder-induced localization, and/or bath engineering. In
Sec.~\ref{sec:ManyBody} below we outline the main principles underlying these approaches, and some
of the key results obtained so far.

%\noindent {\it -- Philosophy of the paper}\\
Over the past decade, work on Floquet systems has developed in a number of interesting directions,
including Floquet engineering of magnetic and other strongly-correlated phases~\cite{Grushin2014,
Klinovaja2016, Liu2018,Gorg2018, Kennes2018}, as well as the formulation %the development
of topological classifications~\cite{Kitagawa2010,Rudner2013,Nathan2015,Roy2017,Roy2017b,Yao2017,
PlateroPRL, Graf2018, Shapiro2018} and notions of
symmetry-breaking~\cite{Khemani2016,Else2016b,Else2017, Rudner2018,Nag2018,Kinoshita2018,
Harper2019} and symmetry-protected topological phases in non-equilibrium quantum many-body
systems~\cite{Harper2019, vonKeyserlingk2016, Potter2016, Else2016, Harper2017}.
 These developments motivated investigations of fundamental questions in non-equilibrium quantum dynamics
including the manifestations of ergodicity and localization, and the dynamics of
(pre)thermalization in the unitary evolution of quantum many-body systems,
 which naturally arise and can addressed in many-body Floquet systems~\cite{Moessner2017}.
Many of these directions have been explored experimentally in cold
atoms~\cite{Jotzu2014,Flaschner2016}, photonics~\cite{Rechtsman2013}, and solid state systems
~\cite{Wang2013,McIver2018}. In this review, our primary focus is on FTIs and their realization
solid state and cold atomic systems, and we direct  readers interested in other directions to the
cited works and reviews.
%Where appropriate we will make connection to related areas of Floquet engineering, for example in magnetic systems or in more exotic strongly-correlated systems such as Floquet symmetry-protected topological phases (SPTs) and Floquet time crystals. However, our treatment of these topics will necessarily be limited.

Within the realm of FTIs proper, our goal in this review is to provide a pedagogical introduction
to key concepts underlying the formation and topological characterization of Floquet-Bloch bands,
and their physical manifestations. In illustrating these principles we further aim to give a broad
overview of key results in the field. Many topics have been approached in different ways by
different groups; we will not be able to cover all alternative points of view on each topic, but
will provide relevant references to original sources and other reviews. The accompanying
Supplementary Material~\cite{SM} contains additional key technical details that we expect will be
of interest for those who wish to begin working in this field.

%%%%%%%%%%%%%%%%%%%%%%%%%%%%%%%%%%%%%%%%%%%%%%%%%%%%%%%%%%%%%%%%%%%%%%%%%%%%%%%%%%

%%%%%%%%%%%%%%%%%%%%%%%%%%%%%%%%%%%%%%%%%%%%%%%%%%%%%%%%%%%%%%%%%%%%%%%%%%%%%%%%%%
\section{Floquet band structure engineering}
\label{sec:FloquetEngineering}

  We now briefly review how time-periodic driving  can be used to modify the band structure of a quantum particle moving in a spatially periodic lattice potential.
We leave the question of how such bands are populated, and a detailed discussion of observables,
for later sections.

Time evolution in the %periodically-
driven system is governed by a time-periodic Hamiltonian $H(t) = H(t + T)$, where $T$ is the driving period. 
We define $\omega = 2\pi/T$ as the driving frequency, but note that many harmonics may be involved. According to Floquet's theorem~\cite{Floquet1883}, the
Schr\"{o}dinger equation $i\hbar\frac{d}{dt}\Ket{\psi(t)} = H(t)\Ket{\psi(t)}$ admits a complete
set of orthogonal solutions of the form $\Ket{\psi(t)} = e^{-i\varepsilon t/\hbar}\Ket{\Phi(t)}$,
with $\Ket{\Phi(t)} = \Ket{\Phi(t + T)}$. Here, the ``quasienergy'' $\varepsilon$ plays a role
analogous to the energy of a Hamiltonian eigenstate in a non-driven system. The quasienergy
spectrum is determined by the one-period evolution operator $U(T) =
\mathcal{T}e^{-(i/\hbar)\int_0^T dt' H(t')}$, where $\mathcal{T}$ denotes time-ordering:
$U(T)\Ket{\psi(0)} = e^{-i\epsilon T/\hbar}\Ket{\psi(0)}$. For a particle moving in a crystalline
lattice potential, the Floquet (quasienergy) spectrum is organized into bands, analogous to those
of a non-driven system (for a pedagogical introduction see, e.g., Ref.~\cite{Holthaus2016}).

In direct analogy with the emergence of a crystal momentum Brillouin zone for a particle in a periodic potential, %due to the discrete time-translation symmetry of the system
all independent solutions of the Schr\"{o}dinger equation may be indexed by quasienergy values that fall within a single ``Floquet-Brillouin zone,'' $\varepsilon_0 \le \varepsilon < \varepsilon_0 + \hbar\omega$. %periodicity of $\Ket{\phi(t)}$,
(To see this, note that $e^{-i(\epsilon + n\hbar\omega)T/\hbar} = e^{-i\epsilon T/\hbar}$ for any
integer $n$.)
Below we take $\varepsilon_0 = 0$ or $\varepsilon_0 = -\hbar\omega/2$, depending on the context.
As we demonstrate throughout this review, the unique topological phenomena that occur
in Floquet systems, without analogues in equilibrium, owe their existence to the periodicity of
quasienergy.

For many cases, it is useful to relate the Floquet operator $U(T)$ to the evolution of a system
with a {\it static}, effective Hamiltonian, $H_{\rm eff}$: $U(T) \equiv e^{-iH_{\rm eff}T/\hbar}$.
The stroboscopic evolution described by $U(T)$ is identical to that described by evolution with
$H_{\rm eff}$ for time $T$. In this section we will examine a few simple cases where the effects of
the driving field can be considered perturbatively, allowing us to heuristically infer the
properties of $H_{\rm eff}$. As we will see, even a weak drive can be used to construct an
effective Hamiltonian with topological properties that differ from those of the system in the
absence of driving.

To illustrate the principles of Floquet band engineering we consider a two-dimensional (2D) Dirac
mode subjected to a uniform, circularly polarized driving field. This model, which captures
dynamics for crystal momenta in the vicinity of a gapped or gapless Dirac point, provides a
building block for describing Floquet engineering in  a variety of settings such as graphene and
transition metal dichalcogenides (TMDs), as well as surface states of three-dimensional topological
insulators. For any given system, the net topological effects of the drive are determined by its
action throughout the entire Brillouin zone (which may include several such Dirac points).

For each value of the momentum $\vec{k} = (k_x, k_y)$, the evolution is described by the (Bloch)
Hamiltonian \be \label{eq:Dirac2D} H_{\rm 2D}(\vec{k}, t) = v \left[\hbar \vec{k} -
e\vec{A}(t)\right] \cdot \boldsymbol{\sigma} + \frac{\Delta}{2}\sigma_z, \ee
where $v$ is the velocity of the Dirac electrons, %Fermi velocity,
$\boldsymbol{\sigma} = (\sigma_x, \sigma_y)$ is a vector of Pauli matrices describing either spin
or orbital pseudospin indices, and $\Delta$ is the ``mass gap'' at $\vec{k} = 0$. For simplicity we
specialize to a left-hand circularly polarized (LHP) driving field described by
 the vector potential $\vec{A}(t) = A_0(\cos \omega t, \sin \omega t)$, with driving amplitude $A_0$.
In the absence of driving, $H_{\rm 2D}$ gives rise to a pseudo-relativistic dispersion relation,
$E(k) = \sqrt{(\hbar v k)^2 + (\Delta/2)^2}$, where $k = |\vec{k}|$. To endow the model with a
finite bandwidth, $W$,  we impose a simple cutoff on $k$ such that $E(k) < W/2$. In practice, the
form of the Hamiltonian and its eigenstates at large values of $k$, which are important for
topological properties, must be found by stitching together multiple Dirac valleys or through a
proper termination that respects the periodicity of the Brillouin zone.

For a system with two bands, as described in Eq.~(\ref{eq:Dirac2D}), band topology can be
visualized geometrically in terms of a unit vector $\hat{\vec{n}}(\vec{k})$ that relates the
(static or Floquet) band eigenstates $\Ket{\psi(\vec{k})}$ to points on the Bloch sphere:
$\hat{n}_\alpha(\vec{k})  = \MatEl{\psi(\vec{k})}{\sigma_\alpha}{\psi(\vec{k})}$. In terms of
$\hat{\vec{n}}(\vec{k})$, the integer-valued Chern number $\mathcal{C}$ that characterizes the
topology of a given band simply corresponds to the net number of times that
$\hat{\vec{n}}(\vec{k})$ covers the Bloch sphere as $\vec{k}$ scans over the entire Brillouin zone
(BZ), $\mathcal{C} = \frac{1}{4\pi} \oint_{\rm BZ} d^2k\, \hat{\vec{n}} \cdot (\partial_{k_x}
\hat{\vec{n}} \times \partial_{k_y} \hat{\vec{n}})$.
The goal of topological Floquet band engineering is to
dynamically manipulate the configurations of $\hat{\vec{n}}(\vec{k})$ in momentum space, and in
particular to induce transitions where, e.g., $\hat{\vec{n}}(\vec{k})$ exhibits a trivial
configuration for the nondriven system (corresponding to $\mathcal{C} = 0$), 
but is nontrivial for the system's Floquet bands in the presence of the drive ($\mathcal{C}\neq 0$).

%%%%%%%%%%%%%%%%%%%%%%%%%%%%%%%%%%%%%%%%%%%%%%%%%%%%%%%%%%%%%%%%%%%%%%%%%%%
\begin{figure}[t]
\includegraphics[width=\columnwidth]{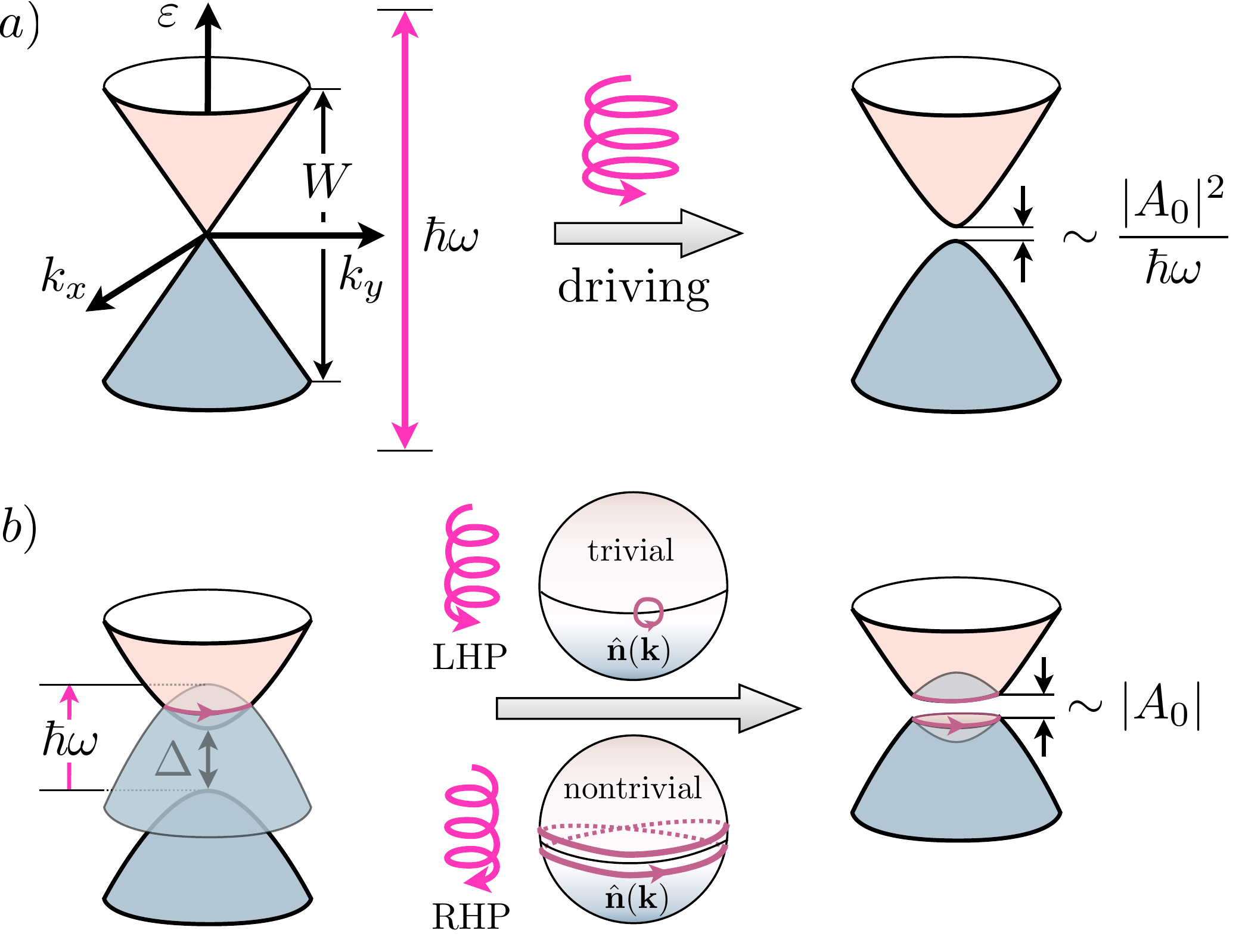}
\caption{Floquet band engineering. (a) Circularly polarized light induces a gap in a 2D Dirac dispersion via
an off-resonant process involving the absorption and re-emission of a drive photon.
The induced gap is proportional to the intensity of the driving field, and the induced Berry curvature is concentrated near the band extrema.
For graphene, the $K$ and $K'$ valleys exhibit Berry curvatures with identical signs (determined by the handedness of the drive), leading to Chern numbers of $\pm 1$ of the Floquet bands in the absence of resonances.
(b) A spin-orbit coupled direct-bandgap semiconductor subjected to circularly polarized light.
In a rotating frame, the drive opens a gap along the contour of resonant transitions between the valence and conduction bands.
The bands obtained in the rotating wave approximation exhibit a band inversion.
For a two band system, the contribution of the Berry curvature arising from this band inversion to the
Chern numbers of the Floquet bands is equal to the winding of the pseudospin $\hat{\bf{n}}(\bk)$ as $\vec{k}$ traverses the resonance contour (see text for definitions).
When the pseudospin winding of the original bands and the polarization of the driving field have the same chirality, $\hat{\bf{n}}(\bk)$ acquires a winding of $\pm 2$.
When the chiralities are opposite, the winding (and hence Chern number) vanishes.}
\label{fig:BandEngineering}
\vspace{-0.2 in}
\end{figure}
%%%%%%%%%%%%%%%%%%%%%%%%%%%%%%%%%%%%%%%%%%%%%%%%%%%%%%%%%%%%%%%%%%%%%%%%%%%

We now introduce two key paradigms for inducing topological transitions through i) off-resonant and
ii) resonant driving fields.
For gapless systems such as graphene, an {\it off-resonant} drive can be used to induce nontrivial
topology in the system's band structure~\cite{Oka2009, Kitagawa2011}. Near zero energy, graphene's
band structure is characterized by two Dirac valleys centered at the $K$ and $K'$ points at
opposite corners of the Brillouin zone.
In valley $K$ the band structure is described by $H_{\rm 2D}$ in Eq.~(\ref{eq:Dirac2D}) with $\Delta = 0$; the Hamiltonian near $K'$ is similar, with $\sigma_y \rightarrow -\sigma_y$. 

When the drive frequency is large compared to the single-particle bandwidth, $\hbar\omega \gg W$, the drive cannot resonantly couple states in the valence and conduction bands for any value of $\vec{k}$ in the Brillouin zone (see Fig.~\ref{fig:BandEngineering}a). 
Here the essential effect of the drive is to lift the degeneracy at the Dirac points where the
valence and conduction bands touch.
At $\vec{k} = 0$ [relative to the Dirac points in the $K$ ($+$) and $K'$ ($-$) valleys], the Hamiltonian takes the simple form of a pure rotating field: $H_{\rm 2D}^\pm(\vec{k} = 0, t) = A_0 \cos(\omega t)\, \sigma_x \pm A_0 \sin ( \omega t)\, \sigma_y$. 
Although the time-average of $H_{\rm 2D}^\pm(\vec{k} = 0, t)$ vanishes, the lack of commutativity
of the Hamiltonian with itself at different times leads to the emergence of a term proportional to
$\sigma_z$ in the corresponding effective Hamiltonian~\cite{Kitagawa2011}.
As discussed below Eq.~(\ref{eq:Dirac2D}) and depicted in Fig.~\ref{fig:BandEngineering}a, such a term open gaps at the Dirac points.
In the Supplementary Material (SM), we provide a simple, explicit calculation that gives further insight into this mechanism of gap opening through an off-resonant drive.

Importantly, the signs of the $\sigma_z$ terms induced by the circularly polarized field are
opposite in the two valleys. Due to these opposite signs, the induced Berry curvatures {\it add}
together to yield Floquet bands with nonvanishing Chern numbers $\mathcal{C} = \pm
1$~\cite{Oka2009, Kitagawa2011}. The approximate magnitude of the induced gap can be inferred from
second-order perturbation theory: through virtual absorption and emission of a photon from the
driving field, a splitting of magnitude $\Delta_{F} = 2 \frac{v^2e^2A_0^2}{\hbar\omega}$ is
induced~\cite{Kitagawa2011}.

For a band insulator or semiconductor for which the bands are separated by an energy gap $\Delta$
in the absence of a drive, changing band topology requires inducing a band inversion.
As proposed in Ref.~\cite{Lindner2011}, 
such a band inversion can be created in the Floquet spectrum 
by using a {\it
resonant} drive with frequency larger than the band gap, $\Delta$, and smaller than the bandwidth,
$W$: $\Delta < \hbar \omega < W$.
For the 2D Dirac model of Eq.~(\ref{eq:Dirac2D}), in this regime the drive resonantly couples states on the $\vec{k}$-space ring $2E(\bk) = \hbar \omega$. 

The effect of the resonant drive is most easily visualized by moving to a rotating frame:
$\Ket{\psi_R(t)} = R(t)\Ket{\psi(t)}$, with $R(t) = e^{-i\omega t P_-}$, where $P_-$ is a projector
onto the (unperturbed) negative energy band of $H_{\rm 2D}$ in Eq.~(\ref{eq:Dirac2D}) with $\vec{A}
= 0$. As illustrated in Fig.~\ref{fig:BandEngineering}b, in the rotating frame the lower band is
shifted up in energy by $\hbar\omega$.
The driving field (which obtains a dc component in the rotating frame) induces a Rabi-like
splitting between the two bands, opening a Floquet gap proportional to $A_0$ all the way around the
resonant ring. Within the rotating wave approximation, remaining oscillating terms in the rotating
frame are discarded; these residual terms are responsible for multi-photon resonances, which we
will discuss in more detail below.

Importantly, as is evident in Fig.~\ref{fig:BandEngineering}b, the characters (valence-band-like or conduction-band-like) of the states near $\vec{k} = 0$ in the newly reconfigured upper and lower bands are swapped compared to those of the
non-driven system.
However, to determine whether or not a topological band inversion has occurred, it is necessary to check the behavior of $\hat{\vec{n}}(\vec{k})$ over the entire Brillouin zone. % by explicit calculation.
In order for a topological transition to occur, the Floquet band gap must close somewhere in the
Brillouin zone as frequency is swept from values $\hbar\omega < \Delta$, where no resonances are
encountered, to $\hbar\omega > \Delta$, where the bands may become inverted.
A gap closing occurs for the Dirac model (\ref{eq:Dirac2D}) if the matrix element coupling the two bands in the rotating frame vanishes at $\vec{k} = 0$. % in at least one valley.
For a gapped graphene or TMD-like system with two valleys as described above, with $\Delta > 0$ and
a LHP drive, a nonvanishing coupling persists at $\vec{k} = 0$ in valley $K$, such that the two
bands always repel and the gap never closes. In valley $K'$, however, the coupling vanishes at
$\vec{k} = 0$, thus allowing a gap closing and true band inversion to occur~\cite{Sie2015}. By
explicit calculation one can check that in this case the phase of the coupling winds {\it twice}
(i.e., through $4\pi$) as $\vec{k}$ goes around the resonance circle; as a result, the Chern
numbers of the Floquet bands differ by $\pm 2$ from those of the non-driven
system~\cite{Lindner2011}.
Similar considerations apply for $\Delta < 0$ or for a right-hand polarized drive. \\

\noindent \textit{Extensions and applications.---} Many aspects of Floquet engineering of
topological bands have been studied both theoretically and experimentally. Graphene
irradiated with mid-infrared circularly polarized light was shown to exhibit both of the mechanisms
described above \cite{Usaj2014,Quelle2016}. At lower frequencies, multi-photon resonances between
the valence and conduction bands affect both the magnitudes of and the numbers of edge states
traversing the Floquet gaps at $\varepsilon=0$ and $\varepsilon=\hbar\omega/2$ \cite{Gu2011, Rodriguez-Vega2018}; note
that, unlike in equilibrium, the Chern numbers of the bulk Floquet bands do not fully determine the
net chiralities of edge modes in each gap (see Sec.~\ref{sec:FloquetTopology}).
 Strong drive amplitudes may lead to
new topological transitions, which are not captured by the perturbative arguments above
~\cite{Delplace2013, Sentef2015}.
Further works have described new driving mechanisms, such as via Kekul\'{e} terms in honeycomb lattices~\cite{Iadecola2013}, as well as topological Floquet engineering in one~\cite{Jiang2011, Thakurathi2017, Kennes2018b} and three dimensions~\cite{Lindner2013, Wang2014, Chan2016a, Chan2016b, Huebener2017}. 
In cold atoms, topological gap opening in honeycomb lattices has been realized experimentally using circular ``shaking'' of the lattice~\cite{Jotzu2014, Flaschner2016}.
A variety of other approaches for inducing topological states in cold atoms, e.g., via imprinting artificial gauge fields, have been explored (for an overview, see Refs.~\cite{Dalibard2011, Goldman2014, Cooper2019}).
%%%%%%%%%%%%%%%%%%%%%%%%%%%%%%%%%%%%%%%%%%%%%%%%%%%%%%%%%%%%%%%%%%%%%%%%%%%%%%%%%%

%%%%%%%%%%%%%%%%%%%%%%%%%%%%%%%%%%%%%%%%%%%%%%%%%%%%%%%%%%%%%%%%%%%%%%%%%%%%%%%%%%
\section{Topology of Floquet bands}
\label{sec:FloquetTopology}

In simple cases, such as in the limit of high-frequency driving, the effective Hamiltonian approach
described above fully captures the topological features of Floquet-Bloch bands. For example, the
effective Hamiltonian may allow one to correctly infer the appearance of topological edge states at
sample boundaries by computing the standard non-driven system invariants for the Floquet
bands~\cite{Kitaev2009, Ryu2010}. However, there are many important and generic cases where this
approach fails; from a conceptual point of view, these interesting cases reveal the possibilities
for realizing qualitatively new types of topological phenomena in periodically-driven systems.

%%%%%%%%%%%%%%%%%%%%%%%%%%%%%%%%%%%%%%%%%%%%%%%%%%%%%%%%%%%%%%%%%%%%%%%%%%%
\begin{figure}[t]
\includegraphics[width=\columnwidth]{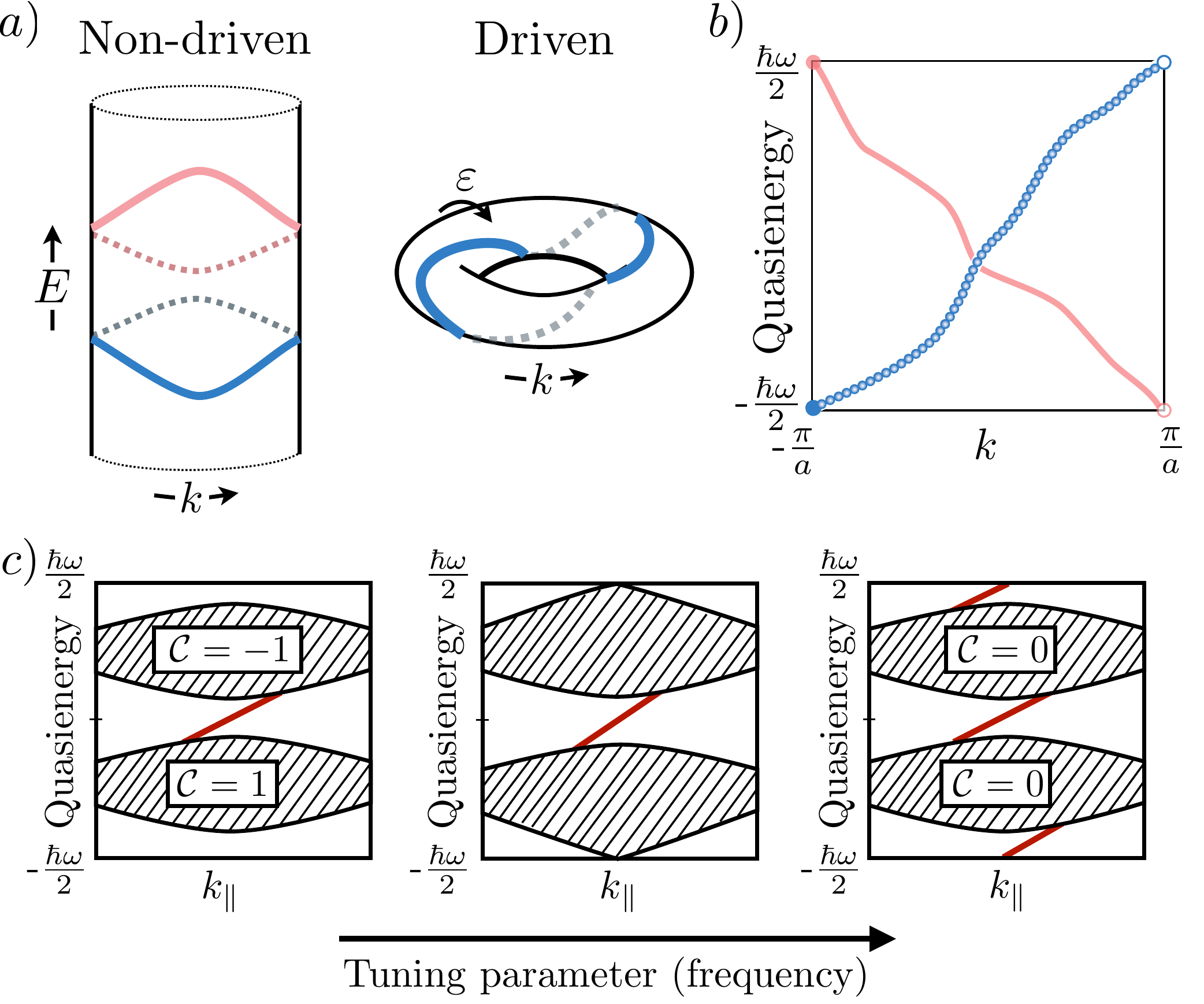}
\caption{Topological features of Floquet bands without analogues in non-driven systems.
a) In a non-driven system, the energy spectrum is real-valued. % and thus the band structure is defined on a cylinder.
With periodic driving, the quasienergy spectrum is periodic and thus % and thus the Floquet band structure is defined on a torus.
%Due to this periodicity,
Floquet bands may exhibit a topological ``quasienergy winding.'' %which has no analogue in non-driven systems.
b) A fully filled Floquet band with non-trivial winding carries a quantized pumped charge per driving cycle (see text).
%Due to the constraint that the net winding of all Floquet bands must vanish, the system necessarily hosts counterprogating modes.
Avoided crossings between counterpropagating chiral Floquet modes vanish exponentially in the adiabatic limit, thus allowing non-trivial pumping to be manifested.
c) Floquet zone-edge topological transition in a 2D system which in the absence of driving hosts bands with Chern numbers $\mathcal{C} = \pm 1$. % for its lower and upper bands.
For a driving frequency larger than the bandwidth, $W$, the drive has only a small quantitative effect on the band structure (left).
As the driving frequency is lowered, %the width of the quasienergy zone shrinks;
the top and bottom of the spectrum meet at the Floquet zone edge when $\hbar \omega = W$ (middle).
When the driving frequency is reduced further, a gap may open at the Floquet zone boundary, resulting in trivial Floquet bands with $\mathcal{C} = 0$ (right).
%For appropriate symmetry of the drive, the Chern numbers of both of the resulting Floquet bands may become trivial, $\mathcal{C}_F = 0$.
Despite the fact that the Chern numbers of all bands vanish, topologically-protected ``anomalous'' chiral edge states propagate along system boundaries.}
%\addMR{Clearance needed for last part of panel c from Winding PRX}}
\label{fig:FloquetTopology}
\vspace{-0.2 in}
\end{figure}
%%%%%%%%%%%%%%%%%%%%%%%%%%%%%%%%%%%%%%%%%%%%%%%%%%%%%%%%%%%%%%%%%%%%%%%%%%%
The main reason why an effective Hamiltonian may fail to capture the topological features of a
Floquet-Bloch system is that the spectrum of a Hamiltonian is defined on the whole real line, while
a Floquet spectrum is defined on a compact Floquet-Brillouin zone.
Due to the periodicity of quasienergy, i) there is no ``top'' and ``bottom'' of the spectrum, and 
ii) a continuous band of quasienergies may {\it wind} around the Floquet-Brillouin zone an integer
number of times as any crystal momentum component traverses the Brillouin zone (see
Fig.~\ref{fig:FloquetTopology}a).

The physical significance of quasienergy winding is transparently manifested for one-dimensional
(1D) systems. In a seminal work in 1983, Thouless showed that the charge pumped through a gapped 1D
system subjected to a cyclic, adiabatic modulation of parameters is quantized in units of the
fundamental charge, per driving cycle~\cite{ThoulessPump}. The quantization is of a topological
nature and is insensitive to the details of the driving cycle, provided that the gap remains open
(and thereby that adiabaticity is preserved) throughout the cycle.

In the language of Floquet bands, Thouless' quantized adiabatic charge pump precisely corresponds to the case where the quasienergies exhibit a nontrivial winding as the crystal momentum $k$ runs from $-\pi/a$ to $\pi/a$, where $a$ is the lattice constant of the system, see Fig.~\ref{fig:FloquetTopology}b. 
The quantization of pumped charge follows from the fact that the average group velocity of a
Floquet band, $\bar{v}_g$, is proportional to the quasienergy winding number of the band,
$\mathcal{W}$: $\bar{v}_g = \frac{a}{2\pi\hbar}\oint dk\, \frac{d\epsilon}{dk} = a\mathcal{W}/T$.
Thus, for a fully-filled band, the average particle displacement over one period, $\overline{\Delta
x} = \bar{v}_g T$, is equal to the integer $\mathcal{W}$ times the lattice constant, $a$: in each
cycle, $\mathcal{W}$ units of charge are pumped through the system~\cite{Kitagawa2010}.

The notion of quasienergy winding suggests a natural form for a topological invariant of the
Floquet operator $U_{\rm 1D}(k, T) \equiv \mathcal{T}e^{-(i/\hbar)\int_0^T dt\, H_{\rm 1D}(k,t)}$,
where $H_{\rm 1D}(k, t)$ is the time-periodic Bloch Hamiltonian of the 1D
system~\cite{Kitagawa2010}: $\nu_1 = \frac{1}{2\pi}\oint dk\, {\rm Tr}[U_{\rm 1D}(k, T)^\dagger i
\partial_k U_{\rm 1D}(k, T)]$. This invariant is a special case of the ``GNVW'' index, defined for
a generic local unitary time step operator in a 1D system (see Ref.~\cite{GNVW} for details).
Physically, $\nu_1$ counts the net winding number of all of the Floquet bands of $U_{\rm 1D}(k,
T)$; according to the arguments above, it thus captures the net flow of particles generated by the
evolution in a fully-filled system. Interestingly, the three dimensional generalization of this
winding number~\cite{Kitagawa2010} is linked to magneto-electric pumping~\cite{Higashikawa2018,
Sun2018}.

Importantly, while there are local unitary operators for which $\nu_1$ (or the GNVW index) can take
nonzero values, these indices must vanish for any unitary evolution  arising from %continuous
time evolution under a finite, local Hamiltonian (in strictly one dimension), see SM~\cite{SM}. For
the Thouless pump, this implies that the spectrum must host bands with winding numbers of 
opposite sign (e.g., see Fig.~\ref{fig:FloquetTopology}b), which necessarily cross.
Generically, these crossings are avoided due to hybridization, yielding bands with trivial winding ($\mathcal{W} = 0$). 
However, these hybridization gaps close {\it exponentially} in
$1/\omega$ as $\omega \to 0$, allowing chiral bands (with $\mathcal{W} \neq 0$) 
to emerge in the adiabatic limit.

In addition to opening the possibility for quasienergy winding of individual Floquet bands, the
periodicity of quasienergy also provides new channels through which topological transitions can
occur. In a non-driven system, a topological transition occurs through the closing and reopening of
the band gap as a parameter (such as quantum well width in the model of Bernevig, Hughes, and
Zhang~\cite{Bernevig2006}) is tuned through a critical value. In Floquet systems there is an
additional gap at the Floquet-Brillouin zone edge ($\epsilon = \pm \hbar \omega/2$), which allows
the closing and reopening of a degeneracy between the {\it bottom} of the lowest Floquet band and
the {\it top} of the highest band, for a given choice of Floquet zone.

In Fig.~\ref{fig:FloquetTopology}c we illustrate a topological transition that occurs through the
Floquet zone edge.
In the absence of driving, the system hosts topologically non-trivial bands with Chern numbers $\pm
1$. With driving, the Chern numbers of both of the resulting Floquet bands become trivial,
$\mathcal{C} = 0$.
Crucially, despite finding trivial topological indices for the Floquet bands of $U(T)$, the system
remains topologically nontrivial. To see this, consider the fate of topological edge states that
existed in the gap of the system before the driving was introduced. Throughout the sequence of
driving frequencies considered in Fig.~\ref{fig:FloquetTopology}c, the original gap (between the
bottom of the conduction band and the top of the valence band) never closed; therefore, the
transition to the configuration with $\mathcal{C} = 0$ cannot cause the original edge states to
disappear. From general arguments about spectral flow, the net chiralities of edge states above and
below a given band must be equal to the Chern number of the band~\cite{Halperin1982,Hatsugai1993}.
Thus we conclude that the system must host chiral edge states in {\it both} of its gaps, near $\epsilon
= 0$ and $\epsilon = \pm \hbar\omega/2$, despite having trivial Chern indices. In such cases, the
nontrivial topology of the Floquet system is encoded not in the Floquet spectrum itself, but rather
in the micromotion that occurs within the driving period~\cite{Rudner2013, Carpentier2015,
Nathan2015, Roy2017, Roy2017b}. Topological edge states that appear in systems with trivial Floquet
operators are referred to as ``anomalous edge states.''

As the example above shows, the Chern numbers of Floquet bands do not provide sufficient
topological information to predict the absolute numbers of chiral edge states that appear within
each gap. This is true both for ``anomalous'' phases with vanishing Chern numbers, described above,
as well for systems with non-vanishing Chern numbers (in particular, this subtlety generically
arises when single or multi-photon resonances are involved, see Sec.~\ref{sec:FloquetEngineering}).

Analogous to the 2D case described above, topological transitions due to gap closings at the
Floquet zone boundary in 1D systems can lead to new types of topological bound states at sample
edges, with protected quasienergy values of 0 and/or $\hbar\omega/2$~\cite{Broome2010, Jiang2011,
Liu2013, Asboth2014, DalLago2015}. When both of such ``0'' and ``$\pi$'' modes are present simultaneously, the
spectrum exhibits a protected quasienergy splitting of $\hbar\omega/2$ that gives rise to robust
oscillations with precisely {\it twice} the driving period (see Sec.~\ref{sec:FloquetMBL} for the
connection to ``Floquet time crystals''). In the superconducting case, these ``$\pi$-Majorana''
modes may yield new routes to braiding non-Abelian particles in a strictly one-dimensional
system~\cite{Bauer2018}.

As in non-driven systems, the time-reversal, particle-hole, and chiral symmetries that distinguish
topological classes can also be implemented in Floquet systems~\cite{Kohler2005, Kitagawa2010,
Asboth2013, Asboth2014, PlateroPRL, Lababidi2014, Zhou2014, Carpentier2015, Nathan2015, Roy2017,
Wang2017, Harper2019}.
A natural way to ensure that the Floquet operator $U(T)$ 
exhibits particle-hole symmetry is to require that the instantaneous Hamiltonian (and hence the
evolution operator $U(t)$ for all $0 \le t < T$) itself possesses particle-hole
symmetry~\cite{Jiang2011, Nathan2015}. This condition is automatically satisfied for the
Bogoliubov-de Gennes evolution describing any superconducting system. Time-reversal and chiral
symmetries can be implemented by enforcing ``time non-local'' symmetries on the
evolution~\cite{Kohler2005, Asboth2013, Carpentier2015, Nathan2015}: $U(t)  = S U(t_* -
t)U^\dagger(t_*)S^{-1}$, where $t_*$ denotes a special point in the driving cycle, and chiral
(time-reversal) symmetry is ensured by taking $S$ is to be a unitary (anti-unitary) operator. With
the Altland-Zirnbauer symmetry classes implemented in this way for the driven case, the same
classes allow for topologically nontrivial bands with and without driving~\cite{Nathan2015,
Roy2017}. Importantly, however, driven systems support a more rich set of possibilities within each
nontrivial symmetry class, due to the possibility of different micromotion phases which may host
anomalous edge states (see above)~\cite{Rudner2013, Carpentier2015, Nathan2015, Roy2017}.
Moreover, novel types of nonsymmorphic spacetime symmetries (combining time translations with spatial symmetry operations) may also protect topological features of Floquet-Bloch bands~\cite{Morimoto2017, Xu2018, Peng2018}.\\

\noindent {\it Realizations of anomalous topological Floquet bands.---} The unique aspects of
topology in Floquet systems, without analogues in equilibrium, have been investigated for optical,
cold atomic, and solid state systems. In the optical domain, one-dimensional discrete time quantum
walks with chiral symmetry provided the first experimental demonstration of anomalous ``0'' and
``$\pi$'' edge states~\cite{Broome2010}. Waveguide arrays with periodic modulations
along the propagation axis~\cite{Rechtsman2013}, as well as microwave resonator
arrays~\cite{Hu2015} have also been used to demonstrate the emergence of anomalous topological
Floquet edge modes in one- and two-dimensions~\cite{Hu2015, Mukherjee2017, Maczewsky2017, Cheng2019}. In cold atoms, one-dimensional
Thouless pumps realizing nontrivial quasienergy winding were recently realized, in
fermionic~\cite{Nakajima2016} and bosonic~\cite{Lohse2016} systems. Realizations of 2D anomalous
phases (Fig.~\ref{fig:FloquetTopology}c) via periodic modulation of an optical honeycomb lattice
were proposed in~\cite{Quelle2017}. Solid state realizations of ``0'' and ``$\pi$'' Majorana modes
have also been proposed for superconducting nanowires~\cite{Jiang2011} and Josephson
junctions~\cite{Liu2019}.

%%%%%%%%%%%%%%%%%%%%%%%%%%%%%%%%%%%%%%%%%%%%%%%%%%%%%%%%%%%%%%%%%%%%%%%%%%%%%%%%%%

%%%%%%%%%%%%%%%%%%%%%%%%%%%%%%%%%%%%%%%%%%%%%%%%%%%%%%%%%%%%%%%%%%%%%%%%%%%%%%%%%%

\section{Many-body dynamics and observables in Floquet systems}
\label{sec:ManyBody}

The discussion in Sections~\ref{sec:FloquetEngineering}--\ref{sec:FloquetTopology} focused on
single particle Floquet-Bloch bands in the presence of a drive.
Whether the Floquet bands provide a useful starting point for describing the dynamics of the system depends on the extent to which the physical properties of  %many-body
the system are simply described within this basis. The remainder of this review is devoted to
characterizing the physical properties and observables of periodically driven many-body systems,
and the conditions under which driving allows novel topological phenomena to be observed via
transport or spectral measurements.

\subsection{Mesoscopic transport in  Floquet systems} %topological?
\label{sec:Transport}
In this subsection we discuss mesoscopic transport through %non-interacting
Floquet systems. We consider a setup
where a driven electronic system is connected to two or more leads, where the driving acts  only on
the system and does not directly affect the leads. The electrons within each lead $\lambda$ are
assumed to be described by a Fermi-Dirac distribution with chemical potential $\mu_\lambda$ and
temperature $T$.
We focus on the regime where the transit time for an electron through the system is short compared with the timescales % system size is small compared with the mean free paths
 associated with electron-electron scattering (effectively assumed to be absent) and inelastic scattering due to coupling to phonons or other environmental degrees of freedom.
Under these conditions, the electronic state within the system is fully controlled by the
distributions in the reservoirs, and the system's two- or multi-terminal conductance is given by a
``Floquet-Landauer'' formula (see below).
For a general review of driven transport in this regime, see Ref.~\onlinecite{Kohler2005}. In
Sec.~\ref{sec:OpenSystems}, we will discuss the local {\it conductivity} of FTIs (obtained from the
Kubo formula); in that approach, the steady state occupation of the Floquet states must be
determined explicitly by considering the coupling of the driven system to its environment.

The Floquet-Landauer approach has been used to study transport through a variety of topological driven systems~\cite{Gu2011, Kitagawa2011, Kundu2013, Usaj2014, FoaTorresMultiTerminal, Farrell2015, Pereg-Barnea2016, Kundu2017}. 
For a nondriven system, the Landauer formula relates current to the Fermi distributions of the leads and to the energy-dependent transmission probability (for the case of two-terminal transport), $T(\mathcal{E})$, where $\mathcal{E}$ is the energy of the incoming (and outgoing) states. 
Importantly, in a driven system, electrons may coherently absorb or emit energy in multiples of the
driving field quantum $\hbar \omega$ as they pass through the system. In a Floquet system, the
transmission probabilities ${T}_{\rm RL}^{(k)}(\mathcal{E})$ and ${T}_{\rm LR}^{(k)}(\mathcal{E})$
between the left (L) and right (R) leads thus depend on the incident energy $\mathcal{E}$ and an
additional integer-valued index, $k$, which counts the net number of photons absorbed during the
electron's transit through the system. (The difference between energies of the incoming and
outgoing states is equal to $k\hbar\omega$.) In terms of the Floquet transmission probabilities,
the two terminal current is given by:
\begin{eqnarray}
I = 2\pi\! \int_{-\infty}^{\infty}\!\! d\mathcal{E}\sum_k \left\{ T^{(k)}_{\rm RL}(\mathcal{E}) f_{\rm L}(\mathcal{E}) - T^{(k)}_{\rm LR}(\mathcal{E}) f_{\rm R}(\mathcal{E})\right\}\!, 
\end{eqnarray}
where $f_{\rm L}(\mathcal{E})$ and $f_{\rm R}(\mathcal{E})$ are the Fermi-Dirac distributions in
the left and right leads, respectively.

As described in Ref.~\onlinecite{Kohler2005}, the transmission probabilities $T^{(k)}_{\rm RL}(\mathcal{E})$ 
and $T^{(k)}_{\rm LR}(\mathcal{E})$ can be expressed in terms of the Floquet Green's function of the open, driven system (including the effects of its coupling to the leads). 
Through this connection, transport through the driven system can be related to its quasienergy
spectrum; in particular, this formalism (and its generalization for multi-terminal transport)
provides the basis for describing the transport signatures of Floquet gap opening and the
appearance of topological Floquet edge modes~\cite{Gu2011, Kitagawa2011, Kundu2013, Usaj2014,
FoaTorresMultiTerminal, Farrell2015, Pereg-Barnea2016, Kundu2017}.

In Ref.~\onlinecite{Gu2011}, Gu and coworkers investigated how topological gap opening in
graphene~\cite{Oka2009, Kitagawa2011} is reflected in two-terminal transport. They considered the
case of a low-frequency driving field, with $\hbar\omega$ much less than the bandwidth of the
system; in this case, the graphene band structure must be ``folded down'' several times in order to
fit within a single quasienergy Brillouin zone. Therefore, in addition to the Floquet gap opened at
the original Dirac point of the non-driven system, a hierarchy of increasingly small gaps
(corresponding to even-order multiphoton resonances) and associated edge states appear near zero
quasienergy~\cite{Perez-Piskunow2015}. Transport through the edge states appearing at zero
quasienergy yields a non-zero conductance in the large system size limit. However, in contrast to
equilibrium quantum Hall systems, this conductance is not quantized~\cite{Gu2011}.

%%%%%%%%%%%%%%%%%%%%%%%%%%%%%%%%%%%%%%%%%%%%%%%%%%%%%%%%%%%%%%%%%%%%%%%%%%%
\begin{figure}[t]
\includegraphics[width=\columnwidth]{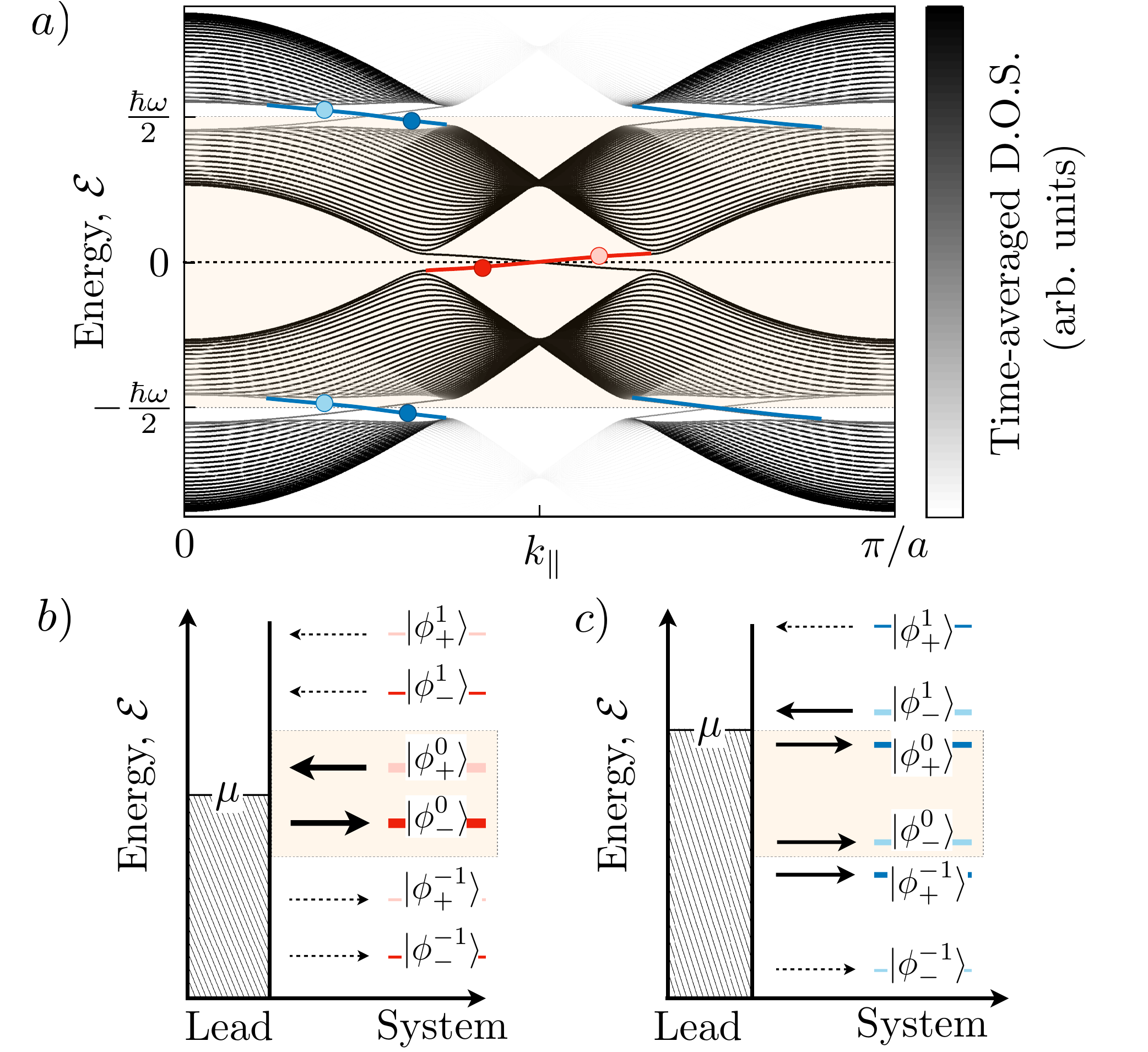}
\caption{Edge state transport in FTIs.
a) Time-averaged spectral function of a graphene strip driven by circularly polarized light, as a function of the crystal momentum component $k_\parallel$ parallel to the strip. 
The highlighted region marks the Floquet Brillouin zone.
The time-averaged spectral function helps to visualize the frequency content of the Floquet states, which governs their coupling to external leads.
b) Tunneling between a Fermi reservoir and edge states $\Ket{\psi_-}$ and $\Ket{\psi_+}$ at two values of $k_\parallel$ in the gap around $\mathcal{E} = 0$ (dark and light red dots in panel a). The dominant Fourier components of $\Ket{\psi_-}$ and $\Ket{\psi_+}$ are centered near the chemical potential, $\mu$, which is set to $\mathcal{E} = 0$.
The filled and empty states of the lead near $\mathcal{E} = 0$ predominantly fill (empty) the Floquet states $\Ket{\psi_-}$ and $\Ket{\psi_+}$ at negative (positive) quasienergies.
However, the harmonic components $\Ket{\phi^{(\pm 1)}_-}$ and $\Ket{\phi^{(\pm 1)}_+}$ near $\mathcal{E} \approx \pm \hbar\omega$ [see Eq.~(\ref{eq: Floquet expansion})] open channels for electrons to tunnel into (out of) the nominally empty (filled) states, as indicated by the small dotted arrows.
These additional photon-assisted tunneling processes spoil the ideal filling of the Floquet edge modes and the perfect quantization of edge transport.
c) Edge states in the gap around $\mathcal{E} = \pm \hbar\omega/2$ (dark and light blue dots in panel a) 
are comprised of roughly equal superpositions of the valence and conduction bands of the non-driven system.
When the chemical potential of the lead is centered in the gap at $\mathcal{E} = \hbar\omega/2$,
%Due to their large valence-band components,
all states in the edge mode experience significant coupling to filled states of the lead due to their large valence-band components.
As a result, 
these chiral Floquet edge modes do not simply mediate quantized transport~\cite{FoaTorresMultiTerminal, Pereg-Barnea2016}.}
\label{fig:FloquetTransport}
\vspace{-0.2 in}
\end{figure}
%%%%%%%%%%%%%%%%%%%%%%%%%%%%%%%%%%%%%%%%%%%%%%%%%%%%%%%%%%%%%%%%%%%%%%%%%%%
This lack of quantization was further elucidated in Refs.~\cite{FoaTorresMultiTerminal,
Pereg-Barnea2016}. As pointed out in these works, the ability of a Floquet edge mode to transmit a
current between source and drain contacts depends crucially on the matching of energies between the
(nondriven) states in the leads and the energies of the harmonic (sideband) components
$\ket{\phi_{\varepsilon}^m}$ in the expansion \be \ket{\psi_\varepsilon(t)}=e^{-i \varepsilon
t}\sum_n e^{-im \Omega t}\ket{\phi_{\varepsilon}^m} \label{eq: Floquet expansion} \ee
for the Floquet edge mode $\ket{\psi_\varepsilon(t)}$. 

The time-averaged spectral function (see Refs.~\onlinecite{Usaj2014, Uhrig2019}, and Supplementary Material
\cite{SM} for definition and discussion) provides a helpful way of visualizing the spectral weight of the Floquet
states and how they couple to states at specific {\it energies} in the non-driven leads.
We illustrate the time-averaged spectral function for graphene under circularly polarized light in Fig.~\ref{fig:FloquetTransport}.
As shown, the Floquet edge states in the gap opened at the Dirac points are built primarily
from states near zero energy and can efficiently couple to states in the lead near $\mathcal{E} =
0$. However, some of the spectral weight of these Floquet edge states is shifted to sidebands near
energies $n \hbar \omega$, for integer $n \neq 0$ (see Fig.~\ref{fig:FloquetTransport}b). 
The weights of these sidebands are controlled by the ratio of the drive amplitude to the
photon energy.

Importantly, sidebands corresponding to $n \neq 0$ primarily couple to states of the leads whose
energies are far from the chemical potential (set near $\mathcal{E} = 0$ to probe transport through
the drive-induced topological edge states). As a result, the differential conductance resulting
from transport through the edge states deviates from the quantized value of $2 e^2/h$ per mode~\cite{Gu2011}.
Similar considerations apply to edge states in the driving-induced gap appearing at
the Floquet zone edge, $\epsilon = \pm \hbar\omega/2$, and to ``higher-order'' edge states that
form in mini-gaps in the quasienergy spectrum due to multiphoton resonances; such states carry
significant spectral weight at energies far from the chemical potentials of the leads and therefore
give nonuniversal contributions to the differential conductance.
Thus 
in general there is no simple relation between differential conductance at a given value of lead
chemical potential and the number of chiral edge states in a corresponding quasienergy
gap~\cite{FoaTorresMultiTerminal, Pereg-Barnea2016}.

In some cases, new paradigms for quantized transport (beyond simple zero-bias conductance) have
been found for driven systems. For example, the appearance of Floquet-Majorana edge modes at the
end of a periodically-driven one-dimensional superconductor does not give rise to a quantized
differential conductance at zero bias, as in the equilibrium case~\cite{Sengupta2001, Law2009};
rather, for a system with a Floquet-Majorana mode at quasienergy $\epsilon = 0$ or $\hbar\omega/2$
(i.e., a ``Floquet $\pi$-Majorana mode''), the differential conductance $\sigma(\mu)$ {\it summed
over the discrete set of lead chemical potentials (or back gate potentials) $\mu_n = \epsilon + n\hbar\omega$}, for all integers $n$, yields
the quantized value $2e^2/h$~\cite{Kundu2013}.
Inspired by these results, the authors of Ref.~\onlinecite{Pereg-Barnea2016} showed that perfect quantization of the edge state mediated conductance in a 2D FTI is also recovered through an analogous sum rule.

As discussed above, the presence of chiral Floquet edge modes around the perimeter of a 2D system
does not guaranty a quantized Hall conductance at low bias. 
However, the ``anomalous'' 2D phase described in Sec.~\ref{sec:FloquetTopology}
and depicted in the rightmost panel of Fig.~\ref{fig:FloquetTopology}c offers a new possibility:
quantized {\it current} at large source-drain bias. Specifically, in the presence of disorder, the
anomalous 2D system's Floquet bands (all with Chern number zero) become fully localized, while the
system's chiral edge modes persist~\cite{AFAI}. Crucially, whereas in a non-driven system delocalized states
are needed to provide a spectral termination for chiral edge modes, in a Floquet system a chiral
edge state can ``wind'' around the quasienergy Brillouin zone and terminate on itself (analogous to
the 1D winding bands shown in Figs.~\ref{fig:FloquetTopology}a and b). 
At large source-drain bias, one of the chiral Floquet modes of this ``anomalous Floquet Anderson insulator'' (AFAI) 
may become completely filled,
while the mode propagating in the opposite direction is completely empty. In this situation, the
system hosts a quantized current~\cite{AFAI, Kundu2017}, for analogous reasons to those described
for the Thouless pump in Sec.~\ref{sec:FloquetTopology}. In contrast to the situation of a truly 1D
system as shown in Fig.~\ref{fig:FloquetTopology}b, the counterpropagating chiral edge modes of a
2D AFAI are spatially separated and therefore exhibit exponentially small splittings in the sample
{\it width}, even at finite driving frequency. This property allows the 2D AFAI, with a fully
localized bulk, to carry a quantized current, even in the non-adiabatic driving regime~\cite{AFAI, Kundu2017}.
We note that signatures of light-induced chiral edge states may also appear in magnetization~\cite{Dahlhaus2015, Nathan2017_magnetization}.

%%%%%%%%%%%%%%%%%%%%%%%%%%%%%%%%%%%%%%%%%%%%%%%%%%%%%%%%%%%%%%%%%%%%%%%%%%%%%%%%%%

%%%%%%%%%%%%%%%%%%%%%%%%%%%%%%%%%%%%%%%%%%%%%%%%%%%%%%%%%%%%%%%%%%%%%%%%%%%%%%%%%%
\subsection{Transient dynamics and prethermalization}
\label{sec:Prethermalization}

Early experiments on FTIs in solid state systems have focused on short-time dynamics through
pump-probe measurements~\cite{Wang2013,Mahmood2016, McIver2018}. By focusing on short times, the
formation of Floquet-Bloch states can be probed separately from the question of the system's fate
at long times. In Ref.~\cite{Wang2013}, Wang and coworkers demonstrated topological Floquet gap
opening on the surface of the three-dimensional TI Bi$_2$Se$_3$ via time-resolved angular-resolved
photoemission spectroscopy (tr-ARPES). The temporal emergence of Floquet bands in such pump-probe
settings, as manifested in the tr-ARPES signal, has been theoretically studied in
Refs.~\cite{Fregoso2013, Sentef2015, Farrell2016, Kandelaki2017}.
Very recently, McIver and coworkers observed the photo-induced Hall effect in graphene subjected to circularly polarized light~\cite{Oka2009, Kitagawa2011} using a newly-developed high speed time-resolved transport setup~\cite{McIver2018}. % \mpar{TMDs \cite{Giovannini2016}?}\\

Although at long times closed, interacting Floquet systems generically heat towards featureless high entropy-density states, 
 the question of {\it when} this heating sets in is of crucial importance.
Under appropriate conditions, the timescale over which runaway heating occurs may become extremely
long~\cite{BukovReview, Eckardt2015, Kuwahara2016, Abanin2017, Else2017, Lindner2017, Mori2018}. In
such cases, on short to intermediate timescales the system may attain a (nearly) time-periodic
``quasisteady state'' that hosts topological phenomena arising from the underlying Floquet band
structure.

%%%%%%%%%%%%%%%%%%%%%%%%%%%%%%%%%%%%%%%%%%%%%%%%%%%%%%%%%%%%%%%%%%%%%%%%%%%
\begin{figure}[t]
\includegraphics[width=\columnwidth]{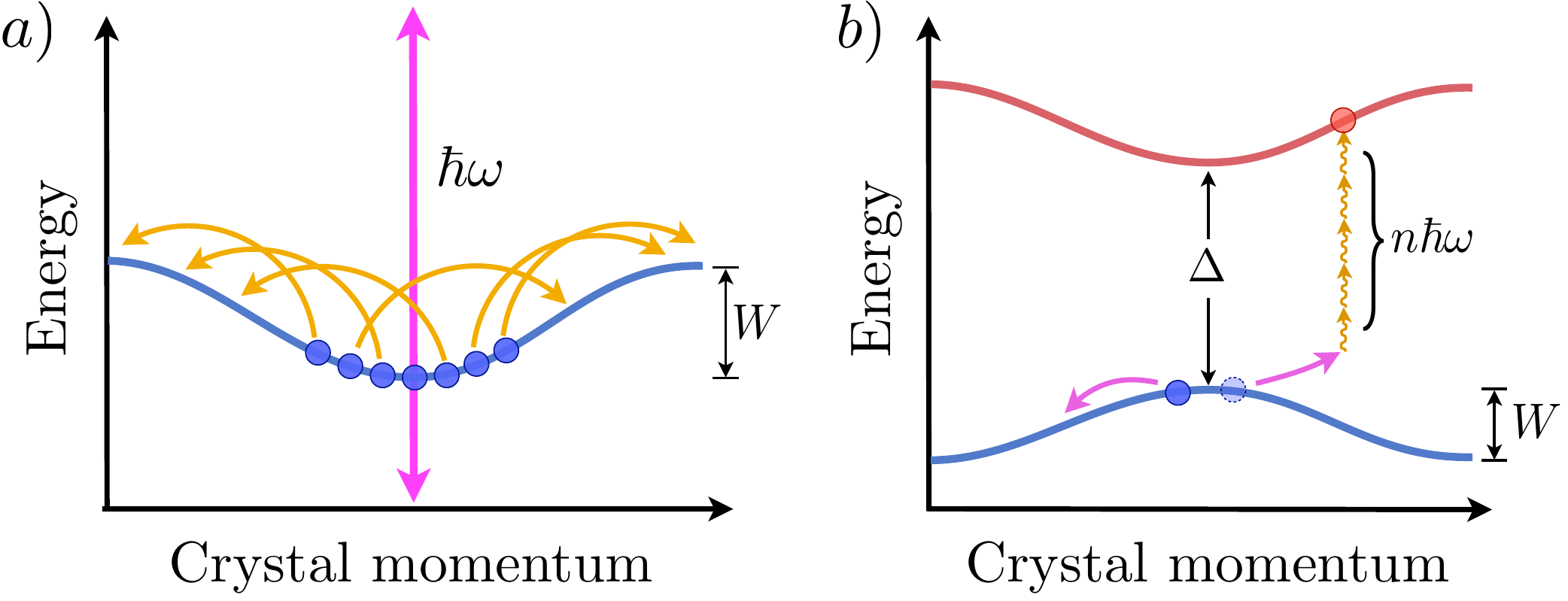}
\caption{Heating via energy absorption from the drive.
a) When the drive frequency is large compared with the single-particle bandwidth, $W$, and the energy scale of local interactions, $U$, absorption of one driving field photon of energy $\hbar\omega$ requires a high-order rearrangement of particles in the system.
For weak interactions, $U \ll W$, the order of the process is controlled by the ratio $\hbar\omega/W$, which appears as a {\it power} in the corresponding transition rate.
Therefore the absorption rate is exponentially suppressed in $\omega$ for $\hbar\omega \gg W, U$.
b) Consider a system comprised of low-energy and high-energy bands, separated by a gap $\Delta$ that is much larger than the bandwidth of the low-lying bands, $W$. 
For $W, \hbar \omega, U \ll \Delta$, transitions across the single-particle gap $\Delta$ can only be excited through processes involving high-order multiphoton absorption and/or rearrangements within the system.
The minimal order of the excitation process is governed by the ratio $\Delta / {\rm max}\{\hbar\omega, W, U\}$; for $W,U \lesssim \hbar\omega$ this implies that the rate for exciting the high-energy degrees of freedom is exponentially suppressed in $1/\omega$.
}
\label{fig:FloquetHeating}
\vspace{-0.2 in}
\end{figure}
%%%%%%%%%%%%%%%%%%%%%%%%%%%%%%%%%%%%%%%%%%%%%%%%%%%%%%%%%%%%%%%%%%%%%%%%%%%

One important regime where heating rates can be suppressed is the limit where the drive frequency
is large compared with the relevant energy scales in the system.
As illustrated in Fig.~\ref{fig:FloquetHeating}a, in this regime the driving field photon energy is much larger than the energy change in any single scattering event in the system; energy absorption from the drive requires high-order processes involving many-particle rearrangements (so-called ``many-body resonances''~\cite{Bukov2016}), and is therefore heavily suppressed. % for high frequencies.

In an idealized setting where the system has a finite single-particle bandwidth (e.g., in a lattice
with a finite number of orbitals per unit cell), the heating rate may become exponentially
suppressed in $\omega$~\cite{Abanin2015}.
However, 
in any real physical system the kinetic energy is not bounded from above and higher bands will
inevitably be present.
The idealized high frequency limit is therefore never truly realized: the photon energy
$\hbar\omega$ (or its integer multiples) necessarily directly connects low-lying  bands of interest
with higher energy states. The lifetime of the quasisteady state is therefore also limited by
interband transitions to these high energy states. At high driving frequencies the rates of these
interband transitions are expected to be at least suppressed as a power law in $1/\omega$ (see,
e.g., Ref.~\cite{Reitter2017}), ensuring that a useful high frequency regime does exist (as
prominently exploited, e.g., in experiments on cold atoms in optical lattices).

When the high-lying bands are separated from a set of low-lying bands by a large single-particle
energy gap $\Delta$, the rates of interband transitions (i.e., with particles excited across the
gap) may be {\it exponentially} suppressed for {\it low} driving frequencies, $\hbar\omega/\Delta
\ll 1$~\footnote{Note that this low-frequency driving regime does not imply adiabaticity; while a
large gap in the single spectrum is required, the system need not have a {\it many-body} gap.}.
In this regime, many photons must be absorbed (or many particles rearranged within the low-energy
sector) in order for one excitation in the high-energy subspace to be created
(Fig.~\ref{fig:FloquetHeating}b).
The minimal order in perturbation theory at which a real transition can occur is thus determined by
the ratio of the gap, $\Delta$, to the energy change associated with a single scattering event.
This latter scale is governed by the local interaction energy scale, $U$, the single-particle bandwidth of
the low-lying bands, $W$, and the driving field photon energy, $\hbar \omega$. Thus we may expect
an exponential suppression of the rate of excitations across the gap $\Delta$ to the high-energy
subspace, with a power controlled by the ratio $\Delta/{\rm max}\{W, \hbar\omega,
U\}$~\cite{Lindner2017}.

During a time window in which scattering processes involving energy absorption from the drive are
absent, one may find a rotating frame in which the system evolves according to a {\it static},
effective Hamiltonian~\cite{BukovReview, Eckardt2015, Kuwahara2016, Abanin2017, Abanin2017a, Else2017}, $H_*$.
In this rotating frame, an ergodic system may thus ``pre-thermalize'' to an equilibrium-like state
with respect to $H_*$.
Crucially, the prethermal effective Hamiltonian is not simply the time-averaged system Hamiltonian,
and may exhibit important qualitative differences from the Hamiltonian of the non-driven system
(e.g., gap opening in graphene).

Ensuring a long lifetime for the quasisteady state is important for the realization of Floquet
topological insulators in the presence of interactions, and is particularly crucial for realizing
strongly-correlated topological phases of matter in Floquet systems. A number of proposals have
been made in this direction, including the realization of a fractional Chern
insulator~\cite{Regnault2011} phase in strongly driven graphene-like systems~\cite{Grushin2014} as
well as interesting magnetic phases in optically-driven Mott insulators~\cite{Claassen2017,
Balents2018}. In such settings, the ramp-up of the drive should furthermore be optimized to achieve
prethermal states with suitably low energy density~\cite{Ho2016,Kennes2018}.

For systems with a clear separation of scales, $W,U \ll \Delta$, the lifetime of the prethermal state can be optimized by working in the regime $W,U \ll \hbar\omega \ll \Delta$ where both the high- and low-frequency conditions (with respect to $W,U$, and to $\Delta$, respectively) are simultaneously satisfied. 
For cold atoms, such a situation may be arranged, for example, by working with a very deep optical
lattice. 
Heating rates for cold atoms in
Floquet-Bloch bands in optical lattices have recently been experimentally characterized, e.g.,  in Refs.~\cite{Jotzu2014, Reitter2017, Gorg2018, Singh2018}.

A qualitatively different quasisteady regime occurs 
for $\hbar \omega \lesssim W, U  \ll \Delta$. 
The conditions $\hbar \omega, W, U \ll \Delta$ ensure that {\it interband} transitions are
suppressed.
However, the system may still rapidly absorb energy and heat up {\it within} the low- and high-energy sectors. 
After an initial relaxation time, the system reaches a ``restricted'' infinite-temperature-like quasisteady state, with maximal local entropy density subject to the constraint of fixed particle number in each sector. 
Remarkably, such states may display robust, topological behavior resulting from the uniform
occupation of modes in each Floquet band, which follows from entropy density maximization.
Non-universal features of the system's Floquet spectrum 
are washed out, leaving the global, topological properties on display.
For example, in a 1D system with winding quasienergy bands as in Fig.~\ref{fig:FloquetTopology}b, the uniform occupation distribution implies that each Floquet band carries a universal current $I = \rho \mathcal{W}/T$, where $\rho$ and $\mathcal{W}$ are the particle density and quasienergy winding number of the band, and $T$ is the driving period~\cite{Lindner2017, Gulden2019}. 

%%%%%%%%%%%%%%%%%%%%%%%%%%%%%%%%%%%%%%%%%%%%%%%%%%%%%%%%%%%%%%%%%%%%%%%%%%%%%%%%%%

%%%%%%%%%%%%%%%%%%%%%%%%%%%%%%%%%%%%%%%%%%%%%%%%%%%%%%%%%%%%%%%%%%%%%%%%%%%%%%%%%%
\subsection{Floquet-MBL}
\label{sec:FloquetMBL}

In well-isolated systems such as cold atoms, the phenomenon of many-body localization
(MBL)~\cite{Basko2006, Oganesyan2007, Pal2010, NandkishoreReview, Lazarides2015, Ponte2015,
Bordia2017} may completely eliminate the system's tendency to absorb energy from the drive,
allowing nontrivial stationary states to persist even in the long time limit.
From a conceptual point of view, the possibility of stabilizing long-time steady states of closed, periodically-driven many-body systems enables the sharp delineation of {\it intrinsic} phases in Floquet systems~\cite{Khemani2016}; importantly, %in both driven and non-driven systems,
MBL is compatible with a variety of symmetry-breaking and topological
orders~\cite{NandkishoreReview, Moessner2017,vonKeyserlingk2016, Potter2016, Else2016, Harper2017}.
Interestingly, FTIs which exhibit protected edge states at both ``0'' and ``$\pi$'' quasienergy are
closely connected to Floquet-MBL phases (dubbed ``Floquet time crystals'') that break the discrete
time-translation symmetry of the system down to an integer multiple of the driving
period~\cite{Wilczek2012, Khemani2016, Else2016b, Else2017, Zeng2017, Zhang2017, Choi2017, Dykman2018}.

For FTIs in two dimensions and above, the necessity of compatibility with MBL strongly restricts the possibilities for finding stable intrinsic FTI phases in closed systems. In equilibrium, systems with nontrivial topological indices (such as Chern numbers) necessarily host delocalized states at certain energies. 
These delocalized states are fundamentally connected with the chiral or helical modes that appear
at system boundaries, and cannot be destroyed by disorder (without inducing a topological
transition into a trivial phase). Crucially, in Ref.~\onlinecite{Nandkishore2014}, Nandkishore and
Potter showed that the presence of delocalized states, even at high energy, generically destabilizes
MBL at all energy densities. This leads to the conclusion that FTIs in 2D and 3D whose Floquet
bands carry non-trivial topological indices are intrinsically unstable in the absence of a bath.
However, this argument does not rule out the stability of {\it anomalous Floquet
phases}~\cite{vonKeyserlingk2016, Potter2016, Else2016, Harper2017, Po2016, Po2017}. These phases
feature topologically-protected edge modes coexisting with trivial Floquet bands, and therefore need
not host delocalized bulk states, see Sec.~\ref{sec:Transport} and Ref.~\onlinecite{AFAI}.
In Ref.~\cite{Nathan2017}, the
nontrivial intra-period micromotion that characterizes these phases was shown to be consistent with MBL.

%%%%%%%%%%%%%%%%%%%%%%%%%%%%%%%%%%%%%%%%%%%%%%%%%%%%%%%%%%%%%%%%%%%%%%%%%%%%%%%%%%

%%%%%%%%%%%%%%%%%%%%%%%%%%%%%%%%%%%%%%%%%%%%%%%%%%%%%%%%%%%%%%%%%%%%%%%%%%%%%%%%%%
\subsection{Steady states in open Floquet systems}
\label{sec:OpenSystems}

When a driven many-body system is {\it open}, it will tend to a steady state where energy and entropy
production (heating) due to the drive are balanced by an outflow of these quantities to external
baths. This scenario is particularly relevant in the solid state, where electrons are
inevitably coupled to  environmental degrees of freedom including electromagnetic
radiation, phononic modes of the host crystal, and possibly particle reservoirs (leads).

In equilibrium, the steady state of an open system takes the universal form of a Gibbs state, 
fully determined by the temperature and chemical potential of the bath to which it is coupled.
Under special conditions, a Floquet system may attain an analogous ``Floquet-Gibbs'' steady state,
defined via $\rho_{\mathrm{FG}}=e^{-\beta H_{\mathrm{eff}}}/\Tr[e^{-\beta H_{\mathrm{eff}}}]$,
where $1/\beta$ is the temperature of the bath~\cite{Liu2015,Shirai2015, Shirai2016}. Generically,
however, such a universal statement cannot be made. Rather, the steady states of Floquet systems
are typically sensitive to details of the bath as well as the form of the system-bath coupling.
While this added complexity makes non-equilibrium systems challenging to study theoretically, it
also opens opportunities for controlling steady states through reservoir engineering (see below).

Physically, in order for the induced geometric or topological features of a system's Floquet
 bands to be reflected in its linear response to  applied probe fields, the steady state $\rho_{\rm steady}$
  should take a simple form in terms of populations of the single particle Floquet states.
Suppose for example that a weak probing electric field $\delta\mathbf{E}(\Omega)$ is applied to the
system with frequency ${\Omega}$. For ${\Omega} \ll \omega$, the Floquet-Kubo
formula~\cite{TorresFloquetKubo2005,Oka2009}
 can be used to obtain the period-averaged conductivity $\bar{\sigma}_{\alpha\beta}(\Omega)$ appearing in Ohm's law
   $\delta J_\alpha({\Omega})=\bar{\sigma}_{\alpha\beta}(\Omega)\delta E_\beta(\Omega)$,
 where $\delta\vec{J}(\Omega)$ is the change in the current density at frequency ${\Omega}$ due to the probe (see SM for details~\cite{SM}):
$\bar{\sigma}_{\alpha\beta}({\Omega})=\frac{-1}{{\Omega} T}\int_0^\infty d\tau e^{i{\Omega} \tau}
\int_0^T dt' \Tr\left\{\rho_{\mathrm{steady}}(t')\left[ J_\alpha(t'+\tau),
J_\beta(t')\right]\right\}+\frac{1}{i\Omega}\mathcal{K}^{(0)}_{\alpha\beta}(\Omega)$. Here,
$\mathcal{K}^{(0)}_{\alpha\beta}$ is the time averaged ``diamagnetic'' contribution to the conductivity, see
SM~\cite{SM}.

The expression for the conductivity above reduces to a particularly simple form in the special case where
$\rho_{\mathrm{steady}}$ obeys Wick's theorem and the corresponding single-particle density matrix
is diagonal in the Floquet basis. We will refer to steady states obeying these conditions as
``diagonal.'' (Note that generic Floquet systems do not obey such strict conditions~\cite{Kohn2001,
Hone1997, Hone2009}.) For diagonal steady states, the expression for $\bar{\sigma}_{\alpha\beta}$
takes an analogous form to that obtained for weakly-interacting electrons in equilibrium, with
energies replaced by quasienergies, and the equilibrium Fermi-Dirac distribution replaced by the
non-equilibrium populations of the Floquet states (see SM~\cite{SM}).
In particular, when the steady state takes on a band-insulator-like form in terms of Floquet band populations, 
then, as in equilibrium, the Floquet-Kubo formula implies that the bulk Hall conductivity
$\bar{\sigma}_{xy}$ is quantized while $\bar{\sigma}_{xx}$ vanishes~\cite{Oka2009, Dehghani2015}. Transport in
Floquet systems can also be treated within a Floquet-Boltzmann approach \cite{Genske2015,Esin2018},
giving similar conclusions.

Given this complicated situation, under what conditions may we hope to stabilize TI-like behavior in a Floquet system?
To be concrete, for the discussion that follows we will consider a resonantly-driven two-dimensional, single-valley direct band gap semiconductor, with dispersion at low energies described by the 2D Dirac model of Eq.~(\ref{eq:Dirac2D}).
Similar considerations may be applied to gapless (graphene-like) systems, when the drive resonantly couples states away from the Dirac points.

%%%%%%%%%%%%%%%%%%%%%%%%%%%%%%%%%%%%%%%%%%%%%%%%%%%%%%%%%%%%%%%%%%%%%%%%%%%
\begin{figure}[!t]
\includegraphics[width=0.95\columnwidth]{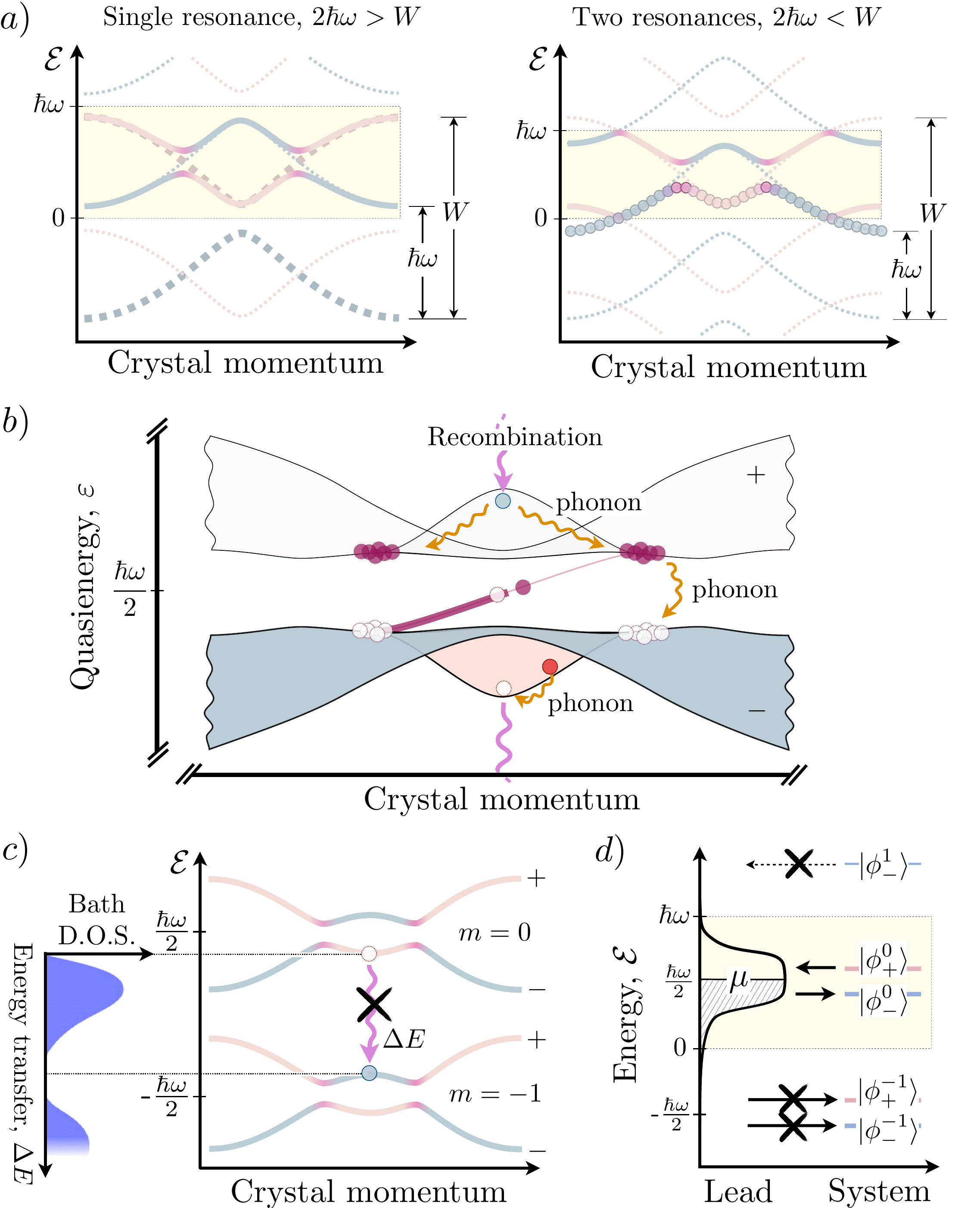}
\caption{Steady states of FTIs coupled to external baths.
a) When $2\hbar\omega$ is greater than the bandwidth $W$, there are no two-photon (or higher-order) resonances, and the Floquet bands (solid lines) exhibit at most a single band inversion compared with the original bands (bold dotted lines).
For $2\hbar\omega < W$, the characters of the Floquet bands switch multiple times between valence-band-like and conduction-band-like.
In the generalized Floquet insulator state, the bottom of the original valence band remains full and the top of the original conduction band remains empty, as indicated by the filled circles.
b) Scattering processes stabilizing a Floquet insulator state with remnant electron and hole excitations (filled and empty circles, respectively). 
Excitations from the lower to the upper bulk Floquet band are mediated by radiative recombination (purple wavy arrows), as well as Floquet-Umklapp phonon scattering and electron-electron collisions.
Relaxation to the lower Floquet band is mediated predominantly by electron-phonon scattering (orange wavy arrows).
c) Reservoir engineering can suppress unwanted Floquet-Umklapp electron-phonon and -photon scattering by suppressing the corresponding densities of states at relevant transition energies (adapted from Ref.~\cite{Iadecola2015}).
d) Coupling FTIs to external electronic leads through energy filters suppresses photon-assisted tunneling.
In the absence of the filter, the population of a Floquet state with quasienergy $\varepsilon$ is influenced by the populations of electrons in the reservoir at several energies $\varepsilon+n\hbar\omega$, through the sideband components $\{\ket{\phi^n_\pm}\}$,  which can be both below and above the Fermi level in the lead.
With an appropriate filter of bandwidth less than $\hbar\omega$, only one sideband component effectively participates %as shown in the figure,
and the population of a Floquet state is determined by electronic states in the lead at a single energy.
As drawn, the Floquet states $\ket{\psi_-}$ and $\ket{\psi_+}$, with sideband components $\{\ket{\phi^n_\pm}\}$, will be filled and empty, respectively.}
\label{fig:OpenFloquet}
\vspace{-0.1 in}
\end{figure}
%%%%%%%%%%%%%%%%%%%%%%%%%%%%%%%%%%%%%%%%%%%%%%%%%%%%%%%%%%%%%%%%%%%%%%%%%%%

As discussed in Sec.~\ref{sec:FloquetEngineering} (see Fig.~\ref{fig:BandEngineering}b), one of the key aims in Floquet band engineering is to use above-band-gap radiation to drive a topological transition by inducing an effective band inversion in the Floquet spectrum.
While the goal may be to induce a single band inversion near $\vec{k} = 0$, as in the left panel of Fig.~\ref{fig:OpenFloquet}a, for generic drive frequencies multiple resonances $2E(\vec{k}) = n\hbar\omega$ with $n > 1$ [see Eq.~(\ref{eq:Dirac2D})] may occur in the Brillouin zone, leading to multiple ``foldings'' of the Floquet bands into the quasienergy Brillouin zone.
While the ``lower'' Floquet band may exhibit the desired structure in the vicinity of the first
resonance, at larger momenta (in the vicinity of the two-photon resonance and beyond) it may
actually contain states arising from high up in the original conduction band (right panel of
Fig.~\ref{fig:OpenFloquet}a). The Floquet-Gibbs-insulator state would then host a finite density of
electrons and holes at very high energies --- a situation that is both highly unstable and far from
the goal of inducing a topological transition by modifying the states only in the vicinity of the
direct gap.

From a physical point of view, for a reasonable drive amplitude one may expect states deep in the valence band to remain occupied in the presence of the drive (with only weak modifications due to off-resonant hybridization with the conduction band).
In contrast to the Floquet-Gibbs-insulator state, a much more natural type of ``generalized Floquet insulator'' state is therefore as shown schematically by the filled circles in Fig.~\ref{fig:OpenFloquet}a: for momenta out to the vicinity of the second (two-photon) resonance the populations correspond to a filled Floquet band, while at larger momenta the electronic populations essentially follow perturbatively modified valence-band-like states (rather than the ``folded'' Floquet bands).

Importantly, in the generalized Floquet insulator state, the single-particle density matrix (at
each momentum $\bk$) is diagonal in the Floquet basis, and features populations only in the
``lower'' Floquet band, for momenta within and in the vicinity of the single-photon resonance ring
$2E(\vec{k}) = \hbar\omega$. In particular, the form of the single-particle density matrix is
maintained at the resonance ring itself, where the original valence and conduction band are
strongly hybridized.  This property is essential for observing topological phenomena in FTIs, as
the drive-induced Berry curvature is strongly localized within the resonance ring. Achieving this
condition requires a driving amplitude large enough that the induced Floquet gap exceeds the
scattering rates in the steady state (i.e., the separation between Floquet bands should exceed the
quasiparticle lifetime broadening in the steady state).  We note that, due to the drive-induced
band inversion, the generalized Floquet insulator state hosts a {\it population inversion} with
respect to the original bands for momenta within the resonance ring. Significant population
inversions (referred to in the literature as electron-hole plasmas) can be maintained even for a
weak drive for which the Floquet gap cannot be resolved~\cite{Goebel1978, Glazman1981,Glazman1983},
and are in fact exploited in semiconductor lasers~\cite{Chow1999, Huang2001, Roder2013}.

To see how such an insulator-like filling can be approximately maintained in the vicinity of a
drive-induced band inversion, we now describe the population kinetics of electrons in an FTI
connected to (zero temperature) heat baths consisting of acoustic phonons and the electromagnetic
environment~\cite{Dehghani2014, Seetharam2015, Iadecola2015}, see Fig.~\ref{fig:OpenFloquet}b.
As we expect higher-order resonances to play only a minor role for relatively weak driving, for simplicity we focus here on a parameter regime like that depicted in the left panel of Fig.~\ref{fig:OpenFloquet}a and as studied for example in Ref.~\cite{Seetharam2015}, where only a single resonance is supported in the system~\footnote{On the level of a theoretical model, the single resonance condition can be imposed by taking a system with narrow bands separated by a large gap.}.
In this situation, the generalized Floquet insulator state and the Floquet-Gibbs-insulator state coincide.
The details of the distribution near higher-order resonances and the corresponding implications for the system's response remain important directions for investigation.

As emphasized in the sections above, a crucial aspect of Floquet systems is that quasienergy is only defined modulo multiples of the driving field photon energy $\hbar\omega$.
Beyond the usual kinematic constraints on scattering processes of non-driven systems, the periodicity of quasienergy allows for scattering processes in which the total initial quasienergy of the particles and bath modes  differs from the total quasienergy of the outgoing particles and modes by integer multiples of $\hbar\omega$.
We refer to these as Floquet-Umklapp scattering processes.
In particular, these ``quantum Floquet heating'' processes spontaneously scatter electrons from the lower to the upper Floquet band % of the rotating frame at
at zero bath temperature, a process which would be impossible in the bands of a nondriven system in thermal equilibrium~\cite{Dykman2011}.
The impact of these processes on the steady state of the system is therefore significant, leading to important deviations from the ``ideal'' Floquet insulator distribution described above.

Note that by shifting the choice of Floquet-Brillouin zone (i.e., the value of $\epsilon_0$ defined in Sec.~\ref{sec:FloquetEngineering}), the ordering of bands can be interchanged; the definition of what is called a Floquet-Umklapp process thus also depends on the choice of Floquet-Brillouin zone.
This is only a matter of terminology: as long as all physical scattering processes are considered, the physical results do not depend on the choice of Floquet-Brillouin zone.

As first shown by Galitskii, Goreslavskii, and Elesin~\cite{Galitskii1970}, in the (hypothetical) absence of Floquet-Umklapp processes, all allowed scattering processes are akin to those of an equilibrium system and the steady state must therefore be of Floquet-Gibbs form. 
Importantly, while some Floquet-Umklapp processes involve photon-assisted scattering that may be suppressed for weak driving, others are simply spontaneous processes that appear inverted due to the folding of energies into the quasienergy Brillouin zone and cannot be neglected.
For example, an electron near $\bk = 0$ in the original conduction band can recombine with a nearby hole in the original valence band via the spontaneous emission of a photon.
Due to the Floquet band inversion, this process corresponds to the scattering of an electron
from the {\it lower} Floquet band to the {\it upper} one, see Fig.~\ref{fig:OpenFloquet}b.
Such Floquet-Umklapp processes add excited
electrons in the upper Floquet band and holes in the lower one. Once such excitations are created,
they quickly relax to the local extrema of the Floquet bands by emitting acoustic phonons. 
Although acoustic phonons typically do not scatter electrons between the conduction and valence bands, the strong drive-induced hybridization of the bands near the resonances allows electrons to traverse the Floquet gap to recombine with holes in the lower Floquet band via further acoustic phonon emission, see Fig.~\ref{fig:OpenFloquet}b.

When the total rate
of Floquet-Umklapp ``heating'' processes is much smaller than the rate of the phonon-assisted ``cooling'' that reduces the number of excitations in the Floquet bands, the steady state resembles the (generalized) Floquet insulator state, with small additional pockets of electron and hole excitations in the upper and lower Floquet bands, respectively. 
Deviations from Floquet insulator state lead to deviations of the period-averaged dc Hall conductivity $\bar{\sigma}_{xy}(\Omega = 0)$ from a quantized value in units of $e^2/h$~\cite{Dehghani2015, Esin2018, Sato2019}.

By properly engineering the  environment of an FTI, deviations from the desired steady state can be minimized.
As shown in Fig.~\ref{fig:OpenFloquet}c, Floquet-Umklapp scattering  involving photon or phonon emission can be
suppressed by suppressing the density of states of these bosonic modes at appropriate energies
\cite{Iadecola2015,Seetharam2015}. For example, a cavity can be used to suppress the density of state of the electromagnetic environment at the frequencies corresponding to electron-hole recombination near $\bk=0$.

The need for reservoir engineering is even more pronounced when electronic reservoirs are
connected to the system. For conventional metallic contacts with a wide bandwidth, %featureless density of states,
electrons (holes) in the contact with energies well below (above) the Fermi level may tunnel into
the upper (lower) Floquet bands via absorption of integer multiples of $\hbar\omega$ from the
drive (cf.~discussion in Sec.~\ref{sec:Transport}). These processes lead to a  proliferation of excitations with respect to the ideal Floquet insulator~\cite{Seetharam2015, Iadecola2015, Esin2018}. An energy-selective filter placed between the contact and the
FTI suppresses the tunneling density of states outside a narrow window defined by the filter. Such
a filter may not only suppress unwanted tunneling events, but in fact may also deplete the number of
excitations and help to establish and control a sharp ``Fermi level'' in the FTI's edge states~\cite{Seetharam2015, Esin2018}, see Fig~\ref{fig:OpenFloquet}c.

The general considerations outlined in this section provide a framework for identifying promising setups for realizing steady-state FTIs in solid state systems.
The next important steps on the horizon are to find specific candidate materials and compatible implementations of the bath engineering schemes discussed above.
Theoretically, the role of higher-order resonances and their impact on transport coefficients also remain to be elucidated.

%%%%%%%%%%%%%%%%%%%%%%%%%%%%%%%%%%%%%%%%%%%%%%%%%%%%%%%%%%%%%%%%%%%%%%%%%%%%%%%%%%

%%%%%%%%%%%%%%%%%%%%%%%%%%%%%%%%%%%%%%%%%%%%%%%%%%%%%%%%%%%%%%%%%%%%%%%%%%%%%%%%%%
\section{Outlook}
 As we have discussed, Floquet engineering provides a versatile set of tools for controlling topological properties of quantum matter.
Experiments in solid state, optical, and cold atomic systems have demonstrated many of the basic features of topological Floquet band engineering at the single-particle level. Important ingredients for Floquet engineering of many-body systems,
including Floquet MBL~\cite{Bordia2017} and drive-tunable exchange interactions~\cite{Gorg2018},
have been experimentally demonstrated with cold atoms. These encouraging early works highlight the
promise of Floquet engineering, and point to a number of important open theoretical and
experimental challenges.

One of the key hallmarks of topological behavior in equilibrium electronic systems, and an
important motivation behind many studies in Floquet engineering, is the appearance of robustly
quantized observables, such as transport coefficients. On a theoretical level, it is therefore
important to investigate the fundamental limits on the degree of robustness that can be achieved in
{\it non-equilibrium} topological matter. Identifying promising new candidate materials and setups
where reservoir engineering can be achieved, as well as devising new strategies for obtaining
stable topological phases in the steady-states of interacting systems are also crucial for further
progress in the field. Experimentally, stabilizing topological steady states and observing robust
quantization of observables in many-body system remain open challenges and prominent goals in the
field.

Going beyond the realm of Floquet topological insulators, we anticipate that Floquet engineering of
stable strongly correlated and topologically ordered phases of matter will play an important role
in future progress in contemporary condensed matter physics and quantum information.

\begin{acknowledgments}
%\section{Acknowledgments}
We thank all of our collaborators on FTI-related work, with whom we have had many stimulating
interactions. In particular, we acknowledge Erez Berg, Eugene Demler, Victor Galitski, Takuya
Kitagawa, Michael Levin and Gil Refael, with whom we began our journey in this field. We also thank
Iliya Esin for help with figures and helpful discussions. N. L. acknowledges support from the
European Research Council (ERC) under the European Union Horizon 2020 Research and Innovation
Programme (Grant Agreement No. 639172), and from the Israeli Center of Research Excellence (I-CORE)
``Circle of Light''. M. R. gratefully acknowledges the support of the European Research Council
(ERC) under the European Union Horizon 2020 Research and
Innovation Programme (Grant Agreement No.678862), the Villum Foundation, and CRC
183 of the Deutsche Forschungsgemeinschaft. 

\end{acknowledgments}

%%%%%%%%%%%%%%%%%%%%%%%%%%%%%%%%%%%%%%%%%%%%%%%%%%%%%%%%%%%%%%%%%%%%%%%%%%%%%%%%%%

\bibliography{FloquetTopologicalInsulator_references}

\begin{thebibliography}{177}%
\makeatletter
\providecommand \@ifxundefined [1]{%
 \@ifx{#1\undefined}
}%
\providecommand \@ifnum [1]{%
 \ifnum #1\expandafter \@firstoftwo
 \else \expandafter \@secondoftwo
 \fi
}%
\providecommand \@ifx [1]{%
 \ifx #1\expandafter \@firstoftwo
 \else \expandafter \@secondoftwo
 \fi
}%
\providecommand \natexlab [1]{#1}%
\providecommand \enquote  [1]{``#1''}%
\providecommand \bibnamefont  [1]{#1}%
\providecommand \bibfnamefont [1]{#1}%
\providecommand \citenamefont [1]{#1}%
\providecommand \href@noop [0]{\@secondoftwo}%
\providecommand \href [0]{\begingroup \@sanitize@url \@href}%
\providecommand \@href[1]{\@@startlink{#1}\@@href}%
\providecommand \@@href[1]{\endgroup#1\@@endlink}%
\providecommand \@sanitize@url [0]{\catcode `\\12\catcode `\$12\catcode
  `\&12\catcode `\#12\catcode `\^12\catcode `\_12\catcode `\%12\relax}%
\providecommand \@@startlink[1]{}%
\providecommand \@@endlink[0]{}%
\providecommand \url  [0]{\begingroup\@sanitize@url \@url }%
\providecommand \@url [1]{\endgroup\@href {#1}{\urlprefix }}%
\providecommand \urlprefix  [0]{URL }%
\providecommand \Eprint [0]{\href }%
\providecommand \doibase [0]{http://dx.doi.org/}%
\providecommand \selectlanguage [0]{\@gobble}%
\providecommand \bibinfo  [0]{\@secondoftwo}%
\providecommand \bibfield  [0]{\@secondoftwo}%
\providecommand \translation [1]{[#1]}%
\providecommand \BibitemOpen [0]{}%
\providecommand \bibitemStop [0]{}%
\providecommand \bibitemNoStop [0]{.\EOS\space}%
\providecommand \EOS [0]{\spacefactor3000\relax}%
\providecommand \BibitemShut  [1]{\csname bibitem#1\endcsname}%
\let\auto@bib@innerbib\@empty
%</preamble>
\bibitem [{\citenamefont {Basov}\ \emph {et~al.}(2017)\citenamefont {Basov},
  \citenamefont {Averitt},\ and\ \citenamefont {Hsieh}}]{Basov2017}%
  \BibitemOpen
  \bibfield  {author} {\bibinfo {author} {\bibfnamefont {D.~N.}\ \bibnamefont
  {Basov}}, \bibinfo {author} {\bibfnamefont {R.~D.}\ \bibnamefont {Averitt}},
  \ and\ \bibinfo {author} {\bibfnamefont {D.}~\bibnamefont {Hsieh}},\
  }\bibfield  {title} {\enquote {\bibinfo {title} {{Towards properties on
  demand in quantum materials}},}\ }\href@noop {} {\bibfield  {journal}
  {\bibinfo  {journal} {Nature Materials}\ }\textbf {\bibinfo {volume} {16}},\
  \bibinfo {pages} {1077} (\bibinfo {year} {2017})}\BibitemShut {NoStop}%
\bibitem [{\citenamefont {Floquet}(1883)}]{Floquet1883}%
  \BibitemOpen
  \bibfield  {author} {\bibinfo {author} {\bibfnamefont {Gaston}\ \bibnamefont
  {Floquet}},\ }\bibfield  {title} {\enquote {\bibinfo {title} {{Sur les
  equations differentielles lineaires a coefficients periodiques}},}\
  }\href@noop {} {\bibfield  {journal} {\bibinfo  {journal} {Annales de l'Ecole
  Normale Superieure}\ }\textbf {\bibinfo {volume} {12}},\ \bibinfo {pages}
  {47} (\bibinfo {year} {1883})}\BibitemShut {NoStop}%
\bibitem [{\citenamefont {Yao}\ \emph {et~al.}(2007)\citenamefont {Yao},
  \citenamefont {MacDonald},\ and\ \citenamefont {Niu}}]{Yao2007}%
  \BibitemOpen
  \bibfield  {author} {\bibinfo {author} {\bibfnamefont {W.}~\bibnamefont
  {Yao}}, \bibinfo {author} {\bibfnamefont {A.~H.}\ \bibnamefont {MacDonald}},
  \ and\ \bibinfo {author} {\bibfnamefont {Q.}~\bibnamefont {Niu}},\ }\bibfield
   {title} {\enquote {\bibinfo {title} {{Optical Control of Topological Quantum
  Transport in Semiconductors}},}\ }\href@noop {} {\bibfield  {journal}
  {\bibinfo  {journal} {Phys. Rev. Lett.}\ }\textbf {\bibinfo {volume} {99}},\
  \bibinfo {pages} {047401} (\bibinfo {year} {2007})}\BibitemShut {NoStop}%
\bibitem [{\citenamefont {Hasan}\ and\ \citenamefont
  {Kane}(2010)}]{HasanTI_RMP}%
  \BibitemOpen
  \bibfield  {author} {\bibinfo {author} {\bibfnamefont {M.~Z.}\ \bibnamefont
  {Hasan}}\ and\ \bibinfo {author} {\bibfnamefont {C.~L.}\ \bibnamefont
  {Kane}},\ }\bibfield  {title} {\enquote {\bibinfo {title} {Colloquium:
  Topological insulators},}\ }\href@noop {} {\bibfield  {journal} {\bibinfo
  {journal} {Rev. Mod. Phys.}\ }\textbf {\bibinfo {volume} {82}},\ \bibinfo
  {pages} {3045--3067} (\bibinfo {year} {2010})}\BibitemShut {NoStop}%
\bibitem [{\citenamefont {Oka}\ and\ \citenamefont {Aoki}(2009)}]{Oka2009}%
  \BibitemOpen
  \bibfield  {author} {\bibinfo {author} {\bibfnamefont {Takashi}\ \bibnamefont
  {Oka}}\ and\ \bibinfo {author} {\bibfnamefont {Hideo}\ \bibnamefont {Aoki}},\
  }\bibfield  {title} {\enquote {\bibinfo {title} {{Photovoltaic Hall effect in
  graphene}},}\ }\href@noop {} {\bibfield  {journal} {\bibinfo  {journal}
  {Phys. Rev. B}\ }\textbf {\bibinfo {volume} {79}},\ \bibinfo {pages} {081406}
  (\bibinfo {year} {2009})}\BibitemShut {NoStop}%
\bibitem [{\citenamefont {Kitagawa}\ \emph {et~al.}(2010)\citenamefont
  {Kitagawa}, \citenamefont {Berg}, \citenamefont {Rudner},\ and\ \citenamefont
  {Demler}}]{Kitagawa2010}%
  \BibitemOpen
  \bibfield  {author} {\bibinfo {author} {\bibfnamefont {Takuya}\ \bibnamefont
  {Kitagawa}}, \bibinfo {author} {\bibfnamefont {Erez}\ \bibnamefont {Berg}},
  \bibinfo {author} {\bibfnamefont {Mark}\ \bibnamefont {Rudner}}, \ and\
  \bibinfo {author} {\bibfnamefont {Eugene}\ \bibnamefont {Demler}},\
  }\bibfield  {title} {\enquote {\bibinfo {title} {Topological characterization
  of periodically driven quantum systems},}\ }\href@noop {} {\bibfield
  {journal} {\bibinfo  {journal} {Phys. Rev. B}\ }\textbf {\bibinfo {volume}
  {82}},\ \bibinfo {pages} {235114} (\bibinfo {year} {2010})}\BibitemShut
  {NoStop}%
\bibitem [{\citenamefont {Lindner}\ \emph {et~al.}(2011)\citenamefont
  {Lindner}, \citenamefont {Refael},\ and\ \citenamefont
  {Galitski}}]{Lindner2011}%
  \BibitemOpen
  \bibfield  {author} {\bibinfo {author} {\bibfnamefont {Netanel~H.}\
  \bibnamefont {Lindner}}, \bibinfo {author} {\bibfnamefont {Gil}\ \bibnamefont
  {Refael}}, \ and\ \bibinfo {author} {\bibfnamefont {Victor}\ \bibnamefont
  {Galitski}},\ }\bibfield  {title} {\enquote {\bibinfo {title} {{Floquet
  topological insulator in semiconductor quantum wells}},}\ }\href@noop {}
  {\bibfield  {journal} {\bibinfo  {journal} {Nature Phys.}\ }\textbf {\bibinfo
  {volume} {7}},\ \bibinfo {pages} {490} (\bibinfo {year} {2011})}\BibitemShut
  {NoStop}%
\bibitem [{\citenamefont {D'Alessio}\ and\ \citenamefont
  {Rigol}(2014)}]{DAlessio2014}%
  \BibitemOpen
  \bibfield  {author} {\bibinfo {author} {\bibfnamefont {Luca}\ \bibnamefont
  {D'Alessio}}\ and\ \bibinfo {author} {\bibfnamefont {Marcos}\ \bibnamefont
  {Rigol}},\ }\bibfield  {title} {\enquote {\bibinfo {title} {{Long-time
  behavior of isolated periodically driven interacting lattice systems}},}\
  }\href@noop {} {\bibfield  {journal} {\bibinfo  {journal} {Phys. Rev. X}\
  }\textbf {\bibinfo {volume} {4}},\ \bibinfo {pages} {041048} (\bibinfo {year}
  {2014})}\BibitemShut {NoStop}%
\bibitem [{\citenamefont {Lazarides}\ \emph {et~al.}(2014)\citenamefont
  {Lazarides}, \citenamefont {Das},\ and\ \citenamefont
  {Moessner}}]{Lazarides2014}%
  \BibitemOpen
  \bibfield  {author} {\bibinfo {author} {\bibfnamefont {A.}~\bibnamefont
  {Lazarides}}, \bibinfo {author} {\bibfnamefont {A.}~\bibnamefont {Das}}, \
  and\ \bibinfo {author} {\bibfnamefont {R.}~\bibnamefont {Moessner}},\
  }\bibfield  {title} {\enquote {\bibinfo {title} {Equilibrium states of
  generic quantum systems subject to periodic driving},}\ }\href@noop {}
  {\bibfield  {journal} {\bibinfo  {journal} {Phys. Rev. E}\ }\textbf {\bibinfo
  {volume} {90}},\ \bibinfo {pages} {012110} (\bibinfo {year}
  {2014})}\BibitemShut {NoStop}%
\bibitem [{Note1()}]{Note1}%
  \BibitemOpen
  \bibinfo {note} {Exceptions to the runaway heating scenario, e.g., based on
  localization in energy space~\cite {Prosen1998, Kukuljan2015, Citro2016,
  Chandran2016, Haldar2018} or finite-size effects~\cite {Seetharam2018}, are
  also being explored.}\BibitemShut {Stop}%
\bibitem [{\citenamefont {Grushin}\ \emph {et~al.}(2014)\citenamefont
  {Grushin}, \citenamefont {G{\'o}mez-Le{\'o}n},\ and\ \citenamefont
  {Neupert}}]{Grushin2014}%
  \BibitemOpen
  \bibfield  {author} {\bibinfo {author} {\bibfnamefont {Adolfo~G.}\
  \bibnamefont {Grushin}}, \bibinfo {author} {\bibfnamefont {{\'A}lvaro}\
  \bibnamefont {G{\'o}mez-Le{\'o}n}}, \ and\ \bibinfo {author} {\bibfnamefont
  {Titus}\ \bibnamefont {Neupert}},\ }\bibfield  {title} {\enquote {\bibinfo
  {title} {{Floquet fractional Chern insulators}},}\ }\href@noop {} {\bibfield
  {journal} {\bibinfo  {journal} {Phys. Rev. Lett.}\ }\textbf {\bibinfo
  {volume} {112}},\ \bibinfo {pages} {156801} (\bibinfo {year}
  {2014})}\BibitemShut {NoStop}%
\bibitem [{\citenamefont {Klinovaja}\ \emph {et~al.}(2016)\citenamefont
  {Klinovaja}, \citenamefont {Stano},\ and\ \citenamefont
  {Loss}}]{Klinovaja2016}%
  \BibitemOpen
  \bibfield  {author} {\bibinfo {author} {\bibfnamefont {Jelena}\ \bibnamefont
  {Klinovaja}}, \bibinfo {author} {\bibfnamefont {Peter}\ \bibnamefont
  {Stano}}, \ and\ \bibinfo {author} {\bibfnamefont {Daniel}\ \bibnamefont
  {Loss}},\ }\bibfield  {title} {\enquote {\bibinfo {title} {{Topological
  Floquet Phases in Driven Coupled Rashba Nanowires}},}\ }\href@noop {}
  {\bibfield  {journal} {\bibinfo  {journal} {Phys. Rev. Lett.}\ }\textbf
  {\bibinfo {volume} {116}},\ \bibinfo {pages} {176401} (\bibinfo {year}
  {2016})}\BibitemShut {NoStop}%
\bibitem [{\citenamefont {Liu}\ \emph {et~al.}(2018{\natexlab{a}})\citenamefont
  {Liu}, \citenamefont {Hejazi},\ and\ \citenamefont {Balents}}]{Liu2018}%
  \BibitemOpen
  \bibfield  {author} {\bibinfo {author} {\bibfnamefont {Jianpeng}\
  \bibnamefont {Liu}}, \bibinfo {author} {\bibfnamefont {Kasra}\ \bibnamefont
  {Hejazi}}, \ and\ \bibinfo {author} {\bibfnamefont {Leon}\ \bibnamefont
  {Balents}},\ }\bibfield  {title} {\enquote {\bibinfo {title} {Floquet
  engineering of multiorbital mott insulators: Applications to orthorhombic
  titanates},}\ }\href@noop {} {\bibfield  {journal} {\bibinfo  {journal}
  {Phys. Rev. Lett.}\ }\textbf {\bibinfo {volume} {121}},\ \bibinfo {pages}
  {107201} (\bibinfo {year} {2018}{\natexlab{a}})}\BibitemShut {NoStop}%
\bibitem [{\citenamefont {G\"{o}rg}\ \emph {et~al.}(2018)\citenamefont
  {G\"{o}rg}, \citenamefont {Messer}, \citenamefont {Sandholzer}, \citenamefont
  {Jotzu}, \citenamefont {Desbuquois},\ and\ \citenamefont
  {Esslinger}}]{Gorg2018}%
  \BibitemOpen
  \bibfield  {author} {\bibinfo {author} {\bibfnamefont {Frederik}\
  \bibnamefont {G\"{o}rg}}, \bibinfo {author} {\bibfnamefont {Michael}\
  \bibnamefont {Messer}}, \bibinfo {author} {\bibfnamefont {Kilian}\
  \bibnamefont {Sandholzer}}, \bibinfo {author} {\bibfnamefont {Gregor}\
  \bibnamefont {Jotzu}}, \bibinfo {author} {\bibfnamefont {Remi}\ \bibnamefont
  {Desbuquois}}, \ and\ \bibinfo {author} {\bibfnamefont {Tilman}\ \bibnamefont
  {Esslinger}},\ }\bibfield  {title} {\enquote {\bibinfo {title} {{Enhancement
  and sign change of magnetic correlations in a driven quantum many-body
  system}},}\ }\href@noop {} {\bibfield  {journal} {\bibinfo  {journal}
  {Nature}\ }\textbf {\bibinfo {volume} {553}},\ \bibinfo {pages} {481}
  (\bibinfo {year} {2018})}\BibitemShut {NoStop}%
\bibitem [{\citenamefont {Kennes}\ \emph {et~al.}(2018)\citenamefont {Kennes},
  \citenamefont {de~la Torre}, \citenamefont {Ron}, \citenamefont {Hsieh},\
  and\ \citenamefont {Millis}}]{Kennes2018}%
  \BibitemOpen
  \bibfield  {author} {\bibinfo {author} {\bibfnamefont {D.~M.}\ \bibnamefont
  {Kennes}}, \bibinfo {author} {\bibfnamefont {A.}~\bibnamefont {de~la Torre}},
  \bibinfo {author} {\bibfnamefont {A.}~\bibnamefont {Ron}}, \bibinfo {author}
  {\bibfnamefont {D.}~\bibnamefont {Hsieh}}, \ and\ \bibinfo {author}
  {\bibfnamefont {A.~J.}\ \bibnamefont {Millis}},\ }\bibfield  {title}
  {\enquote {\bibinfo {title} {{Floquet Engineering in Quantum Chains}},}\
  }\href@noop {} {\bibfield  {journal} {\bibinfo  {journal} {Phys. Rev. Lett.}\
  }\textbf {\bibinfo {volume} {120}},\ \bibinfo {pages} {127601} (\bibinfo
  {year} {2018})}\BibitemShut {NoStop}%
\bibitem [{\citenamefont {Rudner}\ \emph {et~al.}(2013)\citenamefont {Rudner},
  \citenamefont {Lindner}, \citenamefont {Berg},\ and\ \citenamefont
  {Levin}}]{Rudner2013}%
  \BibitemOpen
  \bibfield  {author} {\bibinfo {author} {\bibfnamefont {Mark~S.}\ \bibnamefont
  {Rudner}}, \bibinfo {author} {\bibfnamefont {Netanel~H.}\ \bibnamefont
  {Lindner}}, \bibinfo {author} {\bibfnamefont {Erez}\ \bibnamefont {Berg}}, \
  and\ \bibinfo {author} {\bibfnamefont {Michael}\ \bibnamefont {Levin}},\
  }\bibfield  {title} {\enquote {\bibinfo {title} {Anomalous edge states and
  the bulk-edge correspondence for periodically driven two-dimensional
  systems},}\ }\href@noop {} {\bibfield  {journal} {\bibinfo  {journal} {Phys.
  Rev. X}\ }\textbf {\bibinfo {volume} {3}},\ \bibinfo {pages} {031005}
  (\bibinfo {year} {2013})}\BibitemShut {NoStop}%
\bibitem [{\citenamefont {Nathan}\ and\ \citenamefont
  {Rudner}(2015)}]{Nathan2015}%
  \BibitemOpen
  \bibfield  {author} {\bibinfo {author} {\bibfnamefont {Frederik}\
  \bibnamefont {Nathan}}\ and\ \bibinfo {author} {\bibfnamefont {Mark~S.}\
  \bibnamefont {Rudner}},\ }\bibfield  {title} {\enquote {\bibinfo {title}
  {{Topological singularities and the general classification of Floquet-Bloch
  systems}},}\ }\href@noop {} {\bibfield  {journal} {\bibinfo  {journal} {New
  J. Phys.}\ }\textbf {\bibinfo {volume} {17}},\ \bibinfo {pages} {125014}
  (\bibinfo {year} {2015})}\BibitemShut {NoStop}%
\bibitem [{\citenamefont {Roy}\ and\ \citenamefont
  {Harper}(2017{\natexlab{a}})}]{Roy2017}%
  \BibitemOpen
  \bibfield  {author} {\bibinfo {author} {\bibfnamefont {Rahul}\ \bibnamefont
  {Roy}}\ and\ \bibinfo {author} {\bibfnamefont {Fenner}\ \bibnamefont
  {Harper}},\ }\bibfield  {title} {\enquote {\bibinfo {title} {{Floquet
  topological phases with symmetry in all dimensions}},}\ }\href@noop {}
  {\bibfield  {journal} {\bibinfo  {journal} {Phys. Rev. B}\ }\textbf {\bibinfo
  {volume} {95}},\ \bibinfo {pages} {195128} (\bibinfo {year}
  {2017}{\natexlab{a}})}\BibitemShut {NoStop}%
\bibitem [{\citenamefont {Roy}\ and\ \citenamefont
  {Harper}(2017{\natexlab{b}})}]{Roy2017b}%
  \BibitemOpen
  \bibfield  {author} {\bibinfo {author} {\bibfnamefont {Rahul}\ \bibnamefont
  {Roy}}\ and\ \bibinfo {author} {\bibfnamefont {Fenner}\ \bibnamefont
  {Harper}},\ }\bibfield  {title} {\enquote {\bibinfo {title} {{Periodic table
  for Floquet topological insulators}},}\ }\href@noop {} {\bibfield  {journal}
  {\bibinfo  {journal} {Phys. Rev. B}\ }\textbf {\bibinfo {volume} {96}},\
  \bibinfo {pages} {155118} (\bibinfo {year} {2017}{\natexlab{b}})}\BibitemShut
  {NoStop}%
\bibitem [{\citenamefont {Yao}\ \emph {et~al.}(2017{\natexlab{a}})\citenamefont
  {Yao}, \citenamefont {Yan},\ and\ \citenamefont {Wang}}]{Yao2017}%
  \BibitemOpen
  \bibfield  {author} {\bibinfo {author} {\bibfnamefont {Shunyu}\ \bibnamefont
  {Yao}}, \bibinfo {author} {\bibfnamefont {Zhongbo}\ \bibnamefont {Yan}}, \
  and\ \bibinfo {author} {\bibfnamefont {Zhong}\ \bibnamefont {Wang}},\
  }\bibfield  {title} {\enquote {\bibinfo {title} {{Topological invariants of
  Floquet systems: General formulation, special properties, and Floquet
  topological defects}},}\ }\href@noop {} {\bibfield  {journal} {\bibinfo
  {journal} {Phys. Rev. B}\ }\textbf {\bibinfo {volume} {96}},\ \bibinfo
  {pages} {195303} (\bibinfo {year} {2017}{\natexlab{a}})}\BibitemShut
  {NoStop}%
\bibitem [{\citenamefont {G\'omez-Le\'on}\ and\ \citenamefont
  {Platero}(2013)}]{PlateroPRL}%
  \BibitemOpen
  \bibfield  {author} {\bibinfo {author} {\bibfnamefont {A.}~\bibnamefont
  {G\'omez-Le\'on}}\ and\ \bibinfo {author} {\bibfnamefont {G.}~\bibnamefont
  {Platero}},\ }\bibfield  {title} {\enquote {\bibinfo {title} {{Floquet-Bloch
  Theory and Topology in Periodically Driven Lattices}},}\ }\href@noop {}
  {\bibfield  {journal} {\bibinfo  {journal} {Phys. Rev. Lett.}\ }\textbf
  {\bibinfo {volume} {110}},\ \bibinfo {pages} {200403} (\bibinfo {year}
  {2013})}\BibitemShut {NoStop}%
\bibitem [{\citenamefont {Graf}\ and\ \citenamefont {Tauber}(2018)}]{Graf2018}%
  \BibitemOpen
  \bibfield  {author} {\bibinfo {author} {\bibfnamefont {G.~M.}\ \bibnamefont
  {Graf}}\ and\ \bibinfo {author} {\bibfnamefont {C.}~\bibnamefont {Tauber}},\
  }\bibfield  {title} {\enquote {\bibinfo {title} {{Bulk-Edge Correspondence
  for Two-Dimensional Floquet Topological Insulators}},}\ }\href@noop {}
  {\bibfield  {journal} {\bibinfo  {journal} {Ann. Henri Poincare}\ }\textbf
  {\bibinfo {volume} {19}},\ \bibinfo {pages} {709} (\bibinfo {year}
  {2018})}\BibitemShut {NoStop}%
\bibitem [{\citenamefont {{Shapiro}}\ and\ \citenamefont
  {{Tauber}}(2019)}]{Shapiro2018}%
  \BibitemOpen
  \bibfield  {author} {\bibinfo {author} {\bibfnamefont {Jacob}\ \bibnamefont
  {{Shapiro}}}\ and\ \bibinfo {author} {\bibfnamefont {Cl{\'e}ment}\
  \bibnamefont {{Tauber}}},\ }\bibfield  {title} {\enquote {\bibinfo {title}
  {{Strongly Disordered Floquet Topological Systems}},}\ }\href@noop {}
  {\bibfield  {journal} {\bibinfo  {journal} {Annales Henri Poincare}\ }\textbf
  {\bibinfo {volume} {20}},\ \bibinfo {pages} {1837} (\bibinfo {year}
  {2019})}\BibitemShut {NoStop}%
\bibitem [{\citenamefont {Khemani}\ \emph {et~al.}(2016)\citenamefont
  {Khemani}, \citenamefont {Lazarides}, \citenamefont {Moessner},\ and\
  \citenamefont {Sondhi}}]{Khemani2016}%
  \BibitemOpen
  \bibfield  {author} {\bibinfo {author} {\bibfnamefont {V.}~\bibnamefont
  {Khemani}}, \bibinfo {author} {\bibfnamefont {A.}~\bibnamefont {Lazarides}},
  \bibinfo {author} {\bibfnamefont {R.}~\bibnamefont {Moessner}}, \ and\
  \bibinfo {author} {\bibfnamefont {S.~L.}\ \bibnamefont {Sondhi}},\ }\bibfield
   {title} {\enquote {\bibinfo {title} {{Phase Structure of Driven Quantum
  Systems}},}\ }\href@noop {} {\bibfield  {journal} {\bibinfo  {journal} {Phys.
  Rev. Lett.}\ }\textbf {\bibinfo {volume} {116}},\ \bibinfo {pages} {250401}
  (\bibinfo {year} {2016})}\BibitemShut {NoStop}%
\bibitem [{\citenamefont {Else}\ \emph {et~al.}(2016)\citenamefont {Else},
  \citenamefont {Bauer},\ and\ \citenamefont {Nayak}}]{Else2016b}%
  \BibitemOpen
  \bibfield  {author} {\bibinfo {author} {\bibfnamefont {Dominic~V.}\
  \bibnamefont {Else}}, \bibinfo {author} {\bibfnamefont {Bela}\ \bibnamefont
  {Bauer}}, \ and\ \bibinfo {author} {\bibfnamefont {Chetan}\ \bibnamefont
  {Nayak}},\ }\bibfield  {title} {\enquote {\bibinfo {title} {Floquet time
  crystals},}\ }\href@noop {} {\bibfield  {journal} {\bibinfo  {journal} {Phys.
  Rev. Lett.}\ }\textbf {\bibinfo {volume} {117}},\ \bibinfo {pages} {090402}
  (\bibinfo {year} {2016})}\BibitemShut {NoStop}%
\bibitem [{\citenamefont {Else}\ \emph {et~al.}(2017)\citenamefont {Else},
  \citenamefont {Bauer},\ and\ \citenamefont {Nayak}}]{Else2017}%
  \BibitemOpen
  \bibfield  {author} {\bibinfo {author} {\bibfnamefont {Dominic~V.}\
  \bibnamefont {Else}}, \bibinfo {author} {\bibfnamefont {Bela}\ \bibnamefont
  {Bauer}}, \ and\ \bibinfo {author} {\bibfnamefont {Chetan}\ \bibnamefont
  {Nayak}},\ }\bibfield  {title} {\enquote {\bibinfo {title} {{Prethermal
  Phases of Matter Protected by Time-Translation Symmetry}},}\ }\href@noop {}
  {\bibfield  {journal} {\bibinfo  {journal} {Phys. Rev. X}\ }\textbf {\bibinfo
  {volume} {7}},\ \bibinfo {pages} {011026} (\bibinfo {year}
  {2017})}\BibitemShut {NoStop}%
\bibitem [{\citenamefont {{Rudner}}\ and\ \citenamefont
  {{Song}}(2019)}]{Rudner2018}%
  \BibitemOpen
  \bibfield  {author} {\bibinfo {author} {\bibfnamefont {Mark~S.}\ \bibnamefont
  {{Rudner}}}\ and\ \bibinfo {author} {\bibfnamefont {Justin C.~W.}\
  \bibnamefont {{Song}}},\ }\bibfield  {title} {\enquote {\bibinfo {title}
  {{Self-induced Berry flux and spontaneous non-equilibrium magnetism}},}\
  }\href@noop {} {\bibfield  {journal} {\bibinfo  {journal} {Nature Physics}\ }
  (\bibinfo {year} {2019})}\BibitemShut {NoStop}%
\bibitem [{\citenamefont {{Nag}}\ \emph {et~al.}(2018)\citenamefont {{Nag}},
  \citenamefont {{Slager}}, \citenamefont {{Higuchi}},\ and\ \citenamefont
  {{Oka}}}]{Nag2018}%
  \BibitemOpen
  \bibfield  {author} {\bibinfo {author} {\bibfnamefont {Tanay}\ \bibnamefont
  {{Nag}}}, \bibinfo {author} {\bibfnamefont {Robert-Jan}\ \bibnamefont
  {{Slager}}}, \bibinfo {author} {\bibfnamefont {Takuya}\ \bibnamefont
  {{Higuchi}}}, \ and\ \bibinfo {author} {\bibfnamefont {Takashi}\ \bibnamefont
  {{Oka}}},\ }\bibfield  {title} {\enquote {\bibinfo {title} {{Dynamical
  synchronization transition in interacting electron systems}},}\ }\href@noop
  {} {\bibfield  {journal} {\bibinfo  {journal} {arXiv:1802.02161}\ } (\bibinfo
  {year} {2018})}\BibitemShut {NoStop}%
\bibitem [{\citenamefont {Kinoshita}\ \emph {et~al.}(2018)\citenamefont
  {Kinoshita}, \citenamefont {Murata},\ and\ \citenamefont
  {Oka}}]{Kinoshita2018}%
  \BibitemOpen
  \bibfield  {author} {\bibinfo {author} {\bibfnamefont {Shunichiro}\
  \bibnamefont {Kinoshita}}, \bibinfo {author} {\bibfnamefont {Keiju}\
  \bibnamefont {Murata}}, \ and\ \bibinfo {author} {\bibfnamefont {Takashi}\
  \bibnamefont {Oka}},\ }\bibfield  {title} {\enquote {\bibinfo {title}
  {Holographic floquet states ii: Floquet condensation of vector mesons in
  nonequilibrium phase diagram},}\ }\href@noop {} {\bibfield  {journal}
  {\bibinfo  {journal} {Journal of High Energy Physics}\ }\textbf {\bibinfo
  {volume} {2018}},\ \bibinfo {pages} {96} (\bibinfo {year}
  {2018})}\BibitemShut {NoStop}%
\bibitem [{\citenamefont {Harper}\ \emph {et~al.}(2019)\citenamefont {Harper},
  \citenamefont {Roy}, \citenamefont {Rudner},\ and\ \citenamefont
  {Sondhi}}]{Harper2019}%
  \BibitemOpen
  \bibfield  {author} {\bibinfo {author} {\bibfnamefont {Fenner}\ \bibnamefont
  {Harper}}, \bibinfo {author} {\bibfnamefont {Rahul}\ \bibnamefont {Roy}},
  \bibinfo {author} {\bibfnamefont {Mark~S.}\ \bibnamefont {Rudner}}, \ and\
  \bibinfo {author} {\bibfnamefont {S.~L.}\ \bibnamefont {Sondhi}},\ }\bibfield
   {title} {\enquote {\bibinfo {title} {{Topology and Broken Symmetry in
  Floquet Systems}},}\ }\href@noop {} {\bibfield  {journal} {\bibinfo
  {journal} {arXiv:1905.01317}\ } (\bibinfo {year} {2019})}\BibitemShut
  {NoStop}%
\bibitem [{\citenamefont {von Keyserlingk}\ and\ \citenamefont
  {Sondhi}(2016)}]{vonKeyserlingk2016}%
  \BibitemOpen
  \bibfield  {author} {\bibinfo {author} {\bibfnamefont {C.~W.}\ \bibnamefont
  {von Keyserlingk}}\ and\ \bibinfo {author} {\bibfnamefont {S.~L.}\
  \bibnamefont {Sondhi}},\ }\bibfield  {title} {\enquote {\bibinfo {title}
  {{Phase structure of one-dimensional interacting Floquet systems. I. Abelian
  symmetry-protected topological phases}},}\ }\href@noop {} {\bibfield
  {journal} {\bibinfo  {journal} {Phys. Rev. B}\ }\textbf {\bibinfo {volume}
  {93}},\ \bibinfo {pages} {245145} (\bibinfo {year} {2016})}\BibitemShut
  {NoStop}%
\bibitem [{\citenamefont {Potter}\ \emph {et~al.}(2016)\citenamefont {Potter},
  \citenamefont {Morimoto},\ and\ \citenamefont {Vishwanath}}]{Potter2016}%
  \BibitemOpen
  \bibfield  {author} {\bibinfo {author} {\bibfnamefont {Andrew~C.}\
  \bibnamefont {Potter}}, \bibinfo {author} {\bibfnamefont {Takahiro}\
  \bibnamefont {Morimoto}}, \ and\ \bibinfo {author} {\bibfnamefont {Ashvin}\
  \bibnamefont {Vishwanath}},\ }\bibfield  {title} {\enquote {\bibinfo {title}
  {{Classification of Interacting Topological Floquet Phases in One
  Dimension}},}\ }\href@noop {} {\bibfield  {journal} {\bibinfo  {journal}
  {Phys. Rev. X}\ }\textbf {\bibinfo {volume} {6}},\ \bibinfo {pages} {041001}
  (\bibinfo {year} {2016})}\BibitemShut {NoStop}%
\bibitem [{\citenamefont {Else}\ and\ \citenamefont {Nayak}(2016)}]{Else2016}%
  \BibitemOpen
  \bibfield  {author} {\bibinfo {author} {\bibfnamefont {Dominic~V.}\
  \bibnamefont {Else}}\ and\ \bibinfo {author} {\bibfnamefont {Chetan}\
  \bibnamefont {Nayak}},\ }\bibfield  {title} {\enquote {\bibinfo {title}
  {{Classification of topological phases in periodically driven interacting
  systems}},}\ }\href@noop {} {\bibfield  {journal} {\bibinfo  {journal} {Phys.
  Rev. B}\ }\textbf {\bibinfo {volume} {93}},\ \bibinfo {pages} {201103}
  (\bibinfo {year} {2016})}\BibitemShut {NoStop}%
\bibitem [{\citenamefont {Harper}\ and\ \citenamefont
  {Roy}(2017)}]{Harper2017}%
  \BibitemOpen
  \bibfield  {author} {\bibinfo {author} {\bibfnamefont {Fenner}\ \bibnamefont
  {Harper}}\ and\ \bibinfo {author} {\bibfnamefont {Rahul}\ \bibnamefont
  {Roy}},\ }\bibfield  {title} {\enquote {\bibinfo {title} {{Floquet
  Topological Order in Interacting Systems of Bosons and Fermions}},}\
  }\href@noop {} {\bibfield  {journal} {\bibinfo  {journal} {Phys. Rev. Lett.}\
  }\textbf {\bibinfo {volume} {118}},\ \bibinfo {pages} {115301} (\bibinfo
  {year} {2017})}\BibitemShut {NoStop}%
\bibitem [{\citenamefont {Moessner}\ and\ \citenamefont
  {Sondhi}(2017)}]{Moessner2017}%
  \BibitemOpen
  \bibfield  {author} {\bibinfo {author} {\bibfnamefont {R.}~\bibnamefont
  {Moessner}}\ and\ \bibinfo {author} {\bibfnamefont {S.~L.}\ \bibnamefont
  {Sondhi}},\ }\bibfield  {title} {\enquote {\bibinfo {title} {{Equilibration
  and order in quantum Floquet matter}},}\ }\href@noop {} {\bibfield  {journal}
  {\bibinfo  {journal} {Nature Physics}\ }\textbf {\bibinfo {volume} {13}},\
  \bibinfo {pages} {424} (\bibinfo {year} {2017})}\BibitemShut {NoStop}%
\bibitem [{\citenamefont {Jotzu}\ \emph {et~al.}(2014)\citenamefont {Jotzu},
  \citenamefont {Messer}, \citenamefont {Desbuquois}, \citenamefont {Lebrat},
  \citenamefont {Uehlinger}, \citenamefont {Greif},\ and\ \citenamefont
  {Esslinger}}]{Jotzu2014}%
  \BibitemOpen
  \bibfield  {author} {\bibinfo {author} {\bibfnamefont {G.}~\bibnamefont
  {Jotzu}}, \bibinfo {author} {\bibfnamefont {M.}~\bibnamefont {Messer}},
  \bibinfo {author} {\bibfnamefont {R.}~\bibnamefont {Desbuquois}}, \bibinfo
  {author} {\bibfnamefont {M.}~\bibnamefont {Lebrat}}, \bibinfo {author}
  {\bibfnamefont {T.}~\bibnamefont {Uehlinger}}, \bibinfo {author}
  {\bibfnamefont {D.}~\bibnamefont {Greif}}, \ and\ \bibinfo {author}
  {\bibfnamefont {T.}~\bibnamefont {Esslinger}},\ }\bibfield  {title} {\enquote
  {\bibinfo {title} {{Experimental realization of the topological Haldane model
  with ultracold fermions}},}\ }\href@noop {} {\bibfield  {journal} {\bibinfo
  {journal} {Nature}\ }\textbf {\bibinfo {volume} {515}},\ \bibinfo {pages}
  {237} (\bibinfo {year} {2014})}\BibitemShut {NoStop}%
\bibitem [{\citenamefont {Flaschner}\ \emph {et~al.}(2016)\citenamefont
  {Flaschner}, \citenamefont {Rem}, \citenamefont {Tarnowski}, \citenamefont
  {Vogel}, \citenamefont {L{\"u}hmann}, \citenamefont {Sengstock},\ and\
  \citenamefont {Weitenberg}}]{Flaschner2016}%
  \BibitemOpen
  \bibfield  {author} {\bibinfo {author} {\bibfnamefont {N.}~\bibnamefont
  {Flaschner}}, \bibinfo {author} {\bibfnamefont {B.~S.}\ \bibnamefont {Rem}},
  \bibinfo {author} {\bibfnamefont {M.}~\bibnamefont {Tarnowski}}, \bibinfo
  {author} {\bibfnamefont {D.}~\bibnamefont {Vogel}}, \bibinfo {author}
  {\bibfnamefont {D.-S.}\ \bibnamefont {L{\"u}hmann}}, \bibinfo {author}
  {\bibfnamefont {K.}~\bibnamefont {Sengstock}}, \ and\ \bibinfo {author}
  {\bibfnamefont {C.}~\bibnamefont {Weitenberg}},\ }\bibfield  {title}
  {\enquote {\bibinfo {title} {{Experimental reconstruction of the Berry
  curvature in a Floquet Bloch band}},}\ }\href@noop {} {\bibfield  {journal}
  {\bibinfo  {journal} {Science}\ }\textbf {\bibinfo {volume} {352}},\ \bibinfo
  {pages} {1091} (\bibinfo {year} {2016})}\BibitemShut {NoStop}%
\bibitem [{\citenamefont {Rechtsman}\ \emph {et~al.}(2013)\citenamefont
  {Rechtsman}, \citenamefont {Zeuner}, \citenamefont {Plotnik}, \citenamefont
  {Lumer}, \citenamefont {Podolsky}, \citenamefont {Dreisow}, \citenamefont
  {Nolte}, \citenamefont {Segev},\ and\ \citenamefont
  {Szameit}}]{Rechtsman2013}%
  \BibitemOpen
  \bibfield  {author} {\bibinfo {author} {\bibfnamefont {Mikael~C.}\
  \bibnamefont {Rechtsman}}, \bibinfo {author} {\bibfnamefont {Julia~M.}\
  \bibnamefont {Zeuner}}, \bibinfo {author} {\bibfnamefont {Yonatan}\
  \bibnamefont {Plotnik}}, \bibinfo {author} {\bibfnamefont {Yaakov}\
  \bibnamefont {Lumer}}, \bibinfo {author} {\bibfnamefont {Daniel}\
  \bibnamefont {Podolsky}}, \bibinfo {author} {\bibfnamefont {Felix}\
  \bibnamefont {Dreisow}}, \bibinfo {author} {\bibfnamefont {Stefan}\
  \bibnamefont {Nolte}}, \bibinfo {author} {\bibfnamefont {Mordechai}\
  \bibnamefont {Segev}}, \ and\ \bibinfo {author} {\bibfnamefont {Alexander}\
  \bibnamefont {Szameit}},\ }\bibfield  {title} {\enquote {\bibinfo {title}
  {{Photonic Floquet topological insulators}},}\ }\href@noop {} {\bibfield
  {journal} {\bibinfo  {journal} {Nature}\ }\textbf {\bibinfo {volume} {496}},\
  \bibinfo {pages} {196} (\bibinfo {year} {2013})}\BibitemShut {NoStop}%
\bibitem [{\citenamefont {Wang}\ \emph {et~al.}(2013)\citenamefont {Wang},
  \citenamefont {Steinberg}, \citenamefont {Jarillo-Herrero},\ and\
  \citenamefont {Gedik}}]{Wang2013}%
  \BibitemOpen
  \bibfield  {author} {\bibinfo {author} {\bibfnamefont {Y.~H.}\ \bibnamefont
  {Wang}}, \bibinfo {author} {\bibfnamefont {H.}~\bibnamefont {Steinberg}},
  \bibinfo {author} {\bibfnamefont {P.}~\bibnamefont {Jarillo-Herrero}}, \ and\
  \bibinfo {author} {\bibfnamefont {N.}~\bibnamefont {Gedik}},\ }\bibfield
  {title} {\enquote {\bibinfo {title} {{Observation of Floquet-Bloch states on
  the surface of a topological insulator}},}\ }\href@noop {} {\bibfield
  {journal} {\bibinfo  {journal} {Science}\ }\textbf {\bibinfo {volume}
  {342}},\ \bibinfo {pages} {453} (\bibinfo {year} {2013})}\BibitemShut
  {NoStop}%
\bibitem [{\citenamefont {{McIver}}\ \emph {et~al.}(2018)\citenamefont
  {{McIver}}, \citenamefont {{Schulte}}, \citenamefont {{Stein}}, \citenamefont
  {{Matsuyama}}, \citenamefont {{Jotzu}}, \citenamefont {{Meier}},\ and\
  \citenamefont {{Cavalleri}}}]{McIver2018}%
  \BibitemOpen
  \bibfield  {author} {\bibinfo {author} {\bibfnamefont {J.~W.}\ \bibnamefont
  {{McIver}}}, \bibinfo {author} {\bibfnamefont {B.}~\bibnamefont {{Schulte}}},
  \bibinfo {author} {\bibfnamefont {F.-U.}\ \bibnamefont {{Stein}}}, \bibinfo
  {author} {\bibfnamefont {T.}~\bibnamefont {{Matsuyama}}}, \bibinfo {author}
  {\bibfnamefont {G.}~\bibnamefont {{Jotzu}}}, \bibinfo {author} {\bibfnamefont
  {G.}~\bibnamefont {{Meier}}}, \ and\ \bibinfo {author} {\bibfnamefont
  {A.}~\bibnamefont {{Cavalleri}}},\ }\bibfield  {title} {\enquote {\bibinfo
  {title} {{Light-induced anomalous Hall effect in graphene}},}\ }\href@noop {}
  {\bibfield  {journal} {\bibinfo  {journal} {1811.03522}\ } (\bibinfo {year}
  {2018})}\BibitemShut {NoStop}%
\bibitem [{SM()}]{SM}%
  \BibitemOpen
  \href@noop {} {}\bibinfo {note} {See Supplementary Material for: i) an
  explicit calculation demonstrating Floquet gap opening due an off-resonant
  drive, ii) a discussion of the vanishing of $\nu_1$ for any evolution derived
  from a local Hamiltonian, iii) the definition of the time-averaged spectral
  function discussed in Sec.~\ref{sec:Transport}, and iv) a derivation of the
  period-averaged conductivity via the Floquet-Kubo formula.}\BibitemShut
  {Stop}%
\bibitem [{\citenamefont {Holthaus}(2016)}]{Holthaus2016}%
  \BibitemOpen
  \bibfield  {author} {\bibinfo {author} {\bibfnamefont {Martin}\ \bibnamefont
  {Holthaus}},\ }\bibfield  {title} {\enquote {\bibinfo {title} {{Floquet
  engineering with quasienergy bands of periodically driven optical
  lattices}},}\ }\href@noop {} {\bibfield  {journal} {\bibinfo  {journal}
  {Journal of Physics B: Atomic, Molecular and Optical Physics}\ }\textbf
  {\bibinfo {volume} {49}},\ \bibinfo {pages} {013001} (\bibinfo {year}
  {2016})}\BibitemShut {NoStop}%
\bibitem [{\citenamefont {Kitagawa}\ \emph {et~al.}(2011)\citenamefont
  {Kitagawa}, \citenamefont {Oka}, \citenamefont {Brataas}, \citenamefont
  {Fu},\ and\ \citenamefont {Demler}}]{Kitagawa2011}%
  \BibitemOpen
  \bibfield  {author} {\bibinfo {author} {\bibfnamefont {Takuya}\ \bibnamefont
  {Kitagawa}}, \bibinfo {author} {\bibfnamefont {Takashi}\ \bibnamefont {Oka}},
  \bibinfo {author} {\bibfnamefont {Arne}\ \bibnamefont {Brataas}}, \bibinfo
  {author} {\bibfnamefont {Liang}\ \bibnamefont {Fu}}, \ and\ \bibinfo {author}
  {\bibfnamefont {Eugene}\ \bibnamefont {Demler}},\ }\bibfield  {title}
  {\enquote {\bibinfo {title} {{Transport properties of nonequilibrium systems
  under the application of light: Photoinduced quantum Hall insulators without
  Landau levels}},}\ }\href@noop {} {\bibfield  {journal} {\bibinfo  {journal}
  {Phys. Rev. B}\ }\textbf {\bibinfo {volume} {84}},\ \bibinfo {pages} {235108}
  (\bibinfo {year} {2011})}\BibitemShut {NoStop}%
\bibitem [{\citenamefont {Sie}\ \emph {et~al.}(2015)\citenamefont {Sie},
  \citenamefont {McIver}, \citenamefont {Lee}, \citenamefont {Fu},
  \citenamefont {Kong},\ and\ \citenamefont {Gedik}}]{Sie2015}%
  \BibitemOpen
  \bibfield  {author} {\bibinfo {author} {\bibfnamefont {Edbert~J.}\
  \bibnamefont {Sie}}, \bibinfo {author} {\bibfnamefont {James~W.}\
  \bibnamefont {McIver}}, \bibinfo {author} {\bibfnamefont {Yi-Hsien}\
  \bibnamefont {Lee}}, \bibinfo {author} {\bibfnamefont {Liang}\ \bibnamefont
  {Fu}}, \bibinfo {author} {\bibfnamefont {Jing}\ \bibnamefont {Kong}}, \ and\
  \bibinfo {author} {\bibfnamefont {Nuh}\ \bibnamefont {Gedik}},\ }\bibfield
  {title} {\enquote {\bibinfo {title} {{Valley-selective optical Stark effect
  in monolayer WS$_2$}},}\ }\href@noop {} {\bibfield  {journal} {\bibinfo
  {journal} {Nature Materials}\ }\textbf {\bibinfo {volume} {14}},\ \bibinfo
  {pages} {290} (\bibinfo {year} {2015})}\BibitemShut {NoStop}%
\bibitem [{\citenamefont {Usaj}\ \emph {et~al.}(2014)\citenamefont {Usaj},
  \citenamefont {Perez-Piskunow}, \citenamefont {Foa~Torres},\ and\
  \citenamefont {Balseiro}}]{Usaj2014}%
  \BibitemOpen
  \bibfield  {author} {\bibinfo {author} {\bibfnamefont {G.}~\bibnamefont
  {Usaj}}, \bibinfo {author} {\bibfnamefont {P.~M.}\ \bibnamefont
  {Perez-Piskunow}}, \bibinfo {author} {\bibfnamefont {L.~E.~F.}\ \bibnamefont
  {Foa~Torres}}, \ and\ \bibinfo {author} {\bibfnamefont {C.~A.}\ \bibnamefont
  {Balseiro}},\ }\bibfield  {title} {\enquote {\bibinfo {title} {{Irradiated
  graphene as a tunable Floquet topological insulator}},}\ }\href@noop {}
  {\bibfield  {journal} {\bibinfo  {journal} {Phys. Rev. B}\ }\textbf {\bibinfo
  {volume} {90}},\ \bibinfo {pages} {115423} (\bibinfo {year}
  {2014})}\BibitemShut {NoStop}%
\bibitem [{\citenamefont {Quelle}\ \emph {et~al.}(2016)\citenamefont {Quelle},
  \citenamefont {Goerbig},\ and\ \citenamefont {Smith}}]{Quelle2016}%
  \BibitemOpen
  \bibfield  {author} {\bibinfo {author} {\bibfnamefont {A.}~\bibnamefont
  {Quelle}}, \bibinfo {author} {\bibfnamefont {M.~O.}\ \bibnamefont {Goerbig}},
  \ and\ \bibinfo {author} {\bibfnamefont {C.~Morais}\ \bibnamefont {Smith}},\
  }\bibfield  {title} {\enquote {\bibinfo {title} {{Bandwidth-resonant Floquet
  states in honeycomb optical lattices}},}\ }\href@noop {} {\bibfield
  {journal} {\bibinfo  {journal} {New J. Phys.}\ }\textbf {\bibinfo {volume}
  {18}},\ \bibinfo {pages} {015006} (\bibinfo {year} {2016})}\BibitemShut
  {NoStop}%
\bibitem [{\citenamefont {Gu}\ \emph {et~al.}(2011)\citenamefont {Gu},
  \citenamefont {Fertig}, \citenamefont {Arovas},\ and\ \citenamefont
  {Auerbach}}]{Gu2011}%
  \BibitemOpen
  \bibfield  {author} {\bibinfo {author} {\bibfnamefont {Zhenghao}\
  \bibnamefont {Gu}}, \bibinfo {author} {\bibfnamefont {H.~A.}\ \bibnamefont
  {Fertig}}, \bibinfo {author} {\bibfnamefont {Daniel~P.}\ \bibnamefont
  {Arovas}}, \ and\ \bibinfo {author} {\bibfnamefont {Assa}\ \bibnamefont
  {Auerbach}},\ }\bibfield  {title} {\enquote {\bibinfo {title} {{Floquet
  Spectrum and Transport through an Irradiated Graphene Ribbon}},}\ }\href@noop
  {} {\bibfield  {journal} {\bibinfo  {journal} {Phys. Rev. Lett.}\ }\textbf
  {\bibinfo {volume} {107}},\ \bibinfo {pages} {216601} (\bibinfo {year}
  {2011})}\BibitemShut {NoStop}%
\bibitem [{\citenamefont {Rodriguez-Vega}\ and\ \citenamefont
  {Seradjeh}(2018)}]{Rodriguez-Vega2018}%
  \BibitemOpen
  \bibfield  {author} {\bibinfo {author} {\bibfnamefont {M.}~\bibnamefont
  {Rodriguez-Vega}}\ and\ \bibinfo {author} {\bibfnamefont {B.}~\bibnamefont
  {Seradjeh}},\ }\bibfield  {title} {\enquote {\bibinfo {title} {{Universal
  Fluctuations of Floquet Topological Invariants at Low Frequencies}},}\
  }\href@noop {} {\bibfield  {journal} {\bibinfo  {journal} {Phys. Rev. Lett.}\
  }\textbf {\bibinfo {volume} {121}},\ \bibinfo {pages} {036402} (\bibinfo
  {year} {2018})}\BibitemShut {NoStop}%
\bibitem [{\citenamefont {Delplace}\ \emph {et~al.}(2013)\citenamefont
  {Delplace}, \citenamefont {G\'omez-Le\'on},\ and\ \citenamefont
  {Platero}}]{Delplace2013}%
  \BibitemOpen
  \bibfield  {author} {\bibinfo {author} {\bibfnamefont {Pierre}\ \bibnamefont
  {Delplace}}, \bibinfo {author} {\bibfnamefont {\'Alvaro}\ \bibnamefont
  {G\'omez-Le\'on}}, \ and\ \bibinfo {author} {\bibfnamefont {Gloria}\
  \bibnamefont {Platero}},\ }\bibfield  {title} {\enquote {\bibinfo {title}
  {{Merging of Dirac points and Floquet topological transitions in ac-driven
  graphene}},}\ }\href@noop {} {\bibfield  {journal} {\bibinfo  {journal}
  {Phys. Rev. B}\ }\textbf {\bibinfo {volume} {88}},\ \bibinfo {pages} {245422}
  (\bibinfo {year} {2013})}\BibitemShut {NoStop}%
\bibitem [{\citenamefont {Sentef}\ \emph {et~al.}(2015)\citenamefont {Sentef},
  \citenamefont {Claassen}, \citenamefont {Kemper}, \citenamefont {Moritz},
  \citenamefont {Oka}, \citenamefont {Freericks},\ and\ \citenamefont
  {Devereaux}}]{Sentef2015}%
  \BibitemOpen
  \bibfield  {author} {\bibinfo {author} {\bibfnamefont {M.~A.}\ \bibnamefont
  {Sentef}}, \bibinfo {author} {\bibfnamefont {M.}~\bibnamefont {Claassen}},
  \bibinfo {author} {\bibfnamefont {A.~F.}\ \bibnamefont {Kemper}}, \bibinfo
  {author} {\bibfnamefont {B.}~\bibnamefont {Moritz}}, \bibinfo {author}
  {\bibfnamefont {T.}~\bibnamefont {Oka}}, \bibinfo {author} {\bibfnamefont
  {J.~K.}\ \bibnamefont {Freericks}}, \ and\ \bibinfo {author} {\bibfnamefont
  {T.~P.}\ \bibnamefont {Devereaux}},\ }\bibfield  {title} {\enquote {\bibinfo
  {title} {{Theory of Floquet band formation and local pseudospin textures in
  pump-probe photoemission of graphene}},}\ }\href@noop {} {\bibfield
  {journal} {\bibinfo  {journal} {Nature Comm.}\ }\textbf {\bibinfo {volume}
  {6}},\ \bibinfo {pages} {7047} (\bibinfo {year} {2015})}\BibitemShut
  {NoStop}%
\bibitem [{\citenamefont {Iadecola}\ \emph {et~al.}(2013)\citenamefont
  {Iadecola}, \citenamefont {Campbell}, \citenamefont {Chamon}, \citenamefont
  {Hou}, \citenamefont {Jackiw}, \citenamefont {Pi},\ and\ \citenamefont
  {Kusminskiy}}]{Iadecola2013}%
  \BibitemOpen
  \bibfield  {author} {\bibinfo {author} {\bibfnamefont {Thomas}\ \bibnamefont
  {Iadecola}}, \bibinfo {author} {\bibfnamefont {David}\ \bibnamefont
  {Campbell}}, \bibinfo {author} {\bibfnamefont {Claudio}\ \bibnamefont
  {Chamon}}, \bibinfo {author} {\bibfnamefont {Chang-Yu}\ \bibnamefont {Hou}},
  \bibinfo {author} {\bibfnamefont {Roman}\ \bibnamefont {Jackiw}}, \bibinfo
  {author} {\bibfnamefont {So-Young}\ \bibnamefont {Pi}}, \ and\ \bibinfo
  {author} {\bibfnamefont {Silvia~Viola}\ \bibnamefont {Kusminskiy}},\
  }\bibfield  {title} {\enquote {\bibinfo {title} {{Materials Design from
  Nonequilibrium Steady States: Driven Graphene as a Tunable Semiconductor with
  Topological Properties}},}\ }\href@noop {} {\bibfield  {journal} {\bibinfo
  {journal} {Phys. Rev. Lett.}\ }\textbf {\bibinfo {volume} {110}},\ \bibinfo
  {pages} {176603} (\bibinfo {year} {2013})}\BibitemShut {NoStop}%
\bibitem [{\citenamefont {Jiang}\ \emph {et~al.}(2011)\citenamefont {Jiang},
  \citenamefont {Kitagawa}, \citenamefont {Alicea}, \citenamefont {Akhmerov},
  \citenamefont {Pekker}, \citenamefont {Refael}, \citenamefont {Cirac},
  \citenamefont {Demler}, \citenamefont {Lukin},\ and\ \citenamefont
  {Zoller}}]{Jiang2011}%
  \BibitemOpen
  \bibfield  {author} {\bibinfo {author} {\bibfnamefont {Liang}\ \bibnamefont
  {Jiang}}, \bibinfo {author} {\bibfnamefont {Takuya}\ \bibnamefont
  {Kitagawa}}, \bibinfo {author} {\bibfnamefont {Jason}\ \bibnamefont
  {Alicea}}, \bibinfo {author} {\bibfnamefont {A.~R.}\ \bibnamefont
  {Akhmerov}}, \bibinfo {author} {\bibfnamefont {David}\ \bibnamefont
  {Pekker}}, \bibinfo {author} {\bibfnamefont {Gil}\ \bibnamefont {Refael}},
  \bibinfo {author} {\bibfnamefont {J.~Ignacio}\ \bibnamefont {Cirac}},
  \bibinfo {author} {\bibfnamefont {Eugene}\ \bibnamefont {Demler}}, \bibinfo
  {author} {\bibfnamefont {Mikhail~D.}\ \bibnamefont {Lukin}}, \ and\ \bibinfo
  {author} {\bibfnamefont {Peter}\ \bibnamefont {Zoller}},\ }\bibfield  {title}
  {\enquote {\bibinfo {title} {{Majorana Fermions in Equilibrium and in Driven
  Cold-Atom Quantum Wires}},}\ }\href@noop {} {\bibfield  {journal} {\bibinfo
  {journal} {Phys. Rev. Lett.}\ }\textbf {\bibinfo {volume} {106}},\ \bibinfo
  {pages} {220402} (\bibinfo {year} {2011})}\BibitemShut {NoStop}%
\bibitem [{\citenamefont {Thakurathi}\ \emph {et~al.}(2017)\citenamefont
  {Thakurathi}, \citenamefont {Loss},\ and\ \citenamefont
  {Klinovaja}}]{Thakurathi2017}%
  \BibitemOpen
  \bibfield  {author} {\bibinfo {author} {\bibfnamefont {M.}~\bibnamefont
  {Thakurathi}}, \bibinfo {author} {\bibfnamefont {D.}~\bibnamefont {Loss}}, \
  and\ \bibinfo {author} {\bibfnamefont {J.}~\bibnamefont {Klinovaja}},\
  }\bibfield  {title} {\enquote {\bibinfo {title} {{Floquet Majorana fermions
  and parafermions in driven Rashba nanowires}},}\ }\href@noop {} {\bibfield
  {journal} {\bibinfo  {journal} {Phys. Rev. B}\ }\textbf {\bibinfo {volume}
  {95}},\ \bibinfo {pages} {155407} (\bibinfo {year} {2017})}\BibitemShut
  {NoStop}%
\bibitem [{\citenamefont {{Kennes}}\ \emph {et~al.}(2018)\citenamefont
  {{Kennes}}, \citenamefont {{M{\"u}ller}}, \citenamefont {{Pletyukhov}},
  \citenamefont {{Weber}}, \citenamefont {{Bruder}}, \citenamefont {{Hassler}},
  \citenamefont {{Klinovaja}}, \citenamefont {{Loss}},\ and\ \citenamefont
  {{Schoeller}}}]{Kennes2018b}%
  \BibitemOpen
  \bibfield  {author} {\bibinfo {author} {\bibfnamefont {Dante~M.}\
  \bibnamefont {{Kennes}}}, \bibinfo {author} {\bibfnamefont {Niclas}\
  \bibnamefont {{M{\"u}ller}}}, \bibinfo {author} {\bibfnamefont {Mikhail}\
  \bibnamefont {{Pletyukhov}}}, \bibinfo {author} {\bibfnamefont {Clara}\
  \bibnamefont {{Weber}}}, \bibinfo {author} {\bibfnamefont {Christoph}\
  \bibnamefont {{Bruder}}}, \bibinfo {author} {\bibfnamefont {Fabian}\
  \bibnamefont {{Hassler}}}, \bibinfo {author} {\bibfnamefont {Jelena}\
  \bibnamefont {{Klinovaja}}}, \bibinfo {author} {\bibfnamefont {Daniel}\
  \bibnamefont {{Loss}}}, \ and\ \bibinfo {author} {\bibfnamefont {Herbert}\
  \bibnamefont {{Schoeller}}},\ }\bibfield  {title} {\enquote {\bibinfo {title}
  {{Chiral 1D Floquet topological insulators beyond rotating wave
  approximation}},}\ }\href@noop {} {\bibfield  {journal} {\bibinfo  {journal}
  {arXiv:1811.12062}\ } (\bibinfo {year} {2018})}\BibitemShut {NoStop}%
\bibitem [{\citenamefont {Lindner}\ \emph {et~al.}(2013)\citenamefont
  {Lindner}, \citenamefont {Bergman}, \citenamefont {Refael},\ and\
  \citenamefont {Galitski}}]{Lindner2013}%
  \BibitemOpen
  \bibfield  {author} {\bibinfo {author} {\bibfnamefont {Netanel~H.}\
  \bibnamefont {Lindner}}, \bibinfo {author} {\bibfnamefont {Doron~L.}\
  \bibnamefont {Bergman}}, \bibinfo {author} {\bibfnamefont {Gil}\ \bibnamefont
  {Refael}}, \ and\ \bibinfo {author} {\bibfnamefont {Victor}\ \bibnamefont
  {Galitski}},\ }\bibfield  {title} {\enquote {\bibinfo {title} {{Topological
  Floquet spectrum in three dimensions via a two-photon resonance}},}\
  }\href@noop {} {\bibfield  {journal} {\bibinfo  {journal} {Phys. Rev. B}\
  }\textbf {\bibinfo {volume} {87}},\ \bibinfo {pages} {235131} (\bibinfo
  {year} {2013})}\BibitemShut {NoStop}%
\bibitem [{\citenamefont {Wang}\ \emph {et~al.}(2014)\citenamefont {Wang},
  \citenamefont {Wang}, \citenamefont {Shen}, \citenamefont {Sheng},\ and\
  \citenamefont {Xing}}]{Wang2014}%
  \BibitemOpen
  \bibfield  {author} {\bibinfo {author} {\bibfnamefont {Rui}\ \bibnamefont
  {Wang}}, \bibinfo {author} {\bibfnamefont {Baigeng}\ \bibnamefont {Wang}},
  \bibinfo {author} {\bibfnamefont {Rui}\ \bibnamefont {Shen}}, \bibinfo
  {author} {\bibfnamefont {L.}~\bibnamefont {Sheng}}, \ and\ \bibinfo {author}
  {\bibfnamefont {D.~Y.}\ \bibnamefont {Xing}},\ }\bibfield  {title} {\enquote
  {\bibinfo {title} {{Floquet Weyl semimetal induced by off-resonant light}},}\
  }\href@noop {} {\bibfield  {journal} {\bibinfo  {journal} {Europhys. Lett.}\
  }\textbf {\bibinfo {volume} {105}},\ \bibinfo {pages} {17004} (\bibinfo
  {year} {2014})}\BibitemShut {NoStop}%
\bibitem [{\citenamefont {Chan}\ \emph
  {et~al.}(2016{\natexlab{a}})\citenamefont {Chan}, \citenamefont {Lee},
  \citenamefont {Burch}, \citenamefont {Han},\ and\ \citenamefont
  {Ran}}]{Chan2016a}%
  \BibitemOpen
  \bibfield  {author} {\bibinfo {author} {\bibfnamefont {Ching-Kit}\
  \bibnamefont {Chan}}, \bibinfo {author} {\bibfnamefont {Patrick~A.}\
  \bibnamefont {Lee}}, \bibinfo {author} {\bibfnamefont {Kenneth~S.}\
  \bibnamefont {Burch}}, \bibinfo {author} {\bibfnamefont {Jung~Hoon}\
  \bibnamefont {Han}}, \ and\ \bibinfo {author} {\bibfnamefont {Ying}\
  \bibnamefont {Ran}},\ }\bibfield  {title} {\enquote {\bibinfo {title} {{When
  chiral photons meet chiral fermions: photoinduced anomalous Hall effects in
  Weyl semimetals}},}\ }\href@noop {} {\bibfield  {journal} {\bibinfo
  {journal} {Phys. Rev. Lett.}\ }\textbf {\bibinfo {volume} {116}},\ \bibinfo
  {pages} {026805} (\bibinfo {year} {2016}{\natexlab{a}})}\BibitemShut
  {NoStop}%
\bibitem [{\citenamefont {Chan}\ \emph
  {et~al.}(2016{\natexlab{b}})\citenamefont {Chan}, \citenamefont {Oh},
  \citenamefont {Han},\ and\ \citenamefont {Lee}}]{Chan2016b}%
  \BibitemOpen
  \bibfield  {author} {\bibinfo {author} {\bibfnamefont {Ching-Kit}\
  \bibnamefont {Chan}}, \bibinfo {author} {\bibfnamefont {Yun-Tak}\
  \bibnamefont {Oh}}, \bibinfo {author} {\bibfnamefont {Jung~Hoon}\
  \bibnamefont {Han}}, \ and\ \bibinfo {author} {\bibfnamefont {Patrick~A.}\
  \bibnamefont {Lee}},\ }\bibfield  {title} {\enquote {\bibinfo {title}
  {{Type-II Weyl cone transitions in driven semimetals}},}\ }\href@noop {}
  {\bibfield  {journal} {\bibinfo  {journal} {Phys. Rev. B}\ }\textbf {\bibinfo
  {volume} {94}},\ \bibinfo {pages} {121106} (\bibinfo {year}
  {2016}{\natexlab{b}})}\BibitemShut {NoStop}%
\bibitem [{\citenamefont {H{\"u}bener}\ \emph {et~al.}(2017)\citenamefont
  {H{\"u}bener}, \citenamefont {Sentef}, \citenamefont {de~Giovannini},
  \citenamefont {Kemper},\ and\ \citenamefont {Rubio}}]{Huebener2017}%
  \BibitemOpen
  \bibfield  {author} {\bibinfo {author} {\bibfnamefont {H.}~\bibnamefont
  {H{\"u}bener}}, \bibinfo {author} {\bibfnamefont {M.~A.}\ \bibnamefont
  {Sentef}}, \bibinfo {author} {\bibfnamefont {U.}~\bibnamefont
  {de~Giovannini}}, \bibinfo {author} {\bibfnamefont {A.~F.}\ \bibnamefont
  {Kemper}}, \ and\ \bibinfo {author} {\bibfnamefont {A.}~\bibnamefont
  {Rubio}},\ }\bibfield  {title} {\enquote {\bibinfo {title} {{Creating stable
  Floquet-Weyl semimetals by laser-driving of 3D Dirac materials}},}\
  }\href@noop {} {\bibfield  {journal} {\bibinfo  {journal} {Nature Comm.}\
  }\textbf {\bibinfo {volume} {8}},\ \bibinfo {pages} {13940} (\bibinfo {year}
  {2017})}\BibitemShut {NoStop}%
\bibitem [{\citenamefont {Dalibard}\ \emph {et~al.}(2011)\citenamefont
  {Dalibard}, \citenamefont {Gerbier}, \citenamefont
  {Juzeli\ifmmode~\bar{u}\else \={u}\fi{}nas},\ and\ \citenamefont
  {\"Ohberg}}]{Dalibard2011}%
  \BibitemOpen
  \bibfield  {author} {\bibinfo {author} {\bibfnamefont {Jean}\ \bibnamefont
  {Dalibard}}, \bibinfo {author} {\bibfnamefont {Fabrice}\ \bibnamefont
  {Gerbier}}, \bibinfo {author} {\bibfnamefont {Gediminas}\ \bibnamefont
  {Juzeli\ifmmode~\bar{u}\else \={u}\fi{}nas}}, \ and\ \bibinfo {author}
  {\bibfnamefont {Patrik}\ \bibnamefont {\"Ohberg}},\ }\bibfield  {title}
  {\enquote {\bibinfo {title} {Colloquium: Artificial gauge potentials for
  neutral atoms},}\ }\href@noop {} {\bibfield  {journal} {\bibinfo  {journal}
  {Rev. Mod. Phys.}\ }\textbf {\bibinfo {volume} {83}},\ \bibinfo {pages}
  {1523--1543} (\bibinfo {year} {2011})}\BibitemShut {NoStop}%
\bibitem [{\citenamefont {Goldman}\ \emph {et~al.}(2014)\citenamefont
  {Goldman}, \citenamefont {Juzeli{\={u}}nas}, \citenamefont {Öhberg},\ and\
  \citenamefont {Spielman}}]{Goldman2014}%
  \BibitemOpen
  \bibfield  {author} {\bibinfo {author} {\bibfnamefont {N}~\bibnamefont
  {Goldman}}, \bibinfo {author} {\bibfnamefont {G}~\bibnamefont
  {Juzeli{\={u}}nas}}, \bibinfo {author} {\bibfnamefont {P}~\bibnamefont
  {Öhberg}}, \ and\ \bibinfo {author} {\bibfnamefont {I~B}\ \bibnamefont
  {Spielman}},\ }\bibfield  {title} {\enquote {\bibinfo {title} {Light-induced
  gauge fields for ultracold atoms},}\ }\href@noop {} {\bibfield  {journal}
  {\bibinfo  {journal} {Reports on Progress in Physics}\ }\textbf {\bibinfo
  {volume} {77}},\ \bibinfo {pages} {126401} (\bibinfo {year}
  {2014})}\BibitemShut {NoStop}%
\bibitem [{\citenamefont {Cooper}\ \emph {et~al.}(2019)\citenamefont {Cooper},
  \citenamefont {Dalibard},\ and\ \citenamefont {Spielman}}]{Cooper2019}%
  \BibitemOpen
  \bibfield  {author} {\bibinfo {author} {\bibfnamefont {N.~R.}\ \bibnamefont
  {Cooper}}, \bibinfo {author} {\bibfnamefont {J.}~\bibnamefont {Dalibard}}, \
  and\ \bibinfo {author} {\bibfnamefont {I.~B.}\ \bibnamefont {Spielman}},\
  }\bibfield  {title} {\enquote {\bibinfo {title} {Topological bands for
  ultracold atoms},}\ }\href@noop {} {\bibfield  {journal} {\bibinfo  {journal}
  {Rev. Mod. Phys.}\ }\textbf {\bibinfo {volume} {91}},\ \bibinfo {pages}
  {015005} (\bibinfo {year} {2019})}\BibitemShut {NoStop}%
\bibitem [{\citenamefont {Kitaev}(2009)}]{Kitaev2009}%
  \BibitemOpen
  \bibfield  {author} {\bibinfo {author} {\bibfnamefont {A.}~\bibnamefont
  {Kitaev}},\ }\bibfield  {title} {\enquote {\bibinfo {title} {{Periodic table
  for topological insulators and superconductors}},}\ }\href@noop {} {\bibfield
   {journal} {\bibinfo  {journal} {AIP Conf. Proc.}\ }\textbf {\bibinfo
  {volume} {1134}},\ \bibinfo {pages} {22} (\bibinfo {year}
  {2009})}\BibitemShut {NoStop}%
\bibitem [{\citenamefont {Ryu}\ \emph {et~al.}(2010)\citenamefont {Ryu},
  \citenamefont {Schnyder}, \citenamefont {Furusaki},\ and\ \citenamefont
  {Ludwig}}]{Ryu2010}%
  \BibitemOpen
  \bibfield  {author} {\bibinfo {author} {\bibfnamefont {Shinsei}\ \bibnamefont
  {Ryu}}, \bibinfo {author} {\bibfnamefont {Andreas~P.}\ \bibnamefont
  {Schnyder}}, \bibinfo {author} {\bibfnamefont {Akira}\ \bibnamefont
  {Furusaki}}, \ and\ \bibinfo {author} {\bibfnamefont {Andreas W.~W.}\
  \bibnamefont {Ludwig}},\ }\bibfield  {title} {\enquote {\bibinfo {title}
  {{Topological insulators and superconductors: tenfold way and dimensional
  hierarchy}},}\ }\href@noop {} {\bibfield  {journal} {\bibinfo  {journal} {New
  J. Phys.}\ }\textbf {\bibinfo {volume} {12}},\ \bibinfo {pages} {065010}
  (\bibinfo {year} {2010})}\BibitemShut {NoStop}%
\bibitem [{\citenamefont {Thouless}(1983)}]{ThoulessPump}%
  \BibitemOpen
  \bibfield  {author} {\bibinfo {author} {\bibfnamefont {D.~J.}\ \bibnamefont
  {Thouless}},\ }\bibfield  {title} {\enquote {\bibinfo {title} {Quantization
  of particle transport},}\ }\href@noop {} {\bibfield  {journal} {\bibinfo
  {journal} {Phys. Rev. B}\ }\textbf {\bibinfo {volume} {27}},\ \bibinfo
  {pages} {6083} (\bibinfo {year} {1983})}\BibitemShut {NoStop}%
\bibitem [{\citenamefont {Gross}\ \emph {et~al.}(2012)\citenamefont {Gross},
  \citenamefont {Nesme}, \citenamefont {Vogts},\ and\ \citenamefont
  {Werner}}]{GNVW}%
  \BibitemOpen
  \bibfield  {author} {\bibinfo {author} {\bibfnamefont {D.}~\bibnamefont
  {Gross}}, \bibinfo {author} {\bibfnamefont {V.}~\bibnamefont {Nesme}},
  \bibinfo {author} {\bibfnamefont {H.}~\bibnamefont {Vogts}}, \ and\ \bibinfo
  {author} {\bibfnamefont {R.~F.}\ \bibnamefont {Werner}},\ }\bibfield  {title}
  {\enquote {\bibinfo {title} {Index theory of one dimensional quantum walks
  and cellular automata},}\ }\href@noop {} {\bibfield  {journal} {\bibinfo
  {journal} {Communications in Mathematical Physics}\ }\textbf {\bibinfo
  {volume} {310}},\ \bibinfo {pages} {419} (\bibinfo {year}
  {2012})}\BibitemShut {NoStop}%
\bibitem [{\citenamefont {Higashikawa}\ \emph {et~al.}(2019)\citenamefont
  {Higashikawa}, \citenamefont {Nakagawa},\ and\ \citenamefont
  {Ueda}}]{Higashikawa2018}%
  \BibitemOpen
  \bibfield  {author} {\bibinfo {author} {\bibfnamefont {Sho}\ \bibnamefont
  {Higashikawa}}, \bibinfo {author} {\bibfnamefont {Masaya}\ \bibnamefont
  {Nakagawa}}, \ and\ \bibinfo {author} {\bibfnamefont {Masahito}\ \bibnamefont
  {Ueda}},\ }\bibfield  {title} {\enquote {\bibinfo {title} {Floquet chiral
  magnetic effect},}\ }\href@noop {} {\bibfield  {journal} {\bibinfo  {journal}
  {Phys. Rev. Lett.}\ }\textbf {\bibinfo {volume} {123}},\ \bibinfo {pages}
  {066403} (\bibinfo {year} {2019})}\BibitemShut {NoStop}%
\bibitem [{\citenamefont {Sun}\ \emph {et~al.}(2018)\citenamefont {Sun},
  \citenamefont {Xiao}, \citenamefont {Bzdu\ifmmode~\check{s}\else
  \v{s}\fi{}ek}, \citenamefont {Zhang},\ and\ \citenamefont {Fan}}]{Sun2018}%
  \BibitemOpen
  \bibfield  {author} {\bibinfo {author} {\bibfnamefont {Xiao-Qi}\ \bibnamefont
  {Sun}}, \bibinfo {author} {\bibfnamefont {Meng}\ \bibnamefont {Xiao}},
  \bibinfo {author} {\bibfnamefont {Tom\'a\ifmmode \check{s}\else~\v{s}\fi{}}\
  \bibnamefont {Bzdu\ifmmode~\check{s}\else \v{s}\fi{}ek}}, \bibinfo {author}
  {\bibfnamefont {Shou-Cheng}\ \bibnamefont {Zhang}}, \ and\ \bibinfo {author}
  {\bibfnamefont {Shanhui}\ \bibnamefont {Fan}},\ }\bibfield  {title} {\enquote
  {\bibinfo {title} {{Three-Dimensional Chiral Lattice Fermion in Floquet
  Systems}},}\ }\href@noop {} {\bibfield  {journal} {\bibinfo  {journal} {Phys.
  Rev. Lett.}\ }\textbf {\bibinfo {volume} {121}},\ \bibinfo {pages} {196401}
  (\bibinfo {year} {2018})}\BibitemShut {NoStop}%
\bibitem [{\citenamefont {Bernevig}\ \emph {et~al.}(2006)\citenamefont
  {Bernevig}, \citenamefont {Hughes},\ and\ \citenamefont
  {Zhang}}]{Bernevig2006}%
  \BibitemOpen
  \bibfield  {author} {\bibinfo {author} {\bibfnamefont {B.~Andrei}\
  \bibnamefont {Bernevig}}, \bibinfo {author} {\bibfnamefont {Taylor~L.}\
  \bibnamefont {Hughes}}, \ and\ \bibinfo {author} {\bibfnamefont {Shou-Cheng}\
  \bibnamefont {Zhang}},\ }\bibfield  {title} {\enquote {\bibinfo {title}
  {{Quantum Spin Hall Effect and Topological Phase Transition in HgTe Quantum
  Wells}},}\ }\href@noop {} {\bibfield  {journal} {\bibinfo  {journal}
  {Science}\ }\textbf {\bibinfo {volume} {314}},\ \bibinfo {pages} {1757}
  (\bibinfo {year} {2006})}\BibitemShut {NoStop}%
\bibitem [{\citenamefont {Halperin}(1982)}]{Halperin1982}%
  \BibitemOpen
  \bibfield  {author} {\bibinfo {author} {\bibfnamefont {B.~I.}\ \bibnamefont
  {Halperin}},\ }\bibfield  {title} {\enquote {\bibinfo {title} {{Quantized
  Hall conductance, current-carrying edge states, and the existence of extended
  states in a two-dimensional disordered potential}},}\ }\href@noop {}
  {\bibfield  {journal} {\bibinfo  {journal} {Phys. Rev. B}\ }\textbf {\bibinfo
  {volume} {25}},\ \bibinfo {pages} {2185--2190} (\bibinfo {year}
  {1982})}\BibitemShut {NoStop}%
\bibitem [{\citenamefont {Hatsugai}(1993)}]{Hatsugai1993}%
  \BibitemOpen
  \bibfield  {author} {\bibinfo {author} {\bibfnamefont {Yasuhiro}\
  \bibnamefont {Hatsugai}},\ }\bibfield  {title} {\enquote {\bibinfo {title}
  {{Chern number and edge states in the integer quantum Hall effect}},}\
  }\href@noop {} {\bibfield  {journal} {\bibinfo  {journal} {Phys. Rev. Lett.}\
  }\textbf {\bibinfo {volume} {71}},\ \bibinfo {pages} {3697} (\bibinfo {year}
  {1993})}\BibitemShut {NoStop}%
\bibitem [{\citenamefont {Carpentier}\ \emph {et~al.}(2015)\citenamefont
  {Carpentier}, \citenamefont {Delplace}, \citenamefont {Fruchart},\ and\
  \citenamefont {Gawedzki}}]{Carpentier2015}%
  \BibitemOpen
  \bibfield  {author} {\bibinfo {author} {\bibfnamefont {David}\ \bibnamefont
  {Carpentier}}, \bibinfo {author} {\bibfnamefont {Pierre}\ \bibnamefont
  {Delplace}}, \bibinfo {author} {\bibfnamefont {Michel}\ \bibnamefont
  {Fruchart}}, \ and\ \bibinfo {author} {\bibfnamefont {Krzysztof}\
  \bibnamefont {Gawedzki}},\ }\bibfield  {title} {\enquote {\bibinfo {title}
  {Topological index for periodically driven time-reversal invariant 2d
  systems},}\ }\href@noop {} {\bibfield  {journal} {\bibinfo  {journal} {Phys.
  Rev. Lett.}\ }\textbf {\bibinfo {volume} {114}},\ \bibinfo {pages} {106806}
  (\bibinfo {year} {2015})}\BibitemShut {NoStop}%
\bibitem [{\citenamefont {Broome}\ \emph {et~al.}(2010)\citenamefont {Broome},
  \citenamefont {Fedrizzi}, \citenamefont {Lanyon}, \citenamefont {Kassal},
  \citenamefont {Aspuru-Guzik},\ and\ \citenamefont {White}}]{Broome2010}%
  \BibitemOpen
  \bibfield  {author} {\bibinfo {author} {\bibfnamefont {M.~A.}\ \bibnamefont
  {Broome}}, \bibinfo {author} {\bibfnamefont {A.}~\bibnamefont {Fedrizzi}},
  \bibinfo {author} {\bibfnamefont {B.~P.}\ \bibnamefont {Lanyon}}, \bibinfo
  {author} {\bibfnamefont {I.}~\bibnamefont {Kassal}}, \bibinfo {author}
  {\bibfnamefont {A.}~\bibnamefont {Aspuru-Guzik}}, \ and\ \bibinfo {author}
  {\bibfnamefont {A.~G.}\ \bibnamefont {White}},\ }\bibfield  {title} {\enquote
  {\bibinfo {title} {Discrete single-photon quantum walks with tunable
  decoherence},}\ }\href@noop {} {\bibfield  {journal} {\bibinfo  {journal}
  {Phys. Rev. Lett.}\ }\textbf {\bibinfo {volume} {104}},\ \bibinfo {pages}
  {153602} (\bibinfo {year} {2010})}\BibitemShut {NoStop}%
\bibitem [{\citenamefont {Liu}\ \emph {et~al.}(2013)\citenamefont {Liu},
  \citenamefont {Levchenko},\ and\ \citenamefont {Baranger}}]{Liu2013}%
  \BibitemOpen
  \bibfield  {author} {\bibinfo {author} {\bibfnamefont {Dong~E.}\ \bibnamefont
  {Liu}}, \bibinfo {author} {\bibfnamefont {Alex}\ \bibnamefont {Levchenko}}, \
  and\ \bibinfo {author} {\bibfnamefont {Harold~U.}\ \bibnamefont {Baranger}},\
  }\bibfield  {title} {\enquote {\bibinfo {title} {{Floquet Majorana Fermions
  for Topological Qubits in Superconducting Devices and Cold-Atom Systems}},}\
  }\href@noop {} {\bibfield  {journal} {\bibinfo  {journal} {Phys. Rev. Lett.}\
  }\textbf {\bibinfo {volume} {111}},\ \bibinfo {pages} {047002} (\bibinfo
  {year} {2013})}\BibitemShut {NoStop}%
\bibitem [{\citenamefont {Asb\'oth}\ \emph {et~al.}(2014)\citenamefont
  {Asb\'oth}, \citenamefont {Tarasinski},\ and\ \citenamefont
  {Delplace}}]{Asboth2014}%
  \BibitemOpen
  \bibfield  {author} {\bibinfo {author} {\bibfnamefont {J.~K.}\ \bibnamefont
  {Asb\'oth}}, \bibinfo {author} {\bibfnamefont {B.}~\bibnamefont
  {Tarasinski}}, \ and\ \bibinfo {author} {\bibfnamefont {P.}~\bibnamefont
  {Delplace}},\ }\bibfield  {title} {\enquote {\bibinfo {title} {Chiral
  symmetry and bulk-boundary correspondence in periodically driven
  one-dimensional systems},}\ }\href@noop {} {\bibfield  {journal} {\bibinfo
  {journal} {Phys. Rev. B}\ }\textbf {\bibinfo {volume} {90}},\ \bibinfo
  {pages} {125143} (\bibinfo {year} {2014})}\BibitemShut {NoStop}%
\bibitem [{\citenamefont {Dal~Lago}\ \emph {et~al.}(2015)\citenamefont
  {Dal~Lago}, \citenamefont {Atala},\ and\ \citenamefont
  {Foa~Torres}}]{DalLago2015}%
  \BibitemOpen
  \bibfield  {author} {\bibinfo {author} {\bibfnamefont {V.}~\bibnamefont
  {Dal~Lago}}, \bibinfo {author} {\bibfnamefont {M.}~\bibnamefont {Atala}}, \
  and\ \bibinfo {author} {\bibfnamefont {L.~E.~F.}\ \bibnamefont
  {Foa~Torres}},\ }\bibfield  {title} {\enquote {\bibinfo {title} {Floquet
  topological transitions in a driven one-dimensional topological insulator},}\
  }\href@noop {} {\bibfield  {journal} {\bibinfo  {journal} {Phys. Rev. A}\
  }\textbf {\bibinfo {volume} {92}},\ \bibinfo {pages} {023624} (\bibinfo
  {year} {2015})}\BibitemShut {NoStop}%
\bibitem [{\citenamefont {Bauer}\ \emph {et~al.}(2019)\citenamefont {Bauer},
  \citenamefont {Pereg-Barnea}, \citenamefont {Karzig}, \citenamefont {Rieder},
  \citenamefont {Refael}, \citenamefont {Berg},\ and\ \citenamefont
  {Oreg}}]{Bauer2018}%
  \BibitemOpen
  \bibfield  {author} {\bibinfo {author} {\bibfnamefont {Bela}\ \bibnamefont
  {Bauer}}, \bibinfo {author} {\bibfnamefont {T.}~\bibnamefont {Pereg-Barnea}},
  \bibinfo {author} {\bibfnamefont {Torsten}\ \bibnamefont {Karzig}}, \bibinfo
  {author} {\bibfnamefont {Maria-Theresa}\ \bibnamefont {Rieder}}, \bibinfo
  {author} {\bibfnamefont {Gil}\ \bibnamefont {Refael}}, \bibinfo {author}
  {\bibfnamefont {Erez}\ \bibnamefont {Berg}}, \ and\ \bibinfo {author}
  {\bibfnamefont {Yuval}\ \bibnamefont {Oreg}},\ }\bibfield  {title} {\enquote
  {\bibinfo {title} {{Topologically protected braiding in a single wire using
  Floquet Majorana modes}},}\ }\href@noop {} {\bibfield  {journal} {\bibinfo
  {journal} {Phys. Rev. B}\ }\textbf {\bibinfo {volume} {100}},\ \bibinfo
  {pages} {041102} (\bibinfo {year} {2019})}\BibitemShut {NoStop}%
\bibitem [{\citenamefont {Kohler}\ \emph {et~al.}(2005)\citenamefont {Kohler},
  \citenamefont {Lehmann},\ and\ \citenamefont {Hanggi}}]{Kohler2005}%
  \BibitemOpen
  \bibfield  {author} {\bibinfo {author} {\bibfnamefont {Sigmund}\ \bibnamefont
  {Kohler}}, \bibinfo {author} {\bibfnamefont {Jorg}\ \bibnamefont {Lehmann}},
  \ and\ \bibinfo {author} {\bibfnamefont {Peter}\ \bibnamefont {Hanggi}},\
  }\bibfield  {title} {\enquote {\bibinfo {title} {Driven quantum transport on
  the nanoscale},}\ }\href@noop {} {\bibfield  {journal} {\bibinfo  {journal}
  {Physics Reports}\ }\textbf {\bibinfo {volume} {406}},\ \bibinfo {pages}
  {379} (\bibinfo {year} {2005})}\BibitemShut {NoStop}%
\bibitem [{\citenamefont {Asb\'oth}\ and\ \citenamefont
  {Obuse}(2013)}]{Asboth2013}%
  \BibitemOpen
  \bibfield  {author} {\bibinfo {author} {\bibfnamefont {J\'anos~K.}\
  \bibnamefont {Asb\'oth}}\ and\ \bibinfo {author} {\bibfnamefont {Hideaki}\
  \bibnamefont {Obuse}},\ }\bibfield  {title} {\enquote {\bibinfo {title}
  {Bulk-boundary correspondence for chiral symmetric quantum walks},}\
  }\href@noop {} {\bibfield  {journal} {\bibinfo  {journal} {Phys. Rev. B}\
  }\textbf {\bibinfo {volume} {88}},\ \bibinfo {pages} {121406} (\bibinfo
  {year} {2013})}\BibitemShut {NoStop}%
\bibitem [{\citenamefont {Lababidi}\ \emph {et~al.}(2014)\citenamefont
  {Lababidi}, \citenamefont {Satija},\ and\ \citenamefont
  {Zhao}}]{Lababidi2014}%
  \BibitemOpen
  \bibfield  {author} {\bibinfo {author} {\bibfnamefont {Mahmoud}\ \bibnamefont
  {Lababidi}}, \bibinfo {author} {\bibfnamefont {Indubala~I.}\ \bibnamefont
  {Satija}}, \ and\ \bibinfo {author} {\bibfnamefont {Erhai}\ \bibnamefont
  {Zhao}},\ }\bibfield  {title} {\enquote {\bibinfo {title}
  {{Counter-propagating Edge Modes and Topological Phases of a Kicked Quantum
  Hall System}},}\ }\href@noop {} {\bibfield  {journal} {\bibinfo  {journal}
  {Phys. Rev. Lett.}\ }\textbf {\bibinfo {volume} {112}},\ \bibinfo {pages}
  {026805} (\bibinfo {year} {2014})}\BibitemShut {NoStop}%
\bibitem [{\citenamefont {Zhou}\ \emph {et~al.}(2014)\citenamefont {Zhou},
  \citenamefont {Satija},\ and\ \citenamefont {Zhao}}]{Zhou2014}%
  \BibitemOpen
  \bibfield  {author} {\bibinfo {author} {\bibfnamefont {Zhenyu}\ \bibnamefont
  {Zhou}}, \bibinfo {author} {\bibfnamefont {Indubala~I.}\ \bibnamefont
  {Satija}}, \ and\ \bibinfo {author} {\bibfnamefont {Erhai}\ \bibnamefont
  {Zhao}},\ }\bibfield  {title} {\enquote {\bibinfo {title} {{Floquet edge
  states in a harmonically driven integer quantum Hall system}},}\ }\href@noop
  {} {\bibfield  {journal} {\bibinfo  {journal} {Phys. Rev. B}\ }\textbf
  {\bibinfo {volume} {90}},\ \bibinfo {pages} {205108} (\bibinfo {year}
  {2014})}\BibitemShut {NoStop}%
\bibitem [{\citenamefont {Yao}\ \emph {et~al.}(2017{\natexlab{b}})\citenamefont
  {Yao}, \citenamefont {Yan},\ and\ \citenamefont {Wang}}]{Wang2017}%
  \BibitemOpen
  \bibfield  {author} {\bibinfo {author} {\bibfnamefont {Shunyu}\ \bibnamefont
  {Yao}}, \bibinfo {author} {\bibfnamefont {Zhongbo}\ \bibnamefont {Yan}}, \
  and\ \bibinfo {author} {\bibfnamefont {Zhong}\ \bibnamefont {Wang}},\
  }\bibfield  {title} {\enquote {\bibinfo {title} {{Topological invariants of
  Floquet systems: General formulation, special properties, and Floquet
  topological defects}},}\ }\href@noop {} {\bibfield  {journal} {\bibinfo
  {journal} {Phys. Rev. B}\ }\textbf {\bibinfo {volume} {96}},\ \bibinfo
  {pages} {195303} (\bibinfo {year} {2017}{\natexlab{b}})}\BibitemShut
  {NoStop}%
\bibitem [{\citenamefont {Morimoto}\ \emph {et~al.}(2017)\citenamefont
  {Morimoto}, \citenamefont {Po},\ and\ \citenamefont
  {Vishwanath}}]{Morimoto2017}%
  \BibitemOpen
  \bibfield  {author} {\bibinfo {author} {\bibfnamefont {Takahiro}\
  \bibnamefont {Morimoto}}, \bibinfo {author} {\bibfnamefont {Hoi~Chun}\
  \bibnamefont {Po}}, \ and\ \bibinfo {author} {\bibfnamefont {Ashvin}\
  \bibnamefont {Vishwanath}},\ }\bibfield  {title} {\enquote {\bibinfo {title}
  {Floquet topological phases protected by time glide symmetry},}\ }\href@noop
  {} {\bibfield  {journal} {\bibinfo  {journal} {Phys. Rev. B}\ }\textbf
  {\bibinfo {volume} {95}},\ \bibinfo {pages} {195155} (\bibinfo {year}
  {2017})}\BibitemShut {NoStop}%
\bibitem [{\citenamefont {Xu}\ and\ \citenamefont {Wu}(2018)}]{Xu2018}%
  \BibitemOpen
  \bibfield  {author} {\bibinfo {author} {\bibfnamefont {Shenglong}\
  \bibnamefont {Xu}}\ and\ \bibinfo {author} {\bibfnamefont {Congjun}\
  \bibnamefont {Wu}},\ }\bibfield  {title} {\enquote {\bibinfo {title}
  {Space-time crystal and space-time group},}\ }\href@noop {} {\bibfield
  {journal} {\bibinfo  {journal} {Phys. Rev. Lett.}\ }\textbf {\bibinfo
  {volume} {120}},\ \bibinfo {pages} {096401} (\bibinfo {year}
  {2018})}\BibitemShut {NoStop}%
\bibitem [{\citenamefont {Peng}\ and\ \citenamefont {Refael}(2019)}]{Peng2018}%
  \BibitemOpen
  \bibfield  {author} {\bibinfo {author} {\bibfnamefont {Yang}\ \bibnamefont
  {Peng}}\ and\ \bibinfo {author} {\bibfnamefont {Gil}\ \bibnamefont
  {Refael}},\ }\bibfield  {title} {\enquote {\bibinfo {title} {Floquet
  second-order topological insulators from nonsymmorphic space-time
  symmetries},}\ }\href@noop {} {\bibfield  {journal} {\bibinfo  {journal}
  {Phys. Rev. Lett.}\ }\textbf {\bibinfo {volume} {123}},\ \bibinfo {pages}
  {016806} (\bibinfo {year} {2019})}\BibitemShut {NoStop}%
\bibitem [{\citenamefont {Hu}\ \emph {et~al.}(2015)\citenamefont {Hu},
  \citenamefont {Pillay}, \citenamefont {Wu}, \citenamefont {Pasek},
  \citenamefont {Shum},\ and\ \citenamefont {Chong}}]{Hu2015}%
  \BibitemOpen
  \bibfield  {author} {\bibinfo {author} {\bibfnamefont {Wenchao}\ \bibnamefont
  {Hu}}, \bibinfo {author} {\bibfnamefont {Jason~C.}\ \bibnamefont {Pillay}},
  \bibinfo {author} {\bibfnamefont {Kan}\ \bibnamefont {Wu}}, \bibinfo {author}
  {\bibfnamefont {Michael}\ \bibnamefont {Pasek}}, \bibinfo {author}
  {\bibfnamefont {Perry~Ping}\ \bibnamefont {Shum}}, \ and\ \bibinfo {author}
  {\bibfnamefont {Y.~D.}\ \bibnamefont {Chong}},\ }\bibfield  {title} {\enquote
  {\bibinfo {title} {{Measurement of a Topological Edge Invariant in a
  Microwave Network}},}\ }\href@noop {} {\bibfield  {journal} {\bibinfo
  {journal} {Phys. Rev. X}\ }\textbf {\bibinfo {volume} {5}},\ \bibinfo {pages}
  {011012} (\bibinfo {year} {2015})}\BibitemShut {NoStop}%
\bibitem [{\citenamefont {Mukherjee}\ \emph {et~al.}(2017)\citenamefont
  {Mukherjee}, \citenamefont {Spracklen}, \citenamefont {Valiente},
  \citenamefont {Andersson}, \citenamefont {Ohberg}, \citenamefont {Goldman},\
  and\ \citenamefont {Thomson}}]{Mukherjee2017}%
  \BibitemOpen
  \bibfield  {author} {\bibinfo {author} {\bibfnamefont {S.}~\bibnamefont
  {Mukherjee}}, \bibinfo {author} {\bibfnamefont {A.}~\bibnamefont
  {Spracklen}}, \bibinfo {author} {\bibfnamefont {M.}~\bibnamefont {Valiente}},
  \bibinfo {author} {\bibfnamefont {E.}~\bibnamefont {Andersson}}, \bibinfo
  {author} {\bibfnamefont {O.}~\bibnamefont {Ohberg}}, \bibinfo {author}
  {\bibfnamefont {N.}~\bibnamefont {Goldman}}, \ and\ \bibinfo {author}
  {\bibfnamefont {R.~R.}\ \bibnamefont {Thomson}},\ }\bibfield  {title}
  {\enquote {\bibinfo {title} {{Experimental observation of anomalous
  topological edge modes in a slowly-driven photonic lattice}},}\ }\href@noop
  {} {\bibfield  {journal} {\bibinfo  {journal} {Nature Comm.}\ }\textbf
  {\bibinfo {volume} {8}},\ \bibinfo {pages} {13918} (\bibinfo {year}
  {2017})}\BibitemShut {NoStop}%
\bibitem [{\citenamefont {Maczewsky}\ \emph {et~al.}(2017)\citenamefont
  {Maczewsky}, \citenamefont {Zeuner}, \citenamefont {Nolte},\ and\
  \citenamefont {Szameit}}]{Maczewsky2017}%
  \BibitemOpen
  \bibfield  {author} {\bibinfo {author} {\bibfnamefont {Lukas~J.}\
  \bibnamefont {Maczewsky}}, \bibinfo {author} {\bibfnamefont {Julia~M.}\
  \bibnamefont {Zeuner}}, \bibinfo {author} {\bibfnamefont {Stefan}\
  \bibnamefont {Nolte}}, \ and\ \bibinfo {author} {\bibfnamefont {Alexander}\
  \bibnamefont {Szameit}},\ }\bibfield  {title} {\enquote {\bibinfo {title}
  {{Observation of photonic anomalous Floquet topological insulators}},}\
  }\href@noop {} {\bibfield  {journal} {\bibinfo  {journal} {Nature Comm.}\
  }\textbf {\bibinfo {volume} {8}},\ \bibinfo {pages} {13756} (\bibinfo {year}
  {2017})}\BibitemShut {NoStop}%
\bibitem [{\citenamefont {Cheng}\ \emph {et~al.}(2019)\citenamefont {Cheng},
  \citenamefont {Pan}, \citenamefont {Wang}, \citenamefont {Zhang},
  \citenamefont {Yu}, \citenamefont {Gover}, \citenamefont {Zhang},
  \citenamefont {Li}, \citenamefont {Zhou},\ and\ \citenamefont
  {Zhu}}]{Cheng2019}%
  \BibitemOpen
  \bibfield  {author} {\bibinfo {author} {\bibfnamefont {Qingqing}\
  \bibnamefont {Cheng}}, \bibinfo {author} {\bibfnamefont {Yiming}\
  \bibnamefont {Pan}}, \bibinfo {author} {\bibfnamefont {Huaiqiang}\
  \bibnamefont {Wang}}, \bibinfo {author} {\bibfnamefont {Chaoshi}\
  \bibnamefont {Zhang}}, \bibinfo {author} {\bibfnamefont {Dong}\ \bibnamefont
  {Yu}}, \bibinfo {author} {\bibfnamefont {Avi}\ \bibnamefont {Gover}},
  \bibinfo {author} {\bibfnamefont {Haijun}\ \bibnamefont {Zhang}}, \bibinfo
  {author} {\bibfnamefont {Tao}\ \bibnamefont {Li}}, \bibinfo {author}
  {\bibfnamefont {Lei}\ \bibnamefont {Zhou}}, \ and\ \bibinfo {author}
  {\bibfnamefont {Shining}\ \bibnamefont {Zhu}},\ }\bibfield  {title} {\enquote
  {\bibinfo {title} {{Observation of Anomalous $\ensuremath{\pi}$ Modes in
  Photonic Floquet Engineering}},}\ }\href@noop {} {\bibfield  {journal}
  {\bibinfo  {journal} {Phys. Rev. Lett.}\ }\textbf {\bibinfo {volume} {122}},\
  \bibinfo {pages} {173901} (\bibinfo {year} {2019})}\BibitemShut {NoStop}%
\bibitem [{\citenamefont {Nakajima}\ \emph {et~al.}(2016)\citenamefont
  {Nakajima}, \citenamefont {Tomita}, \citenamefont {Taie}, \citenamefont
  {Ichinose}, \citenamefont {Ozawa}, \citenamefont {Wang}, \citenamefont
  {Troyer},\ and\ \citenamefont {Takahashi}}]{Nakajima2016}%
  \BibitemOpen
  \bibfield  {author} {\bibinfo {author} {\bibfnamefont {S.}~\bibnamefont
  {Nakajima}}, \bibinfo {author} {\bibfnamefont {T.}~\bibnamefont {Tomita}},
  \bibinfo {author} {\bibfnamefont {S.}~\bibnamefont {Taie}}, \bibinfo {author}
  {\bibfnamefont {T.}~\bibnamefont {Ichinose}}, \bibinfo {author}
  {\bibfnamefont {H.}~\bibnamefont {Ozawa}}, \bibinfo {author} {\bibfnamefont
  {L.}~\bibnamefont {Wang}}, \bibinfo {author} {\bibfnamefont {M.}~\bibnamefont
  {Troyer}}, \ and\ \bibinfo {author} {\bibfnamefont {Y.}~\bibnamefont
  {Takahashi}},\ }\bibfield  {title} {\enquote {\bibinfo {title} {{Topological
  Thouless pumping of ultracold fermions}},}\ }\href@noop {} {\bibfield
  {journal} {\bibinfo  {journal} {Nature Phys.}\ }\textbf {\bibinfo {volume}
  {12}},\ \bibinfo {pages} {296} (\bibinfo {year} {2016})}\BibitemShut
  {NoStop}%
\bibitem [{\citenamefont {Lohse}\ \emph {et~al.}(2016)\citenamefont {Lohse},
  \citenamefont {Schweizer}, \citenamefont {Zilberberg}, \citenamefont
  {Aidelsburger},\ and\ \citenamefont {Bloch}}]{Lohse2016}%
  \BibitemOpen
  \bibfield  {author} {\bibinfo {author} {\bibfnamefont {M.}~\bibnamefont
  {Lohse}}, \bibinfo {author} {\bibfnamefont {C.}~\bibnamefont {Schweizer}},
  \bibinfo {author} {\bibfnamefont {O.}~\bibnamefont {Zilberberg}}, \bibinfo
  {author} {\bibfnamefont {M.}~\bibnamefont {Aidelsburger}}, \ and\ \bibinfo
  {author} {\bibfnamefont {I.}~\bibnamefont {Bloch}},\ }\bibfield  {title}
  {\enquote {\bibinfo {title} {{A Thouless quantum pump with ultracold bosonic
  atoms in an optical superlattice}},}\ }\href@noop {} {\bibfield  {journal}
  {\bibinfo  {journal} {Nature Phys.}\ }\textbf {\bibinfo {volume} {12}},\
  \bibinfo {pages} {350} (\bibinfo {year} {2016})}\BibitemShut {NoStop}%
\bibitem [{\citenamefont {Quelle}\ \emph {et~al.}(2017)\citenamefont {Quelle},
  \citenamefont {Weitenberg}, \citenamefont {Sengstock},\ and\ \citenamefont
  {Morais~Smith}}]{Quelle2017}%
  \BibitemOpen
  \bibfield  {author} {\bibinfo {author} {\bibfnamefont {A.}~\bibnamefont
  {Quelle}}, \bibinfo {author} {\bibfnamefont {C.}~\bibnamefont {Weitenberg}},
  \bibinfo {author} {\bibfnamefont {K.}~\bibnamefont {Sengstock}}, \ and\
  \bibinfo {author} {\bibfnamefont {C.}~\bibnamefont {Morais~Smith}},\
  }\bibfield  {title} {\enquote {\bibinfo {title} {{Driving protocol for a
  Floquet topological phase without static counterpart}},}\ }\href@noop {}
  {\bibfield  {journal} {\bibinfo  {journal} {New Journal of Physics}\ }\textbf
  {\bibinfo {volume} {19}},\ \bibinfo {pages} {113010} (\bibinfo {year}
  {2017})}\BibitemShut {NoStop}%
\bibitem [{\citenamefont {Liu}\ \emph {et~al.}(2019)\citenamefont {Liu},
  \citenamefont {Shabani},\ and\ \citenamefont {Mitra}}]{Liu2019}%
  \BibitemOpen
  \bibfield  {author} {\bibinfo {author} {\bibfnamefont {Dillon~T.}\
  \bibnamefont {Liu}}, \bibinfo {author} {\bibfnamefont {Javad}\ \bibnamefont
  {Shabani}}, \ and\ \bibinfo {author} {\bibfnamefont {Aditi}\ \bibnamefont
  {Mitra}},\ }\bibfield  {title} {\enquote {\bibinfo {title} {{Floquet Majorana
  zero and $\ensuremath{\pi}$ modes in planar Josephson junctions}},}\
  }\href@noop {} {\bibfield  {journal} {\bibinfo  {journal} {Phys. Rev. B}\
  }\textbf {\bibinfo {volume} {99}},\ \bibinfo {pages} {094303} (\bibinfo
  {year} {2019})}\BibitemShut {NoStop}%
\bibitem [{\citenamefont {Kundu}\ and\ \citenamefont
  {Seradjeh}(2013)}]{Kundu2013}%
  \BibitemOpen
  \bibfield  {author} {\bibinfo {author} {\bibfnamefont {Arijit}\ \bibnamefont
  {Kundu}}\ and\ \bibinfo {author} {\bibfnamefont {Babak}\ \bibnamefont
  {Seradjeh}},\ }\bibfield  {title} {\enquote {\bibinfo {title} {{Transport
  signatures of Floquet Majorana fermions in driven topological
  superconductors}},}\ }\href@noop {} {\bibfield  {journal} {\bibinfo
  {journal} {Phys. Rev. Lett.}\ }\textbf {\bibinfo {volume} {111}},\ \bibinfo
  {pages} {136402} (\bibinfo {year} {2013})}\BibitemShut {NoStop}%
\bibitem [{\citenamefont {Foa~Torres}\ \emph {et~al.}(2014)\citenamefont
  {Foa~Torres}, \citenamefont {Perez-Piskunow}, \citenamefont {Balseiro},\ and\
  \citenamefont {Usaj}}]{FoaTorresMultiTerminal}%
  \BibitemOpen
  \bibfield  {author} {\bibinfo {author} {\bibfnamefont {L.~E.~F.}\
  \bibnamefont {Foa~Torres}}, \bibinfo {author} {\bibfnamefont {P.~M.}\
  \bibnamefont {Perez-Piskunow}}, \bibinfo {author} {\bibfnamefont {C.~A.}\
  \bibnamefont {Balseiro}}, \ and\ \bibinfo {author} {\bibfnamefont {Gonzalo}\
  \bibnamefont {Usaj}},\ }\bibfield  {title} {\enquote {\bibinfo {title}
  {{Multiterminal Conductance of a Floquet Topological Insulator}},}\
  }\href@noop {} {\bibfield  {journal} {\bibinfo  {journal} {Phys. Rev. Lett.}\
  }\textbf {\bibinfo {volume} {113}},\ \bibinfo {pages} {266801} (\bibinfo
  {year} {2014})}\BibitemShut {NoStop}%
\bibitem [{\citenamefont {Farrell}\ and\ \citenamefont
  {Pereg-Barnea}(2015)}]{Farrell2015}%
  \BibitemOpen
  \bibfield  {author} {\bibinfo {author} {\bibfnamefont {Aaron}\ \bibnamefont
  {Farrell}}\ and\ \bibinfo {author} {\bibfnamefont {T.}~\bibnamefont
  {Pereg-Barnea}},\ }\bibfield  {title} {\enquote {\bibinfo {title}
  {Photon-inhibited topological transport in quantum well heterostructures},}\
  }\href@noop {} {\bibfield  {journal} {\bibinfo  {journal} {Phys. Rev. Lett.}\
  }\textbf {\bibinfo {volume} {115}},\ \bibinfo {pages} {106403} (\bibinfo
  {year} {2015})}\BibitemShut {NoStop}%
\bibitem [{\citenamefont {Farrell}\ and\ \citenamefont
  {Pereg-Barnea}(2016)}]{Pereg-Barnea2016}%
  \BibitemOpen
  \bibfield  {author} {\bibinfo {author} {\bibfnamefont {Aaron}\ \bibnamefont
  {Farrell}}\ and\ \bibinfo {author} {\bibfnamefont {T.}~\bibnamefont
  {Pereg-Barnea}},\ }\bibfield  {title} {\enquote {\bibinfo {title}
  {{Edge-state transport in Floquet topological insulators}},}\ }\href@noop {}
  {\bibfield  {journal} {\bibinfo  {journal} {Phys. Rev. B}\ }\textbf {\bibinfo
  {volume} {93}},\ \bibinfo {pages} {045121} (\bibinfo {year}
  {2016})}\BibitemShut {NoStop}%
\bibitem [{\citenamefont {Kundu}\ \emph {et~al.}(2017)\citenamefont {Kundu},
  \citenamefont {Rudner}, \citenamefont {Berg},\ and\ \citenamefont
  {Lindner}}]{Kundu2017}%
  \BibitemOpen
  \bibfield  {author} {\bibinfo {author} {\bibfnamefont {Arijit}\ \bibnamefont
  {Kundu}}, \bibinfo {author} {\bibfnamefont {Mark~S.}\ \bibnamefont {Rudner}},
  \bibinfo {author} {\bibfnamefont {Erez}\ \bibnamefont {Berg}}, \ and\
  \bibinfo {author} {\bibfnamefont {Netanel~H.}\ \bibnamefont {Lindner}},\
  }\bibfield  {title} {\enquote {\bibinfo {title} {{Quantized large-bias
  current in the anomalous Floquet-Anderson insulator}},}\ }\href@noop {}
  {\bibfield  {journal} {\bibinfo  {journal} {arXiv:1708.05023}\ } (\bibinfo
  {year} {2017})}\BibitemShut {NoStop}%
\bibitem [{\citenamefont {Perez-Piskunow}\ \emph {et~al.}(2015)\citenamefont
  {Perez-Piskunow}, \citenamefont {Foa~Torres},\ and\ \citenamefont
  {Usaj}}]{Perez-Piskunow2015}%
  \BibitemOpen
  \bibfield  {author} {\bibinfo {author} {\bibfnamefont {P.~M.}\ \bibnamefont
  {Perez-Piskunow}}, \bibinfo {author} {\bibfnamefont {L.~E.~F.}\ \bibnamefont
  {Foa~Torres}}, \ and\ \bibinfo {author} {\bibfnamefont {Gonzalo}\
  \bibnamefont {Usaj}},\ }\bibfield  {title} {\enquote {\bibinfo {title}
  {{Hierarchy of Floquet gaps and edge states for driven honeycomb
  lattices}},}\ }\href@noop {} {\bibfield  {journal} {\bibinfo  {journal}
  {Phys. Rev. A}\ }\textbf {\bibinfo {volume} {91}},\ \bibinfo {pages} {043625}
  (\bibinfo {year} {2015})}\BibitemShut {NoStop}%
\bibitem [{\citenamefont {Uhrig}\ \emph {et~al.}(2019)\citenamefont {Uhrig},
  \citenamefont {Kalthoff},\ and\ \citenamefont {Freericks}}]{Uhrig2019}%
  \BibitemOpen
  \bibfield  {author} {\bibinfo {author} {\bibfnamefont {G\"otz~S.}\
  \bibnamefont {Uhrig}}, \bibinfo {author} {\bibfnamefont {Mona~H.}\
  \bibnamefont {Kalthoff}}, \ and\ \bibinfo {author} {\bibfnamefont {James~K.}\
  \bibnamefont {Freericks}},\ }\bibfield  {title} {\enquote {\bibinfo {title}
  {{Positivity of the Spectral Densities of Retarded Floquet Green
  Functions}},}\ }\href@noop {} {\bibfield  {journal} {\bibinfo  {journal}
  {Phys. Rev. Lett.}\ }\textbf {\bibinfo {volume} {122}},\ \bibinfo {pages}
  {130604} (\bibinfo {year} {2019})}\BibitemShut {NoStop}%
\bibitem [{\citenamefont {Sengupta}\ \emph {et~al.}(2001)\citenamefont
  {Sengupta}, \citenamefont {\ifmmode \check{Z}\else
  \v{Z}\fi{}uti\ifmmode~\acute{c}\else \'{c}\fi{}}, \citenamefont {Kwon},
  \citenamefont {Yakovenko},\ and\ \citenamefont {Das~Sarma}}]{Sengupta2001}%
  \BibitemOpen
  \bibfield  {author} {\bibinfo {author} {\bibfnamefont {K.}~\bibnamefont
  {Sengupta}}, \bibinfo {author} {\bibfnamefont {Igor}\ \bibnamefont {\ifmmode
  \check{Z}\else \v{Z}\fi{}uti\ifmmode~\acute{c}\else \'{c}\fi{}}}, \bibinfo
  {author} {\bibfnamefont {Hyok-Jon}\ \bibnamefont {Kwon}}, \bibinfo {author}
  {\bibfnamefont {Victor~M.}\ \bibnamefont {Yakovenko}}, \ and\ \bibinfo
  {author} {\bibfnamefont {S.}~\bibnamefont {Das~Sarma}},\ }\bibfield  {title}
  {\enquote {\bibinfo {title} {Midgap edge states and pairing symmetry of
  quasi-one-dimensional organic superconductors},}\ }\href@noop {} {\bibfield
  {journal} {\bibinfo  {journal} {Phys. Rev. B}\ }\textbf {\bibinfo {volume}
  {63}},\ \bibinfo {pages} {144531} (\bibinfo {year} {2001})}\BibitemShut
  {NoStop}%
\bibitem [{\citenamefont {Law}\ \emph {et~al.}(2009)\citenamefont {Law},
  \citenamefont {Lee},\ and\ \citenamefont {Ng}}]{Law2009}%
  \BibitemOpen
  \bibfield  {author} {\bibinfo {author} {\bibfnamefont {K.~T.}\ \bibnamefont
  {Law}}, \bibinfo {author} {\bibfnamefont {Patrick~A.}\ \bibnamefont {Lee}}, \
  and\ \bibinfo {author} {\bibfnamefont {T.~K.}\ \bibnamefont {Ng}},\
  }\bibfield  {title} {\enquote {\bibinfo {title} {Majorana fermion induced
  resonant andreev reflection},}\ }\href@noop {} {\bibfield  {journal}
  {\bibinfo  {journal} {Phys. Rev. Lett.}\ }\textbf {\bibinfo {volume} {103}},\
  \bibinfo {pages} {237001} (\bibinfo {year} {2009})}\BibitemShut {NoStop}%
\bibitem [{\citenamefont {Titum}\ \emph {et~al.}(2016)\citenamefont {Titum},
  \citenamefont {Berg}, \citenamefont {Rudner}, \citenamefont {Refael},\ and\
  \citenamefont {Lindner}}]{AFAI}%
  \BibitemOpen
  \bibfield  {author} {\bibinfo {author} {\bibfnamefont {Paraj}\ \bibnamefont
  {Titum}}, \bibinfo {author} {\bibfnamefont {Erez}\ \bibnamefont {Berg}},
  \bibinfo {author} {\bibfnamefont {Mark~S.}\ \bibnamefont {Rudner}}, \bibinfo
  {author} {\bibfnamefont {Gil}\ \bibnamefont {Refael}}, \ and\ \bibinfo
  {author} {\bibfnamefont {Netanel~H.}\ \bibnamefont {Lindner}},\ }\bibfield
  {title} {\enquote {\bibinfo {title} {{Anomalous Floquet-Anderson Insulator as
  a Nonadiabatic Quantized Charge Pump}},}\ }\href@noop {} {\bibfield
  {journal} {\bibinfo  {journal} {Phys. Rev. X}\ }\textbf {\bibinfo {volume}
  {6}},\ \bibinfo {pages} {021013} (\bibinfo {year} {2016})}\BibitemShut
  {NoStop}%
\bibitem [{\citenamefont {Dahlhaus}\ \emph {et~al.}(2015)\citenamefont
  {Dahlhaus}, \citenamefont {Fregoso},\ and\ \citenamefont
  {Moore}}]{Dahlhaus2015}%
  \BibitemOpen
  \bibfield  {author} {\bibinfo {author} {\bibfnamefont {Jan~P.}\ \bibnamefont
  {Dahlhaus}}, \bibinfo {author} {\bibfnamefont {Benjamin~M.}\ \bibnamefont
  {Fregoso}}, \ and\ \bibinfo {author} {\bibfnamefont {Joel~E.}\ \bibnamefont
  {Moore}},\ }\bibfield  {title} {\enquote {\bibinfo {title} {{Magnetization
  Signatures of Light-Induced Quantum Hall Edge States}},}\ }\href@noop {}
  {\bibfield  {journal} {\bibinfo  {journal} {Phys. Rev. Lett.}\ }\textbf
  {\bibinfo {volume} {114}},\ \bibinfo {pages} {246802} (\bibinfo {year}
  {2015})}\BibitemShut {NoStop}%
\bibitem [{\citenamefont {Nathan}\ \emph {et~al.}(2017)\citenamefont {Nathan},
  \citenamefont {Rudner}, \citenamefont {Lindner}, \citenamefont {Berg},\ and\
  \citenamefont {Refael}}]{Nathan2017_magnetization}%
  \BibitemOpen
  \bibfield  {author} {\bibinfo {author} {\bibfnamefont {Frederik}\
  \bibnamefont {Nathan}}, \bibinfo {author} {\bibfnamefont {Mark~S.}\
  \bibnamefont {Rudner}}, \bibinfo {author} {\bibfnamefont {Netanel~H.}\
  \bibnamefont {Lindner}}, \bibinfo {author} {\bibfnamefont {Erez}\
  \bibnamefont {Berg}}, \ and\ \bibinfo {author} {\bibfnamefont {Gil}\
  \bibnamefont {Refael}},\ }\bibfield  {title} {\enquote {\bibinfo {title}
  {Quantized magnetization density in periodically driven systems},}\
  }\href@noop {} {\bibfield  {journal} {\bibinfo  {journal} {Phys. Rev. Lett.}\
  }\textbf {\bibinfo {volume} {119}},\ \bibinfo {pages} {186801} (\bibinfo
  {year} {2017})}\BibitemShut {NoStop}%
\bibitem [{\citenamefont {Mahmood}\ \emph {et~al.}(2016)\citenamefont
  {Mahmood}, \citenamefont {Chan}, \citenamefont {Alpichshev}, \citenamefont
  {Gardner}, \citenamefont {Lee}, \citenamefont {Lee},\ and\ \citenamefont
  {Gedik}}]{Mahmood2016}%
  \BibitemOpen
  \bibfield  {author} {\bibinfo {author} {\bibfnamefont {F.}~\bibnamefont
  {Mahmood}}, \bibinfo {author} {\bibfnamefont {C.-K.}\ \bibnamefont {Chan}},
  \bibinfo {author} {\bibfnamefont {Z.}~\bibnamefont {Alpichshev}}, \bibinfo
  {author} {\bibfnamefont {D.}~\bibnamefont {Gardner}}, \bibinfo {author}
  {\bibfnamefont {Y.}~\bibnamefont {Lee}}, \bibinfo {author} {\bibfnamefont
  {P.~A.}\ \bibnamefont {Lee}}, \ and\ \bibinfo {author} {\bibfnamefont
  {N.}~\bibnamefont {Gedik}},\ }\bibfield  {title} {\enquote {\bibinfo {title}
  {{Selective scattering between Floquet-Bloch and Volkov states in a
  topological insulator}},}\ }\href@noop {} {\bibfield  {journal} {\bibinfo
  {journal} {Nature Phys.}\ }\textbf {\bibinfo {volume} {12}},\ \bibinfo
  {pages} {306} (\bibinfo {year} {2016})}\BibitemShut {NoStop}%
\bibitem [{\citenamefont {Fregoso}\ \emph {et~al.}(2013)\citenamefont
  {Fregoso}, \citenamefont {Wang}, \citenamefont {Gedik},\ and\ \citenamefont
  {Galitski}}]{Fregoso2013}%
  \BibitemOpen
  \bibfield  {author} {\bibinfo {author} {\bibfnamefont {Benjamin~M.}\
  \bibnamefont {Fregoso}}, \bibinfo {author} {\bibfnamefont {Y.~H.}\
  \bibnamefont {Wang}}, \bibinfo {author} {\bibfnamefont {N.}~\bibnamefont
  {Gedik}}, \ and\ \bibinfo {author} {\bibfnamefont {Victor}\ \bibnamefont
  {Galitski}},\ }\bibfield  {title} {\enquote {\bibinfo {title} {Driven
  electronic states at the surface of a topological insulator},}\ }\href@noop
  {} {\bibfield  {journal} {\bibinfo  {journal} {Phys. Rev. B}\ }\textbf
  {\bibinfo {volume} {88}},\ \bibinfo {pages} {155129} (\bibinfo {year}
  {2013})}\BibitemShut {NoStop}%
\bibitem [{\citenamefont {Farrell}\ \emph {et~al.}(2016)\citenamefont
  {Farrell}, \citenamefont {Arsenault},\ and\ \citenamefont
  {Pereg-Barnea}}]{Farrell2016}%
  \BibitemOpen
  \bibfield  {author} {\bibinfo {author} {\bibfnamefont {Aaron}\ \bibnamefont
  {Farrell}}, \bibinfo {author} {\bibfnamefont {A.}~\bibnamefont {Arsenault}},
  \ and\ \bibinfo {author} {\bibfnamefont {T.}~\bibnamefont {Pereg-Barnea}},\
  }\bibfield  {title} {\enquote {\bibinfo {title} {{Dirac cones, Floquet side
  bands, and theory of time-resolved angle-resolved photoemission}},}\
  }\href@noop {} {\bibfield  {journal} {\bibinfo  {journal} {Phys. Rev. B}\
  }\textbf {\bibinfo {volume} {94}},\ \bibinfo {pages} {155304} (\bibinfo
  {year} {2016})}\BibitemShut {NoStop}%
\bibitem [{\citenamefont {Kandelaki}\ and\ \citenamefont
  {Rudner}(2018)}]{Kandelaki2017}%
  \BibitemOpen
  \bibfield  {author} {\bibinfo {author} {\bibfnamefont {Ervand}\ \bibnamefont
  {Kandelaki}}\ and\ \bibinfo {author} {\bibfnamefont {Mark~S.}\ \bibnamefont
  {Rudner}},\ }\bibfield  {title} {\enquote {\bibinfo {title} {Many-body
  dynamics and gap opening in interacting periodically driven systems},}\
  }\href@noop {} {\bibfield  {journal} {\bibinfo  {journal} {Phys. Rev. Lett.}\
  }\textbf {\bibinfo {volume} {121}},\ \bibinfo {pages} {036801} (\bibinfo
  {year} {2018})}\BibitemShut {NoStop}%
\bibitem [{\citenamefont {Bukov}\ \emph {et~al.}(2015)\citenamefont {Bukov},
  \citenamefont {D'Alessio},\ and\ \citenamefont {Polkovnikov}}]{BukovReview}%
  \BibitemOpen
  \bibfield  {author} {\bibinfo {author} {\bibfnamefont {Marin}\ \bibnamefont
  {Bukov}}, \bibinfo {author} {\bibfnamefont {Luca}\ \bibnamefont {D'Alessio}},
  \ and\ \bibinfo {author} {\bibfnamefont {Anatoli}\ \bibnamefont
  {Polkovnikov}},\ }\bibfield  {title} {\enquote {\bibinfo {title} {{Universal
  high-frequency behavior of periodically driven systems: from dynamical
  stabilization to Floquet engineering}},}\ }\href@noop {} {\bibfield
  {journal} {\bibinfo  {journal} {Advances in Physics}\ }\textbf {\bibinfo
  {volume} {64}},\ \bibinfo {pages} {139} (\bibinfo {year} {2015})}\BibitemShut
  {NoStop}%
\bibitem [{\citenamefont {Eckardt}\ and\ \citenamefont
  {Anisimovas}(2015)}]{Eckardt2015}%
  \BibitemOpen
  \bibfield  {author} {\bibinfo {author} {\bibfnamefont {A.}~\bibnamefont
  {Eckardt}}\ and\ \bibinfo {author} {\bibfnamefont {E.}~\bibnamefont
  {Anisimovas}},\ }\bibfield  {title} {\enquote {\bibinfo {title}
  {{High-frequency approximation for periodically driven quantum systems from a
  Floquet-space perspective}},}\ }\href@noop {} {\bibfield  {journal} {\bibinfo
   {journal} {New J. Phys.}\ }\textbf {\bibinfo {volume} {17}},\ \bibinfo
  {pages} {093039} (\bibinfo {year} {2015})}\BibitemShut {NoStop}%
\bibitem [{\citenamefont {Kuwahara}\ \emph {et~al.}(2016)\citenamefont
  {Kuwahara}, \citenamefont {Mori},\ and\ \citenamefont
  {Saito}}]{Kuwahara2016}%
  \BibitemOpen
  \bibfield  {author} {\bibinfo {author} {\bibfnamefont {T.}~\bibnamefont
  {Kuwahara}}, \bibinfo {author} {\bibfnamefont {T.}~\bibnamefont {Mori}}, \
  and\ \bibinfo {author} {\bibfnamefont {K.}~\bibnamefont {Saito}},\ }\bibfield
   {title} {\enquote {\bibinfo {title} {{Floquet-Magnus theory and generic
  transient dynamics in periodically driven many-body quantum systems}},}\
  }\href@noop {} {\bibfield  {journal} {\bibinfo  {journal} {Ann. Phys.}\
  }\textbf {\bibinfo {volume} {367}},\ \bibinfo {pages} {96} (\bibinfo {year}
  {2016})}\BibitemShut {NoStop}%
\bibitem [{\citenamefont {Abanin}\ \emph
  {et~al.}(2017{\natexlab{a}})\citenamefont {Abanin}, \citenamefont {De~Roeck},
  \citenamefont {Ho},\ and\ \citenamefont {Huveneers}}]{Abanin2017}%
  \BibitemOpen
  \bibfield  {author} {\bibinfo {author} {\bibfnamefont {Dmitry~A.}\
  \bibnamefont {Abanin}}, \bibinfo {author} {\bibfnamefont {Wojciech}\
  \bibnamefont {De~Roeck}}, \bibinfo {author} {\bibfnamefont {Wen~Wei}\
  \bibnamefont {Ho}}, \ and\ \bibinfo {author} {\bibfnamefont {Francois}\
  \bibnamefont {Huveneers}},\ }\bibfield  {title} {\enquote {\bibinfo {title}
  {{Effective Hamiltonians, prethermalization, and slow energy absorption in
  periodically driven many-body systems}},}\ }\href@noop {} {\bibfield
  {journal} {\bibinfo  {journal} {Phys. Rev. B}\ }\textbf {\bibinfo {volume}
  {95}},\ \bibinfo {pages} {014112} (\bibinfo {year}
  {2017}{\natexlab{a}})}\BibitemShut {NoStop}%
\bibitem [{\citenamefont {Lindner}\ \emph {et~al.}(2017)\citenamefont
  {Lindner}, \citenamefont {Berg},\ and\ \citenamefont {Rudner}}]{Lindner2017}%
  \BibitemOpen
  \bibfield  {author} {\bibinfo {author} {\bibfnamefont {Netanel~H.}\
  \bibnamefont {Lindner}}, \bibinfo {author} {\bibfnamefont {Erez}\
  \bibnamefont {Berg}}, \ and\ \bibinfo {author} {\bibfnamefont {Mark~S.}\
  \bibnamefont {Rudner}},\ }\bibfield  {title} {\enquote {\bibinfo {title}
  {{Universal Chiral Quasisteady States in Periodically Driven Many-Body
  Systems}},}\ }\href@noop {} {\bibfield  {journal} {\bibinfo  {journal} {Phys.
  Rev. X}\ }\textbf {\bibinfo {volume} {7}},\ \bibinfo {pages} {011018}
  (\bibinfo {year} {2017})}\BibitemShut {NoStop}%
\bibitem [{\citenamefont {Mori}\ \emph {et~al.}(2018)\citenamefont {Mori},
  \citenamefont {Ikeda}, \citenamefont {Kamanishi},\ and\ \citenamefont
  {Ueda}}]{Mori2018}%
  \BibitemOpen
  \bibfield  {author} {\bibinfo {author} {\bibfnamefont {Takashi}\ \bibnamefont
  {Mori}}, \bibinfo {author} {\bibfnamefont {Tatsuhiko~N.}\ \bibnamefont
  {Ikeda}}, \bibinfo {author} {\bibfnamefont {Eriko}\ \bibnamefont
  {Kamanishi}}, \ and\ \bibinfo {author} {\bibfnamefont {Masahito}\
  \bibnamefont {Ueda}},\ }\bibfield  {title} {\enquote {\bibinfo {title}
  {Thermalization and prethermalization in isolated quantum systems: a
  theoretical overview},}\ }\href@noop {} {\bibfield  {journal} {\bibinfo
  {journal} {J. Phys. B: At. Mol. Opt. Phys.}\ }\textbf {\bibinfo {volume}
  {51}},\ \bibinfo {pages} {112001} (\bibinfo {year} {2018})}\BibitemShut
  {NoStop}%
\bibitem [{\citenamefont {Bukov}\ \emph {et~al.}(2016)\citenamefont {Bukov},
  \citenamefont {Heyl}, \citenamefont {Huse},\ and\ \citenamefont
  {Polkovnikov}}]{Bukov2016}%
  \BibitemOpen
  \bibfield  {author} {\bibinfo {author} {\bibfnamefont {Marin}\ \bibnamefont
  {Bukov}}, \bibinfo {author} {\bibfnamefont {Markus}\ \bibnamefont {Heyl}},
  \bibinfo {author} {\bibfnamefont {David~A.}\ \bibnamefont {Huse}}, \ and\
  \bibinfo {author} {\bibfnamefont {Anatoli}\ \bibnamefont {Polkovnikov}},\
  }\bibfield  {title} {\enquote {\bibinfo {title} {{Heating and many-body
  resonances in a periodically driven two-band system}},}\ }\href@noop {}
  {\bibfield  {journal} {\bibinfo  {journal} {Phys. Rev. B}\ }\textbf {\bibinfo
  {volume} {93}},\ \bibinfo {pages} {155132} (\bibinfo {year}
  {2016})}\BibitemShut {NoStop}%
\bibitem [{\citenamefont {Abanin}\ \emph {et~al.}(2015)\citenamefont {Abanin},
  \citenamefont {De~Roeck},\ and\ \citenamefont {Huveneers}}]{Abanin2015}%
  \BibitemOpen
  \bibfield  {author} {\bibinfo {author} {\bibfnamefont {Dmitry~A.}\
  \bibnamefont {Abanin}}, \bibinfo {author} {\bibfnamefont {Wojciech}\
  \bibnamefont {De~Roeck}}, \ and\ \bibinfo {author} {\bibfnamefont
  {Fran{\c{c}}ois}\ \bibnamefont {Huveneers}},\ }\bibfield  {title} {\enquote
  {\bibinfo {title} {{Exponentially Slow Heating in Periodically Driven
  Many-Body Systems}},}\ }\href@noop {} {\bibfield  {journal} {\bibinfo
  {journal} {Phys. Rev. Lett.}\ }\textbf {\bibinfo {volume} {115}},\ \bibinfo
  {pages} {256803} (\bibinfo {year} {2015})}\BibitemShut {NoStop}%
\bibitem [{\citenamefont {Reitter}\ \emph {et~al.}(2017)\citenamefont
  {Reitter}, \citenamefont {N\"ager}, \citenamefont {Wintersperger},
  \citenamefont {Str\"ater}, \citenamefont {Bloch}, \citenamefont {Eckardt},\
  and\ \citenamefont {Schneider}}]{Reitter2017}%
  \BibitemOpen
  \bibfield  {author} {\bibinfo {author} {\bibfnamefont {Martin}\ \bibnamefont
  {Reitter}}, \bibinfo {author} {\bibfnamefont {Jakob}\ \bibnamefont
  {N\"ager}}, \bibinfo {author} {\bibfnamefont {Karen}\ \bibnamefont
  {Wintersperger}}, \bibinfo {author} {\bibfnamefont {Christoph}\ \bibnamefont
  {Str\"ater}}, \bibinfo {author} {\bibfnamefont {Immanuel}\ \bibnamefont
  {Bloch}}, \bibinfo {author} {\bibfnamefont {Andr\'e}\ \bibnamefont
  {Eckardt}}, \ and\ \bibinfo {author} {\bibfnamefont {Ulrich}\ \bibnamefont
  {Schneider}},\ }\bibfield  {title} {\enquote {\bibinfo {title} {Interaction
  dependent heating and atom loss in a periodically driven optical lattice},}\
  }\href@noop {} {\bibfield  {journal} {\bibinfo  {journal} {Phys. Rev. Lett.}\
  }\textbf {\bibinfo {volume} {119}},\ \bibinfo {pages} {200402} (\bibinfo
  {year} {2017})}\BibitemShut {NoStop}%
\bibitem [{Note2()}]{Note2}%
  \BibitemOpen
  \bibinfo {note} {Note that this low-frequency driving regime does not imply
  adiabaticity; while a large gap in the single spectrum is required, the
  system need not have a {\protect \it many-body} gap.}\BibitemShut {Stop}%
\bibitem [{\citenamefont {Abanin}\ \emph
  {et~al.}(2017{\natexlab{b}})\citenamefont {Abanin}, \citenamefont {De~Roeck},
  \citenamefont {Ho},\ and\ \citenamefont {Huveneers}}]{Abanin2017a}%
  \BibitemOpen
  \bibfield  {author} {\bibinfo {author} {\bibfnamefont {D.}~\bibnamefont
  {Abanin}}, \bibinfo {author} {\bibfnamefont {W.}~\bibnamefont {De~Roeck}},
  \bibinfo {author} {\bibfnamefont {W.~W.}\ \bibnamefont {Ho}}, \ and\ \bibinfo
  {author} {\bibfnamefont {F.}~\bibnamefont {Huveneers}},\ }\bibfield  {title}
  {\enquote {\bibinfo {title} {{A Rigorous Theory of Many-Body
  Prethermalization for Periodically Driven and Closed Quantum Systems}},}\
  }\href@noop {} {\bibfield  {journal} {\bibinfo  {journal} {Comm. Math Phys.}\
  }\textbf {\bibinfo {volume} {354}},\ \bibinfo {pages} {809} (\bibinfo {year}
  {2017}{\natexlab{b}})}\BibitemShut {NoStop}%
\bibitem [{\citenamefont {Regnault}\ and\ \citenamefont
  {Bernevig}(2011)}]{Regnault2011}%
  \BibitemOpen
  \bibfield  {author} {\bibinfo {author} {\bibfnamefont {N.}~\bibnamefont
  {Regnault}}\ and\ \bibinfo {author} {\bibfnamefont {B.~Andrei}\ \bibnamefont
  {Bernevig}},\ }\bibfield  {title} {\enquote {\bibinfo {title} {{Fractional
  Chern insulator}},}\ }\href@noop {} {\bibfield  {journal} {\bibinfo
  {journal} {Phys. Rev. X}\ }\textbf {\bibinfo {volume} {1}},\ \bibinfo {pages}
  {021014} (\bibinfo {year} {2011})}\BibitemShut {NoStop}%
\bibitem [{\citenamefont {Claassen}\ \emph {et~al.}(2017)\citenamefont
  {Claassen}, \citenamefont {Jiang}, \citenamefont {Mortiz},\ and\
  \citenamefont {Devereaux}}]{Claassen2017}%
  \BibitemOpen
  \bibfield  {author} {\bibinfo {author} {\bibfnamefont {Martin}\ \bibnamefont
  {Claassen}}, \bibinfo {author} {\bibfnamefont {Hong-Chen}\ \bibnamefont
  {Jiang}}, \bibinfo {author} {\bibfnamefont {Brian}\ \bibnamefont {Mortiz}}, \
  and\ \bibinfo {author} {\bibfnamefont {Thomas~P.}\ \bibnamefont
  {Devereaux}},\ }\bibfield  {title} {\enquote {\bibinfo {title} {Dynamical
  time-reversal symmetry breaking and photo-induced chiral spin liquids in
  frustrated mott insulators},}\ }\href@noop {} {\bibfield  {journal} {\bibinfo
   {journal} {Nature Communications}\ }\textbf {\bibinfo {volume} {8}},\
  \bibinfo {pages} {1192} (\bibinfo {year} {2017})}\BibitemShut {NoStop}%
\bibitem [{\citenamefont {Liu}\ \emph {et~al.}(2018{\natexlab{b}})\citenamefont
  {Liu}, \citenamefont {Hejazi},\ and\ \citenamefont {Balents}}]{Balents2018}%
  \BibitemOpen
  \bibfield  {author} {\bibinfo {author} {\bibfnamefont {Jianpeng}\
  \bibnamefont {Liu}}, \bibinfo {author} {\bibfnamefont {Kasra}\ \bibnamefont
  {Hejazi}}, \ and\ \bibinfo {author} {\bibfnamefont {Leon}\ \bibnamefont
  {Balents}},\ }\bibfield  {title} {\enquote {\bibinfo {title} {{Floquet
  engineering of multiorbital Mott insulators: applications to orthorhombic
  titanates}},}\ }\href@noop {} {\bibfield  {journal} {\bibinfo  {journal}
  {Phys. Rev. Lett.}\ }\textbf {\bibinfo {volume} {121}},\ \bibinfo {pages}
  {107201} (\bibinfo {year} {2018}{\natexlab{b}})}\BibitemShut {NoStop}%
\bibitem [{\citenamefont {Ho}\ and\ \citenamefont {Abanin}(2016)}]{Ho2016}%
  \BibitemOpen
  \bibfield  {author} {\bibinfo {author} {\bibfnamefont {Wen~Wei}\ \bibnamefont
  {Ho}}\ and\ \bibinfo {author} {\bibfnamefont {Dmitry~A.}\ \bibnamefont
  {Abanin}},\ }\bibfield  {title} {\enquote {\bibinfo {title} {{Quasi-adiabatic
  dynamics and state preparation in Floquet many-body systems}},}\ }\href@noop
  {} {\bibfield  {journal} {\bibinfo  {journal} {arXiv:1611.05024}\ } (\bibinfo
  {year} {2016})}\BibitemShut {NoStop}%
\bibitem [{\citenamefont {Singh}\ \emph {et~al.}(2018)\citenamefont {Singh},
  \citenamefont {Fujiwara}, \citenamefont {Geiger}, \citenamefont {Simmons},
  \citenamefont {Lipatov}, \citenamefont {Cao}, \citenamefont {Dotti},
  \citenamefont {Rajagopal}, \citenamefont {Senaratne}, \citenamefont
  {Shimasaki}, \citenamefont {Heyl}, \citenamefont {Eckardt},\ and\
  \citenamefont {Weld}}]{Singh2018}%
  \BibitemOpen
  \bibfield  {author} {\bibinfo {author} {\bibfnamefont {K.}~\bibnamefont
  {Singh}}, \bibinfo {author} {\bibfnamefont {K.~M.}\ \bibnamefont {Fujiwara}},
  \bibinfo {author} {\bibfnamefont {Z.~A.}\ \bibnamefont {Geiger}}, \bibinfo
  {author} {\bibfnamefont {E.~Q.}\ \bibnamefont {Simmons}}, \bibinfo {author}
  {\bibfnamefont {M.}~\bibnamefont {Lipatov}}, \bibinfo {author} {\bibfnamefont
  {A.}~\bibnamefont {Cao}}, \bibinfo {author} {\bibfnamefont {P.}~\bibnamefont
  {Dotti}}, \bibinfo {author} {\bibfnamefont {S.~V.}\ \bibnamefont
  {Rajagopal}}, \bibinfo {author} {\bibfnamefont {R.}~\bibnamefont
  {Senaratne}}, \bibinfo {author} {\bibfnamefont {T.}~\bibnamefont
  {Shimasaki}}, \bibinfo {author} {\bibfnamefont {M.}~\bibnamefont {Heyl}},
  \bibinfo {author} {\bibfnamefont {A.}~\bibnamefont {Eckardt}}, \ and\
  \bibinfo {author} {\bibfnamefont {D.~M.}\ \bibnamefont {Weld}},\ }\bibfield
  {title} {\enquote {\bibinfo {title} {{Controlling and characterizing Floquet
  prethermalization in a driven quantum system}},}\ }\href@noop {} {\bibfield
  {journal} {\bibinfo  {journal} {arXiv:1809.05554}\ } (\bibinfo {year}
  {2018})}\BibitemShut {NoStop}%
\bibitem [{\citenamefont {Gulden}\ \emph {et~al.}(2019)\citenamefont {Gulden},
  \citenamefont {Berg}, \citenamefont {Rudner},\ and\ \citenamefont
  {Lindner}}]{Gulden2019}%
  \BibitemOpen
  \bibfield  {author} {\bibinfo {author} {\bibfnamefont {Tobias}\ \bibnamefont
  {Gulden}}, \bibinfo {author} {\bibfnamefont {Erez}\ \bibnamefont {Berg}},
  \bibinfo {author} {\bibfnamefont {Mark~S.}\ \bibnamefont {Rudner}}, \ and\
  \bibinfo {author} {\bibfnamefont {Netanel~H.}\ \bibnamefont {Lindner}},\
  }\bibfield  {title} {\enquote {\bibinfo {title} {{Exponentially long lifetime
  of universal quasi-steady states in topological Floquet pumps}},}\
  }\href@noop {} {\bibfield  {journal} {\bibinfo  {journal} {arXiv:1901.08385}\
  } (\bibinfo {year} {2019})}\BibitemShut {NoStop}%
\bibitem [{\citenamefont {Basko}\ \emph {et~al.}(2006)\citenamefont {Basko},
  \citenamefont {Aleiner},\ and\ \citenamefont {Altshuler}}]{Basko2006}%
  \BibitemOpen
  \bibfield  {author} {\bibinfo {author} {\bibfnamefont {D.~M.}\ \bibnamefont
  {Basko}}, \bibinfo {author} {\bibfnamefont {I.~L.}\ \bibnamefont {Aleiner}},
  \ and\ \bibinfo {author} {\bibfnamefont {B.~L.}\ \bibnamefont {Altshuler}},\
  }\bibfield  {title} {\enquote {\bibinfo {title} {{Metal insulator transition
  in a weakly interacting many-electron system with localized single-particle
  states}},}\ }\href@noop {} {\bibfield  {journal} {\bibinfo  {journal} {Ann.
  Phys.}\ }\textbf {\bibinfo {volume} {321}},\ \bibinfo {pages} {1126}
  (\bibinfo {year} {2006})}\BibitemShut {NoStop}%
\bibitem [{\citenamefont {Oganesyan}\ and\ \citenamefont
  {Huse}(2007)}]{Oganesyan2007}%
  \BibitemOpen
  \bibfield  {author} {\bibinfo {author} {\bibfnamefont {Vadim}\ \bibnamefont
  {Oganesyan}}\ and\ \bibinfo {author} {\bibfnamefont {David~A.}\ \bibnamefont
  {Huse}},\ }\bibfield  {title} {\enquote {\bibinfo {title} {Localization of
  interacting fermions at high temperature},}\ }\href@noop {} {\bibfield
  {journal} {\bibinfo  {journal} {Phys. Rev. B}\ }\textbf {\bibinfo {volume}
  {75}},\ \bibinfo {pages} {155111} (\bibinfo {year} {2007})}\BibitemShut
  {NoStop}%
\bibitem [{\citenamefont {Pal}\ and\ \citenamefont {Huse}(2010)}]{Pal2010}%
  \BibitemOpen
  \bibfield  {author} {\bibinfo {author} {\bibfnamefont {Arijeet}\ \bibnamefont
  {Pal}}\ and\ \bibinfo {author} {\bibfnamefont {David~A.}\ \bibnamefont
  {Huse}},\ }\bibfield  {title} {\enquote {\bibinfo {title} {Many-body
  localization phase transition},}\ }\href@noop {} {\bibfield  {journal}
  {\bibinfo  {journal} {Phys. Rev. B}\ }\textbf {\bibinfo {volume} {82}},\
  \bibinfo {pages} {174411} (\bibinfo {year} {2010})}\BibitemShut {NoStop}%
\bibitem [{\citenamefont {Nandkishore}\ and\ \citenamefont
  {Huse}(2015)}]{NandkishoreReview}%
  \BibitemOpen
  \bibfield  {author} {\bibinfo {author} {\bibfnamefont {Rahul}\ \bibnamefont
  {Nandkishore}}\ and\ \bibinfo {author} {\bibfnamefont {David~A.}\
  \bibnamefont {Huse}},\ }\bibfield  {title} {\enquote {\bibinfo {title}
  {Many-body localization and thermalization in quantum statistical
  mechanics},}\ }\href@noop {} {\bibfield  {journal} {\bibinfo  {journal}
  {Annual Review of Condensed Matter Physics}\ }\textbf {\bibinfo {volume}
  {6}},\ \bibinfo {pages} {15} (\bibinfo {year} {2015})}\BibitemShut {NoStop}%
\bibitem [{\citenamefont {Lazarides}\ \emph {et~al.}(2015)\citenamefont
  {Lazarides}, \citenamefont {Das},\ and\ \citenamefont
  {Moessner}}]{Lazarides2015}%
  \BibitemOpen
  \bibfield  {author} {\bibinfo {author} {\bibfnamefont {Achilleas}\
  \bibnamefont {Lazarides}}, \bibinfo {author} {\bibfnamefont {Arnab}\
  \bibnamefont {Das}}, \ and\ \bibinfo {author} {\bibfnamefont {Roderich}\
  \bibnamefont {Moessner}},\ }\bibfield  {title} {\enquote {\bibinfo {title}
  {{Fate of Many-Body Localization Under Periodic Driving}},}\ }\href@noop {}
  {\bibfield  {journal} {\bibinfo  {journal} {Phys. Rev. Lett.}\ }\textbf
  {\bibinfo {volume} {115}},\ \bibinfo {pages} {030402} (\bibinfo {year}
  {2015})}\BibitemShut {NoStop}%
\bibitem [{\citenamefont {Ponte}\ \emph {et~al.}(2015)\citenamefont {Ponte},
  \citenamefont {Papi\ifmmode~\acute{c}\else \'{c}\fi{}}, \citenamefont
  {Huveneers},\ and\ \citenamefont {Abanin}}]{Ponte2015}%
  \BibitemOpen
  \bibfield  {author} {\bibinfo {author} {\bibfnamefont {Pedro}\ \bibnamefont
  {Ponte}}, \bibinfo {author} {\bibfnamefont {Z.}~\bibnamefont
  {Papi\ifmmode~\acute{c}\else \'{c}\fi{}}}, \bibinfo {author} {\bibfnamefont
  {Fran{\c{c}}ois}\ \bibnamefont {Huveneers}}, \ and\ \bibinfo {author}
  {\bibfnamefont {Dmitry~A.}\ \bibnamefont {Abanin}},\ }\bibfield  {title}
  {\enquote {\bibinfo {title} {{Many-Body Localization in Periodically Driven
  Systems}},}\ }\href@noop {} {\bibfield  {journal} {\bibinfo  {journal} {Phys.
  Rev. Lett.}\ }\textbf {\bibinfo {volume} {114}},\ \bibinfo {pages} {140401}
  (\bibinfo {year} {2015})}\BibitemShut {NoStop}%
\bibitem [{\citenamefont {Bordia}\ \emph {et~al.}(2017)\citenamefont {Bordia},
  \citenamefont {Luschen}, \citenamefont {Schneider}, \citenamefont {Knap},\
  and\ \citenamefont {Bloch}}]{Bordia2017}%
  \BibitemOpen
  \bibfield  {author} {\bibinfo {author} {\bibfnamefont {P.}~\bibnamefont
  {Bordia}}, \bibinfo {author} {\bibfnamefont {H.}~\bibnamefont {Luschen}},
  \bibinfo {author} {\bibfnamefont {U.}~\bibnamefont {Schneider}}, \bibinfo
  {author} {\bibfnamefont {M.}~\bibnamefont {Knap}}, \ and\ \bibinfo {author}
  {\bibfnamefont {I.}~\bibnamefont {Bloch}},\ }\bibfield  {title} {\enquote
  {\bibinfo {title} {{Periodically driving a many-body localized quantum
  system}},}\ }\href@noop {} {\bibfield  {journal} {\bibinfo  {journal} {Nature
  Phys.}\ }\textbf {\bibinfo {volume} {13}},\ \bibinfo {pages} {460} (\bibinfo
  {year} {2017})}\BibitemShut {NoStop}%
\bibitem [{\citenamefont {Wilczek}(2012)}]{Wilczek2012}%
  \BibitemOpen
  \bibfield  {author} {\bibinfo {author} {\bibfnamefont {Frank}\ \bibnamefont
  {Wilczek}},\ }\bibfield  {title} {\enquote {\bibinfo {title} {Quantum time
  crystals},}\ }\href@noop {} {\bibfield  {journal} {\bibinfo  {journal} {Phys.
  Rev. Lett.}\ }\textbf {\bibinfo {volume} {109}},\ \bibinfo {pages} {160401}
  (\bibinfo {year} {2012})}\BibitemShut {NoStop}%
\bibitem [{\citenamefont {Zeng}\ and\ \citenamefont {Sheng}(2017)}]{Zeng2017}%
  \BibitemOpen
  \bibfield  {author} {\bibinfo {author} {\bibfnamefont {T.-S.}\ \bibnamefont
  {Zeng}}\ and\ \bibinfo {author} {\bibfnamefont {D.~N.}\ \bibnamefont
  {Sheng}},\ }\bibfield  {title} {\enquote {\bibinfo {title} {{Prethermal time
  crystals in one-dimensional periodically driven Floquet system}},}\
  }\href@noop {} {\bibfield  {journal} {\bibinfo  {journal} {arXiv:1707.00404}\
  } (\bibinfo {year} {2017})}\BibitemShut {NoStop}%
\bibitem [{\citenamefont {Zhang}\ \emph {et~al.}(2017)\citenamefont {Zhang},
  \citenamefont {Hess}, \citenamefont {Kyprianidis}, \citenamefont {Becker},
  \citenamefont {Lee}, \citenamefont {Smith}, \citenamefont {Pagano},
  \citenamefont {Potirniche}, \citenamefont {Potter}, \citenamefont
  {Vishwanath}, \citenamefont {Yao},\ and\ \citenamefont {Monroe}}]{Zhang2017}%
  \BibitemOpen
  \bibfield  {author} {\bibinfo {author} {\bibfnamefont {J.}~\bibnamefont
  {Zhang}}, \bibinfo {author} {\bibfnamefont {P.~W.}\ \bibnamefont {Hess}},
  \bibinfo {author} {\bibfnamefont {A.}~\bibnamefont {Kyprianidis}}, \bibinfo
  {author} {\bibfnamefont {P.}~\bibnamefont {Becker}}, \bibinfo {author}
  {\bibfnamefont {A.}~\bibnamefont {Lee}}, \bibinfo {author} {\bibfnamefont
  {J.}~\bibnamefont {Smith}}, \bibinfo {author} {\bibfnamefont
  {G.}~\bibnamefont {Pagano}}, \bibinfo {author} {\bibfnamefont {I.-D.}\
  \bibnamefont {Potirniche}}, \bibinfo {author} {\bibfnamefont {A.~C.}\
  \bibnamefont {Potter}}, \bibinfo {author} {\bibfnamefont {A.}~\bibnamefont
  {Vishwanath}}, \bibinfo {author} {\bibfnamefont {N.~Y.}\ \bibnamefont {Yao}},
  \ and\ \bibinfo {author} {\bibfnamefont {C.}~\bibnamefont {Monroe}},\
  }\bibfield  {title} {\enquote {\bibinfo {title} {{Observation of a discrete
  time crystal}},}\ }\href@noop {} {\bibfield  {journal} {\bibinfo  {journal}
  {Nature}\ }\textbf {\bibinfo {volume} {543}},\ \bibinfo {pages} {217}
  (\bibinfo {year} {2017})}\BibitemShut {NoStop}%
\bibitem [{\citenamefont {Choi}\ \emph {et~al.}(2017)\citenamefont {Choi},
  \citenamefont {Choi}, \citenamefont {Landig}, \citenamefont {Kucsko},
  \citenamefont {Zhou}, \citenamefont {Isoya}, \citenamefont {Jelezko},
  \citenamefont {Onoda}, \citenamefont {Sumiya}, \citenamefont {Khemani},
  \citenamefont {von Keyserlingk}, \citenamefont {Yao}, \citenamefont
  {Demler},\ and\ \citenamefont {Lukin}}]{Choi2017}%
  \BibitemOpen
  \bibfield  {author} {\bibinfo {author} {\bibfnamefont {S.}~\bibnamefont
  {Choi}}, \bibinfo {author} {\bibfnamefont {J.}~\bibnamefont {Choi}}, \bibinfo
  {author} {\bibfnamefont {R.}~\bibnamefont {Landig}}, \bibinfo {author}
  {\bibfnamefont {G.}~\bibnamefont {Kucsko}}, \bibinfo {author} {\bibfnamefont
  {H.}~\bibnamefont {Zhou}}, \bibinfo {author} {\bibfnamefont {J.}~\bibnamefont
  {Isoya}}, \bibinfo {author} {\bibfnamefont {F.}~\bibnamefont {Jelezko}},
  \bibinfo {author} {\bibfnamefont {S.}~\bibnamefont {Onoda}}, \bibinfo
  {author} {\bibfnamefont {H.}~\bibnamefont {Sumiya}}, \bibinfo {author}
  {\bibfnamefont {V.}~\bibnamefont {Khemani}}, \bibinfo {author} {\bibfnamefont
  {C.}~\bibnamefont {von Keyserlingk}}, \bibinfo {author} {\bibfnamefont
  {N.~Y.}\ \bibnamefont {Yao}}, \bibinfo {author} {\bibfnamefont
  {E.}~\bibnamefont {Demler}}, \ and\ \bibinfo {author} {\bibfnamefont {M.~D.}\
  \bibnamefont {Lukin}},\ }\bibfield  {title} {\enquote {\bibinfo {title}
  {{Observation of discrete time-crystalline order in a disordered dipolar
  many-body system}},}\ }\href@noop {} {\bibfield  {journal} {\bibinfo
  {journal} {Nature}\ }\textbf {\bibinfo {volume} {543}},\ \bibinfo {pages}
  {221} (\bibinfo {year} {2017})}\BibitemShut {NoStop}%
\bibitem [{\citenamefont {Dykman}\ \emph {et~al.}(2018)\citenamefont {Dykman},
  \citenamefont {Bruder}, \citenamefont {L\"orch},\ and\ \citenamefont
  {Zhang}}]{Dykman2018}%
  \BibitemOpen
  \bibfield  {author} {\bibinfo {author} {\bibfnamefont {M.~I.}\ \bibnamefont
  {Dykman}}, \bibinfo {author} {\bibfnamefont {Christoph}\ \bibnamefont
  {Bruder}}, \bibinfo {author} {\bibfnamefont {Niels}\ \bibnamefont {L\"orch}},
  \ and\ \bibinfo {author} {\bibfnamefont {Yaxing}\ \bibnamefont {Zhang}},\
  }\bibfield  {title} {\enquote {\bibinfo {title} {Interaction-induced
  time-symmetry breaking in driven quantum oscillators},}\ }\href@noop {}
  {\bibfield  {journal} {\bibinfo  {journal} {Phys. Rev. B}\ }\textbf {\bibinfo
  {volume} {98}},\ \bibinfo {pages} {195444} (\bibinfo {year}
  {2018})}\BibitemShut {NoStop}%
\bibitem [{\citenamefont {Nandkishore}\ and\ \citenamefont
  {Potter}(2014)}]{Nandkishore2014}%
  \BibitemOpen
  \bibfield  {author} {\bibinfo {author} {\bibfnamefont {Rahul}\ \bibnamefont
  {Nandkishore}}\ and\ \bibinfo {author} {\bibfnamefont {Andrew~C.}\
  \bibnamefont {Potter}},\ }\bibfield  {title} {\enquote {\bibinfo {title}
  {{Marginal Anderson localization and many-body delocalization}},}\
  }\href@noop {} {\bibfield  {journal} {\bibinfo  {journal} {Phys.~Rev.~B}\
  }\textbf {\bibinfo {volume} {90}},\ \bibinfo {pages} {195115} (\bibinfo
  {year} {2014})}\BibitemShut {NoStop}%
\bibitem [{\citenamefont {Po}\ \emph {et~al.}(2016)\citenamefont {Po},
  \citenamefont {Fidkowski}, \citenamefont {Morimoto}, \citenamefont {Potter},\
  and\ \citenamefont {Vishwanath}}]{Po2016}%
  \BibitemOpen
  \bibfield  {author} {\bibinfo {author} {\bibfnamefont {Hoi~Chun}\
  \bibnamefont {Po}}, \bibinfo {author} {\bibfnamefont {Lukasz}\ \bibnamefont
  {Fidkowski}}, \bibinfo {author} {\bibfnamefont {Takahiro}\ \bibnamefont
  {Morimoto}}, \bibinfo {author} {\bibfnamefont {Andrew~C.}\ \bibnamefont
  {Potter}}, \ and\ \bibinfo {author} {\bibfnamefont {Ashvin}\ \bibnamefont
  {Vishwanath}},\ }\bibfield  {title} {\enquote {\bibinfo {title} {{Chiral
  Floquet Phases of Many-Body Localized Bosons}},}\ }\href@noop {} {\bibfield
  {journal} {\bibinfo  {journal} {Phys. Rev. X}\ }\textbf {\bibinfo {volume}
  {6}},\ \bibinfo {pages} {041070} (\bibinfo {year} {2016})}\BibitemShut
  {NoStop}%
\bibitem [{\citenamefont {Po}\ \emph {et~al.}(2017)\citenamefont {Po},
  \citenamefont {Fidkowski}, \citenamefont {Vishwanath},\ and\ \citenamefont
  {Potter}}]{Po2017}%
  \BibitemOpen
  \bibfield  {author} {\bibinfo {author} {\bibfnamefont {Hoi~Chun}\
  \bibnamefont {Po}}, \bibinfo {author} {\bibfnamefont {Lukasz}\ \bibnamefont
  {Fidkowski}}, \bibinfo {author} {\bibfnamefont {Ashvin}\ \bibnamefont
  {Vishwanath}}, \ and\ \bibinfo {author} {\bibfnamefont {Andrew~C.}\
  \bibnamefont {Potter}},\ }\bibfield  {title} {\enquote {\bibinfo {title}
  {{Radical chiral Floquet phases in a periodically driven Kitaev model and
  beyond}},}\ }\href@noop {} {\bibfield  {journal} {\bibinfo  {journal} {Phys.
  Rev. B}\ }\textbf {\bibinfo {volume} {96}},\ \bibinfo {pages} {245116}
  (\bibinfo {year} {2017})}\BibitemShut {NoStop}%
\bibitem [{\citenamefont {Nathan}\ \emph {et~al.}(2019)\citenamefont {Nathan},
  \citenamefont {Abanin}, \citenamefont {Berg}, \citenamefont {Lindner},\ and\
  \citenamefont {Rudner}}]{Nathan2017}%
  \BibitemOpen
  \bibfield  {author} {\bibinfo {author} {\bibfnamefont {Frederik}\
  \bibnamefont {Nathan}}, \bibinfo {author} {\bibfnamefont {Dmitry}\
  \bibnamefont {Abanin}}, \bibinfo {author} {\bibfnamefont {Erez}\ \bibnamefont
  {Berg}}, \bibinfo {author} {\bibfnamefont {Netanel~H.}\ \bibnamefont
  {Lindner}}, \ and\ \bibinfo {author} {\bibfnamefont {Mark~S.}\ \bibnamefont
  {Rudner}},\ }\bibfield  {title} {\enquote {\bibinfo {title} {{Anomalous
  Floquet Insulators}},}\ }\href@noop {} {\bibfield  {journal} {\bibinfo
  {journal} {Phys.~Rev.~B}\ }\textbf {\bibinfo {volume} {99}},\ \bibinfo
  {pages} {195133} (\bibinfo {year} {2019})}\BibitemShut {NoStop}%
\bibitem [{\citenamefont {Liu}(2015)}]{Liu2015}%
  \BibitemOpen
  \bibfield  {author} {\bibinfo {author} {\bibfnamefont {Dong~E.}\ \bibnamefont
  {Liu}},\ }\bibfield  {title} {\enquote {\bibinfo {title} {{Classification of
  the Floquet statistical distribution for time-periodic open systems}},}\
  }\href@noop {} {\bibfield  {journal} {\bibinfo  {journal} {Phys. Rev. B}\
  }\textbf {\bibinfo {volume} {91}},\ \bibinfo {pages} {144301} (\bibinfo
  {year} {2015})}\BibitemShut {NoStop}%
\bibitem [{\citenamefont {Shirai}\ \emph {et~al.}(2015)\citenamefont {Shirai},
  \citenamefont {Mori},\ and\ \citenamefont {Miyashita}}]{Shirai2015}%
  \BibitemOpen
  \bibfield  {author} {\bibinfo {author} {\bibfnamefont {T.}~\bibnamefont
  {Shirai}}, \bibinfo {author} {\bibfnamefont {T.}~\bibnamefont {Mori}}, \ and\
  \bibinfo {author} {\bibfnamefont {S.}~\bibnamefont {Miyashita}},\ }\bibfield
  {title} {\enquote {\bibinfo {title} {{Condition for emergence of the
  Floquet-Gibbs state in periodically driven open systems}},}\ }\href@noop {}
  {\bibfield  {journal} {\bibinfo  {journal} {Phys. Rev. E}\ }\textbf {\bibinfo
  {volume} {91}},\ \bibinfo {pages} {030101(R)} (\bibinfo {year}
  {2015})}\BibitemShut {NoStop}%
\bibitem [{\citenamefont {Shirai}\ \emph {et~al.}(2016)\citenamefont {Shirai},
  \citenamefont {Thinga}, \citenamefont {Mori}, \citenamefont {Denisov},
  \citenamefont {Hanggi},\ and\ \citenamefont {Miyashita}}]{Shirai2016}%
  \BibitemOpen
  \bibfield  {author} {\bibinfo {author} {\bibfnamefont {T.}~\bibnamefont
  {Shirai}}, \bibinfo {author} {\bibfnamefont {J.}~\bibnamefont {Thinga}},
  \bibinfo {author} {\bibfnamefont {T.}~\bibnamefont {Mori}}, \bibinfo {author}
  {\bibfnamefont {S.}~\bibnamefont {Denisov}}, \bibinfo {author} {\bibfnamefont
  {P.}~\bibnamefont {Hanggi}}, \ and\ \bibinfo {author} {\bibfnamefont
  {S.}~\bibnamefont {Miyashita}},\ }\bibfield  {title} {\enquote {\bibinfo
  {title} {{Effective Floquet-Gibbs states for dissipative quantum systems}},}\
  }\href@noop {} {\bibfield  {journal} {\bibinfo  {journal} {New J. Phys.}\
  }\textbf {\bibinfo {volume} {18}},\ \bibinfo {pages} {053008} (\bibinfo
  {year} {2016})}\BibitemShut {NoStop}%
\bibitem [{\citenamefont {Torres}\ and\ \citenamefont
  {Kunold}(2005)}]{TorresFloquetKubo2005}%
  \BibitemOpen
  \bibfield  {author} {\bibinfo {author} {\bibfnamefont {Manuel}\ \bibnamefont
  {Torres}}\ and\ \bibinfo {author} {\bibfnamefont {Alejandro}\ \bibnamefont
  {Kunold}},\ }\bibfield  {title} {\enquote {\bibinfo {title} {Kubo formula for
  floquet states and photoconductivity oscillations in a two-dimensional
  electron gas},}\ }\href@noop {} {\bibfield  {journal} {\bibinfo  {journal}
  {Phys. Rev. B}\ }\textbf {\bibinfo {volume} {71}},\ \bibinfo {pages} {115313}
  (\bibinfo {year} {2005})}\BibitemShut {NoStop}%
\bibitem [{\citenamefont {Kohn}(2001)}]{Kohn2001}%
  \BibitemOpen
  \bibfield  {author} {\bibinfo {author} {\bibfnamefont {Walter}\ \bibnamefont
  {Kohn}},\ }\bibfield  {title} {\enquote {\bibinfo {title} {Periodic
  thermodynamics},}\ }\href@noop {} {\bibfield  {journal} {\bibinfo  {journal}
  {Journal of Statistical Physics}\ }\textbf {\bibinfo {volume} {103}},\
  \bibinfo {pages} {417} (\bibinfo {year} {2001})}\BibitemShut {NoStop}%
\bibitem [{\citenamefont {Hone}\ \emph {et~al.}(1997)\citenamefont {Hone},
  \citenamefont {Ketzmerick},\ and\ \citenamefont {Kohn}}]{Hone1997}%
  \BibitemOpen
  \bibfield  {author} {\bibinfo {author} {\bibfnamefont {Daniel~W.}\
  \bibnamefont {Hone}}, \bibinfo {author} {\bibfnamefont {Roland}\ \bibnamefont
  {Ketzmerick}}, \ and\ \bibinfo {author} {\bibfnamefont {Walter}\ \bibnamefont
  {Kohn}},\ }\bibfield  {title} {\enquote {\bibinfo {title} {{Time-dependent
  Floquet theory and absence of an adiabatic limit}},}\ }\href@noop {}
  {\bibfield  {journal} {\bibinfo  {journal} {Phys. Rev. A}\ }\textbf {\bibinfo
  {volume} {56}},\ \bibinfo {pages} {4045} (\bibinfo {year}
  {1997})}\BibitemShut {NoStop}%
\bibitem [{\citenamefont {Hone}\ \emph {et~al.}(2009)\citenamefont {Hone},
  \citenamefont {Ketzmerick},\ and\ \citenamefont {Kohn}}]{Hone2009}%
  \BibitemOpen
  \bibfield  {author} {\bibinfo {author} {\bibfnamefont {Daniel~W.}\
  \bibnamefont {Hone}}, \bibinfo {author} {\bibfnamefont {Roland}\ \bibnamefont
  {Ketzmerick}}, \ and\ \bibinfo {author} {\bibfnamefont {Walter}\ \bibnamefont
  {Kohn}},\ }\bibfield  {title} {\enquote {\bibinfo {title} {{Statistical
  mechanics of Floquet systems: The pervasive problem of near degeneracies}},}\
  }\href@noop {} {\bibfield  {journal} {\bibinfo  {journal} {Phys. Rev. E}\
  }\textbf {\bibinfo {volume} {79}},\ \bibinfo {pages} {051129} (\bibinfo
  {year} {2009})}\BibitemShut {NoStop}%
\bibitem [{\citenamefont {Dehghani}\ \emph {et~al.}(2015)\citenamefont
  {Dehghani}, \citenamefont {Oka},\ and\ \citenamefont {Mitra}}]{Dehghani2015}%
  \BibitemOpen
  \bibfield  {author} {\bibinfo {author} {\bibfnamefont {H.}~\bibnamefont
  {Dehghani}}, \bibinfo {author} {\bibfnamefont {T.}~\bibnamefont {Oka}}, \
  and\ \bibinfo {author} {\bibfnamefont {A.}~\bibnamefont {Mitra}},\ }\bibfield
   {title} {\enquote {\bibinfo {title} {{Out-of-equilibrium electrons and the
  Hall conductance of a Floquet topological insulator}},}\ }\href@noop {}
  {\bibfield  {journal} {\bibinfo  {journal} {Phys. Rev. B}\ }\textbf {\bibinfo
  {volume} {91}},\ \bibinfo {pages} {155422} (\bibinfo {year}
  {2015})}\BibitemShut {NoStop}%
\bibitem [{\citenamefont {Genske}\ and\ \citenamefont
  {Rosch}(2015)}]{Genske2015}%
  \BibitemOpen
  \bibfield  {author} {\bibinfo {author} {\bibfnamefont {Maximilian}\
  \bibnamefont {Genske}}\ and\ \bibinfo {author} {\bibfnamefont {Achim}\
  \bibnamefont {Rosch}},\ }\bibfield  {title} {\enquote {\bibinfo {title}
  {{Floquet-Boltzmann equation for periodically driven Fermi systems}},}\
  }\href@noop {} {\bibfield  {journal} {\bibinfo  {journal} {Phys. Rev. A}\
  }\textbf {\bibinfo {volume} {92}},\ \bibinfo {pages} {062108} (\bibinfo
  {year} {2015})}\BibitemShut {NoStop}%
\bibitem [{\citenamefont {Esin}\ \emph {et~al.}(2018)\citenamefont {Esin},
  \citenamefont {Rudner}, \citenamefont {Refael},\ and\ \citenamefont
  {Lindner}}]{Esin2018}%
  \BibitemOpen
  \bibfield  {author} {\bibinfo {author} {\bibfnamefont {Iliya}\ \bibnamefont
  {Esin}}, \bibinfo {author} {\bibfnamefont {Mark~S.}\ \bibnamefont {Rudner}},
  \bibinfo {author} {\bibfnamefont {Gil}\ \bibnamefont {Refael}}, \ and\
  \bibinfo {author} {\bibfnamefont {Netanel~H.}\ \bibnamefont {Lindner}},\
  }\bibfield  {title} {\enquote {\bibinfo {title} {{Quantized transport and
  steady states of Floquet topological insulators}},}\ }\href@noop {}
  {\bibfield  {journal} {\bibinfo  {journal} {Phys. Rev. B}\ }\textbf {\bibinfo
  {volume} {97}},\ \bibinfo {pages} {245401} (\bibinfo {year}
  {2018})}\BibitemShut {NoStop}%
\bibitem [{\citenamefont {Iadecola}\ \emph {et~al.}(2015)\citenamefont
  {Iadecola}, \citenamefont {Neupert},\ and\ \citenamefont
  {Chamon}}]{Iadecola2015}%
  \BibitemOpen
  \bibfield  {author} {\bibinfo {author} {\bibfnamefont {Thomas}\ \bibnamefont
  {Iadecola}}, \bibinfo {author} {\bibfnamefont {Titus}\ \bibnamefont
  {Neupert}}, \ and\ \bibinfo {author} {\bibfnamefont {Claudio}\ \bibnamefont
  {Chamon}},\ }\bibfield  {title} {\enquote {\bibinfo {title} {{Occupation of
  topological Floquet bands in open systems}},}\ }\href@noop {} {\bibfield
  {journal} {\bibinfo  {journal} {Phys. Rev. B}\ }\textbf {\bibinfo {volume}
  {91}},\ \bibinfo {pages} {235133} (\bibinfo {year} {2015})}\BibitemShut
  {NoStop}%
\bibitem [{\citenamefont {Goebel}\ and\ \citenamefont
  {Hildebrand}(1978)}]{Goebel1978}%
  \BibitemOpen
  \bibfield  {author} {\bibinfo {author} {\bibfnamefont {E.~O.}\ \bibnamefont
  {Goebel}}\ and\ \bibinfo {author} {\bibfnamefont {O.}~\bibnamefont
  {Hildebrand}},\ }\bibfield  {title} {\enquote {\bibinfo {title}
  {{Thermalization of the Electron-Hole Plasma in GaAs}},}\ }\href@noop {}
  {\bibfield  {journal} {\bibinfo  {journal} {Phys.~Stat.~Sol.~(b)}\ }\textbf
  {\bibinfo {volume} {88}},\ \bibinfo {pages} {685} (\bibinfo {year}
  {1978})}\BibitemShut {NoStop}%
\bibitem [{\citenamefont {Glazman}(1981)}]{Glazman1981}%
  \BibitemOpen
  \bibfield  {author} {\bibinfo {author} {\bibfnamefont {L.~I.}\ \bibnamefont
  {Glazman}},\ }\bibfield  {title} {\enquote {\bibinfo {title} {Resonant
  excitation of carriers in a seminconductor by a high-power laser pulse},}\
  }\href@noop {} {\bibfield  {journal} {\bibinfo  {journal} {Sov. Phys. JETP}\
  }\textbf {\bibinfo {volume} {53}},\ \bibinfo {pages} {178--181} (\bibinfo
  {year} {1981})}\BibitemShut {NoStop}%
\bibitem [{\citenamefont {Glazman}(1983)}]{Glazman1983}%
  \BibitemOpen
  \bibfield  {author} {\bibinfo {author} {\bibfnamefont {L.~I.}\ \bibnamefont
  {Glazman}},\ }\bibfield  {title} {\enquote {\bibinfo {title} {Kinetics of
  electrons and holes in direct gap seminconductors photoexcited by high
  intensity pulses},}\ }\href@noop {} {\bibfield  {journal} {\bibinfo
  {journal} {Sov. Phys. Semi.}\ }\textbf {\bibinfo {volume} {17}},\ \bibinfo
  {pages} {494--498} (\bibinfo {year} {1983})}\BibitemShut {NoStop}%
\bibitem [{\citenamefont {Chow}\ and\ \citenamefont {Koch}(1999)}]{Chow1999}%
  \BibitemOpen
  \bibfield  {author} {\bibinfo {author} {\bibfnamefont {Weng~W.}\ \bibnamefont
  {Chow}}\ and\ \bibinfo {author} {\bibfnamefont {Stephan~W.}\ \bibnamefont
  {Koch}},\ }\href@noop {} {\emph {\bibinfo {title} {Semiconductor-Laser
  Fundamentals}}}\ (\bibinfo  {publisher} {Springer},\ \bibinfo {year}
  {1999})\BibitemShut {NoStop}%
\bibitem [{\citenamefont {Huang}\ \emph {et~al.}(2001)\citenamefont {Huang},
  \citenamefont {Mao}, \citenamefont {Feick}, \citenamefont {Yan},
  \citenamefont {Wu}, \citenamefont {Kind}, \citenamefont {Weber},
  \citenamefont {Russo},\ and\ \citenamefont {Yang}}]{Huang2001}%
  \BibitemOpen
  \bibfield  {author} {\bibinfo {author} {\bibfnamefont {Michael~H.}\
  \bibnamefont {Huang}}, \bibinfo {author} {\bibfnamefont {Samuel}\
  \bibnamefont {Mao}}, \bibinfo {author} {\bibfnamefont {Henning}\ \bibnamefont
  {Feick}}, \bibinfo {author} {\bibfnamefont {Haoquan}\ \bibnamefont {Yan}},
  \bibinfo {author} {\bibfnamefont {Yiying}\ \bibnamefont {Wu}}, \bibinfo
  {author} {\bibfnamefont {Hannes}\ \bibnamefont {Kind}}, \bibinfo {author}
  {\bibfnamefont {Eicke}\ \bibnamefont {Weber}}, \bibinfo {author}
  {\bibfnamefont {Richard}\ \bibnamefont {Russo}}, \ and\ \bibinfo {author}
  {\bibfnamefont {Peidong}\ \bibnamefont {Yang}},\ }\bibfield  {title}
  {\enquote {\bibinfo {title} {{Room-Temperature Ultraviolet Nanowire
  Nanolasers}},}\ }\href@noop {} {\bibfield  {journal} {\bibinfo  {journal}
  {Science}\ }\textbf {\bibinfo {volume} {292}},\ \bibinfo {pages} {1897}
  (\bibinfo {year} {2001})}\BibitemShut {NoStop}%
\bibitem [{\citenamefont {R\"{o}der}\ \emph {et~al.}(2013)\citenamefont
  {R\"{o}der}, \citenamefont {Wille}, \citenamefont {Geburt}, \citenamefont
  {Rensberg}, \citenamefont {Zhang}, \citenamefont {Lu}, \citenamefont
  {Capasso}, \citenamefont {Buschlinger}, \citenamefont {Peshcel},\ and\
  \citenamefont {Ronning}}]{Roder2013}%
  \BibitemOpen
  \bibfield  {author} {\bibinfo {author} {\bibfnamefont {Robert}\ \bibnamefont
  {R\"{o}der}}, \bibinfo {author} {\bibfnamefont {Marcel}\ \bibnamefont
  {Wille}}, \bibinfo {author} {\bibfnamefont {Sebastian}\ \bibnamefont
  {Geburt}}, \bibinfo {author} {\bibfnamefont {Jura}\ \bibnamefont {Rensberg}},
  \bibinfo {author} {\bibfnamefont {Mengyao}\ \bibnamefont {Zhang}}, \bibinfo
  {author} {\bibfnamefont {Jia~Grace}\ \bibnamefont {Lu}}, \bibinfo {author}
  {\bibfnamefont {Federico}\ \bibnamefont {Capasso}}, \bibinfo {author}
  {\bibfnamefont {Robert}\ \bibnamefont {Buschlinger}}, \bibinfo {author}
  {\bibfnamefont {Ulf}\ \bibnamefont {Peshcel}}, \ and\ \bibinfo {author}
  {\bibfnamefont {Carsten}\ \bibnamefont {Ronning}},\ }\bibfield  {title}
  {\enquote {\bibinfo {title} {Continuous wave nanowire lasing},}\ }\href@noop
  {} {\bibfield  {journal} {\bibinfo  {journal} {Nano Letters}\ }\textbf
  {\bibinfo {volume} {13}},\ \bibinfo {pages} {3602} (\bibinfo {year}
  {2013})}\BibitemShut {NoStop}%
\bibitem [{\citenamefont {Dehghani}\ \emph {et~al.}(2014)\citenamefont
  {Dehghani}, \citenamefont {Oka},\ and\ \citenamefont {Mitra}}]{Dehghani2014}%
  \BibitemOpen
  \bibfield  {author} {\bibinfo {author} {\bibfnamefont {H.}~\bibnamefont
  {Dehghani}}, \bibinfo {author} {\bibfnamefont {T.}~\bibnamefont {Oka}}, \
  and\ \bibinfo {author} {\bibfnamefont {A.}~\bibnamefont {Mitra}},\ }\bibfield
   {title} {\enquote {\bibinfo {title} {{Dissipative Floquet topological
  systems}},}\ }\href@noop {} {\bibfield  {journal} {\bibinfo  {journal} {Phys.
  Rev. B}\ }\textbf {\bibinfo {volume} {90}},\ \bibinfo {pages} {195429}
  (\bibinfo {year} {2014})}\BibitemShut {NoStop}%
\bibitem [{\citenamefont {Seetharam}\ \emph {et~al.}(2015)\citenamefont
  {Seetharam}, \citenamefont {Bardyn}, \citenamefont {Lindner}, \citenamefont
  {Rudner},\ and\ \citenamefont {Refael}}]{Seetharam2015}%
  \BibitemOpen
  \bibfield  {author} {\bibinfo {author} {\bibfnamefont {K.~I.}\ \bibnamefont
  {Seetharam}}, \bibinfo {author} {\bibfnamefont {C.-E.}\ \bibnamefont
  {Bardyn}}, \bibinfo {author} {\bibfnamefont {N.~H.}\ \bibnamefont {Lindner}},
  \bibinfo {author} {\bibfnamefont {M.~S.}\ \bibnamefont {Rudner}}, \ and\
  \bibinfo {author} {\bibfnamefont {G.}~\bibnamefont {Refael}},\ }\bibfield
  {title} {\enquote {\bibinfo {title} {{Controlled Population of Floquet-Bloch
  States via Coupling to Bose and Fermi Baths}},}\ }\href@noop {} {\bibfield
  {journal} {\bibinfo  {journal} {Phys. Rev. X}\ }\textbf {\bibinfo {volume}
  {5}},\ \bibinfo {pages} {041050} (\bibinfo {year} {2015})}\BibitemShut
  {NoStop}%
\bibitem [{Note3()}]{Note3}%
  \BibitemOpen
  \bibinfo {note} {On the level of a theoretical model, the single resonance
  condition can be imposed by taking a system with narrow bands separated by a
  large gap.}\BibitemShut {Stop}%
\bibitem [{\citenamefont {Dykman}\ \emph {et~al.}(2011)\citenamefont {Dykman},
  \citenamefont {Marthaler},\ and\ \citenamefont {Peano}}]{Dykman2011}%
  \BibitemOpen
  \bibfield  {author} {\bibinfo {author} {\bibfnamefont {M.~I.}\ \bibnamefont
  {Dykman}}, \bibinfo {author} {\bibfnamefont {M.}~\bibnamefont {Marthaler}}, \
  and\ \bibinfo {author} {\bibfnamefont {V.}~\bibnamefont {Peano}},\ }\bibfield
   {title} {\enquote {\bibinfo {title} {Quantum heating of a parametrically
  modulated oscillator: Spectral signatures},}\ }\href@noop {} {\bibfield
  {journal} {\bibinfo  {journal} {Phys. Rev. A}\ }\textbf {\bibinfo {volume}
  {83}},\ \bibinfo {pages} {052115} (\bibinfo {year} {2011})}\BibitemShut
  {NoStop}%
\bibitem [{\citenamefont {Galitskii}\ \emph {et~al.}(1970)\citenamefont
  {Galitskii}, \citenamefont {Goreslavskii},\ and\ \citenamefont
  {Elesin}}]{Galitskii1970}%
  \BibitemOpen
  \bibfield  {author} {\bibinfo {author} {\bibfnamefont {V.~M.}\ \bibnamefont
  {Galitskii}}, \bibinfo {author} {\bibfnamefont {S.~P.}\ \bibnamefont
  {Goreslavskii}}, \ and\ \bibinfo {author} {\bibfnamefont {V.~F.}\
  \bibnamefont {Elesin}},\ }\bibfield  {title} {\enquote {\bibinfo {title}
  {{Electric and Magnetic Properties of a Semiconductor in the Field of a
  Strong Electromagnetic Wave}},}\ }\href@noop {} {\bibfield  {journal}
  {\bibinfo  {journal} {Sov. Phys. JETP}\ }\textbf {\bibinfo {volume} {30}},\
  \bibinfo {pages} {117} (\bibinfo {year} {1970})}\BibitemShut {NoStop}%
\bibitem [{\citenamefont {Sato}\ \emph {et~al.}(2019)\citenamefont {Sato},
  \citenamefont {McIver}, \citenamefont {Nuske}, \citenamefont {Tang},
  \citenamefont {Jotzu}, \citenamefont {Schulte}, \citenamefont {H\"ubener},
  \citenamefont {De~Giovannini}, \citenamefont {Mathey}, \citenamefont
  {Sentef}, \citenamefont {Cavalleri},\ and\ \citenamefont {Rubio}}]{Sato2019}%
  \BibitemOpen
  \bibfield  {author} {\bibinfo {author} {\bibfnamefont {S.~A.}\ \bibnamefont
  {Sato}}, \bibinfo {author} {\bibfnamefont {J.~W.}\ \bibnamefont {McIver}},
  \bibinfo {author} {\bibfnamefont {M.}~\bibnamefont {Nuske}}, \bibinfo
  {author} {\bibfnamefont {P.}~\bibnamefont {Tang}}, \bibinfo {author}
  {\bibfnamefont {G.}~\bibnamefont {Jotzu}}, \bibinfo {author} {\bibfnamefont
  {B.}~\bibnamefont {Schulte}}, \bibinfo {author} {\bibfnamefont
  {H.}~\bibnamefont {H\"ubener}}, \bibinfo {author} {\bibfnamefont
  {U.}~\bibnamefont {De~Giovannini}}, \bibinfo {author} {\bibfnamefont
  {L.}~\bibnamefont {Mathey}}, \bibinfo {author} {\bibfnamefont {M.~A.}\
  \bibnamefont {Sentef}}, \bibinfo {author} {\bibfnamefont {A.}~\bibnamefont
  {Cavalleri}}, \ and\ \bibinfo {author} {\bibfnamefont {A.}~\bibnamefont
  {Rubio}},\ }\bibfield  {title} {\enquote {\bibinfo {title} {{Microscopic
  theory for the light-induced anomalous Hall effect in graphene}},}\
  }\href@noop {} {\bibfield  {journal} {\bibinfo  {journal} {Phys. Rev. B}\
  }\textbf {\bibinfo {volume} {99}},\ \bibinfo {pages} {214302} (\bibinfo
  {year} {2019})}\BibitemShut {NoStop}%
\bibitem [{\citenamefont {Prosen}(1998)}]{Prosen1998}%
  \BibitemOpen
  \bibfield  {author} {\bibinfo {author} {\bibfnamefont {Toma\v{z}}\
  \bibnamefont {Prosen}},\ }\bibfield  {title} {\enquote {\bibinfo {title}
  {Time evolution of a quantum many-body system: Transition from integrability
  to ergodicity in the thermodynamic limit},}\ }\href@noop {} {\bibfield
  {journal} {\bibinfo  {journal} {Phys. Rev. Lett.}\ }\textbf {\bibinfo
  {volume} {80}},\ \bibinfo {pages} {1808} (\bibinfo {year}
  {1998})}\BibitemShut {NoStop}%
\bibitem [{\citenamefont {Kukuljan}\ and\ \citenamefont
  {Prosen}(2016)}]{Kukuljan2015}%
  \BibitemOpen
  \bibfield  {author} {\bibinfo {author} {\bibfnamefont {Ivan}\ \bibnamefont
  {Kukuljan}}\ and\ \bibinfo {author} {\bibfnamefont {Tomaz}\ \bibnamefont
  {Prosen}},\ }\bibfield  {title} {\enquote {\bibinfo {title} {{Corner transfer
  matrices for 2D strongly coupled many-body Floquet systems}},}\ }\href@noop
  {} {\bibfield  {journal} {\bibinfo  {journal} {J.~Stat.Mech.}\ }\textbf
  {\bibinfo {volume} {2016}},\ \bibinfo {pages} {043305} (\bibinfo {year}
  {2016})}\BibitemShut {NoStop}%
\bibitem [{\citenamefont {Citro}\ \emph {et~al.}(2015)\citenamefont {Citro},
  \citenamefont {Dalla~Torre}, \citenamefont {D'Alessio}, \citenamefont
  {Polkovnikov}, \citenamefont {Berbadi}, \citenamefont {Oka},\ and\
  \citenamefont {Demler}}]{Citro2016}%
  \BibitemOpen
  \bibfield  {author} {\bibinfo {author} {\bibfnamefont {Roberta}\ \bibnamefont
  {Citro}}, \bibinfo {author} {\bibfnamefont {Emanuele~G.}\ \bibnamefont
  {Dalla~Torre}}, \bibinfo {author} {\bibfnamefont {Luca}\ \bibnamefont
  {D'Alessio}}, \bibinfo {author} {\bibfnamefont {Anatoli}\ \bibnamefont
  {Polkovnikov}}, \bibinfo {author} {\bibfnamefont {Mehrtash}\ \bibnamefont
  {Berbadi}}, \bibinfo {author} {\bibfnamefont {Takashi}\ \bibnamefont {Oka}},
  \ and\ \bibinfo {author} {\bibfnamefont {Eugene}\ \bibnamefont {Demler}},\
  }\bibfield  {title} {\enquote {\bibinfo {title} {{Dynamical stability of a
  many-body Kapitza pendulum}},}\ }\href@noop {} {\bibfield  {journal}
  {\bibinfo  {journal} {Annals of Physics}\ }\textbf {\bibinfo {volume}
  {360}},\ \bibinfo {pages} {694} (\bibinfo {year} {2015})}\BibitemShut
  {NoStop}%
\bibitem [{\citenamefont {Chandran}\ and\ \citenamefont
  {Sondhi}(2016)}]{Chandran2016}%
  \BibitemOpen
  \bibfield  {author} {\bibinfo {author} {\bibfnamefont {Anushya}\ \bibnamefont
  {Chandran}}\ and\ \bibinfo {author} {\bibfnamefont {S.~L.}\ \bibnamefont
  {Sondhi}},\ }\bibfield  {title} {\enquote {\bibinfo {title}
  {{Interaction-stabilized steady states in the driven $O(N)$ model}},}\
  }\href@noop {} {\bibfield  {journal} {\bibinfo  {journal} {Phys. Rev. B}\
  }\textbf {\bibinfo {volume} {93}},\ \bibinfo {pages} {174305} (\bibinfo
  {year} {2016})}\BibitemShut {NoStop}%
\bibitem [{\citenamefont {Haldar}\ \emph {et~al.}(2018)\citenamefont {Haldar},
  \citenamefont {Moessner},\ and\ \citenamefont {Das}}]{Haldar2018}%
  \BibitemOpen
  \bibfield  {author} {\bibinfo {author} {\bibfnamefont {Asmi}\ \bibnamefont
  {Haldar}}, \bibinfo {author} {\bibfnamefont {Roderich}\ \bibnamefont
  {Moessner}}, \ and\ \bibinfo {author} {\bibfnamefont {Arnab}\ \bibnamefont
  {Das}},\ }\bibfield  {title} {\enquote {\bibinfo {title} {{Onset of Floquet
  thermalization}},}\ }\href@noop {} {\bibfield  {journal} {\bibinfo  {journal}
  {Phys. Rev. B}\ }\textbf {\bibinfo {volume} {97}},\ \bibinfo {pages} {245122}
  (\bibinfo {year} {2018})}\BibitemShut {NoStop}%
\bibitem [{\citenamefont {Seetharam}\ \emph {et~al.}(2018)\citenamefont
  {Seetharam}, \citenamefont {Titum}, \citenamefont {Kolodrubetz},\ and\
  \citenamefont {Refael}}]{Seetharam2018}%
  \BibitemOpen
  \bibfield  {author} {\bibinfo {author} {\bibfnamefont {Karthik}\ \bibnamefont
  {Seetharam}}, \bibinfo {author} {\bibfnamefont {Paraj}\ \bibnamefont
  {Titum}}, \bibinfo {author} {\bibfnamefont {Michael}\ \bibnamefont
  {Kolodrubetz}}, \ and\ \bibinfo {author} {\bibfnamefont {Gil}\ \bibnamefont
  {Refael}},\ }\bibfield  {title} {\enquote {\bibinfo {title} {{Absence of
  thermalization in finite isolated interacting Floquet systems}},}\
  }\href@noop {} {\bibfield  {journal} {\bibinfo  {journal} {Phys. Rev. B}\
  }\textbf {\bibinfo {volume} {97}},\ \bibinfo {pages} {014311} (\bibinfo
  {year} {2018})}\BibitemShut {NoStop}%
\bibitem [{\citenamefont {Bruus}\ and\ \citenamefont
  {Felnsberg}(2004)}]{Bruus2004}%
  \BibitemOpen
  \bibfield  {author} {\bibinfo {author} {\bibfnamefont {Henrik}\ \bibnamefont
  {Bruus}}\ and\ \bibinfo {author} {\bibfnamefont {Karsten}\ \bibnamefont
  {Felnsberg}},\ }\href@noop {} {\emph {\bibinfo {title} {Many-Body Quantum
  Theory in Condensed Matter Physics: An Introduction}}}\ (\bibinfo
  {publisher} {Oxford University Press},\ \bibinfo {year} {2004})\BibitemShut
  {NoStop}%
\bibitem [{\citenamefont {Tsuji}\ \emph {et~al.}(2008)\citenamefont {Tsuji},
  \citenamefont {Oka},\ and\ \citenamefont {Aoki}}]{Tsuji2008}%
  \BibitemOpen
  \bibfield  {author} {\bibinfo {author} {\bibfnamefont {Naoto}\ \bibnamefont
  {Tsuji}}, \bibinfo {author} {\bibfnamefont {Takashi}\ \bibnamefont {Oka}}, \
  and\ \bibinfo {author} {\bibfnamefont {Hideo}\ \bibnamefont {Aoki}},\
  }\bibfield  {title} {\enquote {\bibinfo {title} {Correlated electron systems
  periodically driven out of equilibrium: $\text{Floquet}+\text{DMFT}$
  formalism},}\ }\href@noop {} {\bibfield  {journal} {\bibinfo  {journal}
  {Phys. Rev. B}\ }\textbf {\bibinfo {volume} {78}},\ \bibinfo {pages} {235124}
  (\bibinfo {year} {2008})}\BibitemShut {NoStop}%
\bibitem [{\citenamefont {Qin}\ and\ \citenamefont
  {Hofstetter}(2017)}]{Qin2017}%
  \BibitemOpen
  \bibfield  {author} {\bibinfo {author} {\bibfnamefont {Tao}\ \bibnamefont
  {Qin}}\ and\ \bibinfo {author} {\bibfnamefont {Walter}\ \bibnamefont
  {Hofstetter}},\ }\bibfield  {title} {\enquote {\bibinfo {title} {{Spectral
  functions of a time-periodically driven Falicov-Kimball model: Real-space
  Floquet dynamical mean-field theory study}},}\ }\href@noop {} {\bibfield
  {journal} {\bibinfo  {journal} {Phys. Rev. B}\ }\textbf {\bibinfo {volume}
  {96}},\ \bibinfo {pages} {075134} (\bibinfo {year} {2017})}\BibitemShut
  {NoStop}%
\bibitem [{\citenamefont {Kalthoff}\ \emph {et~al.}(2018)\citenamefont
  {Kalthoff}, \citenamefont {Uhrig},\ and\ \citenamefont
  {Freericks}}]{Kalthoff2018}%
  \BibitemOpen
  \bibfield  {author} {\bibinfo {author} {\bibfnamefont {Mona~H.}\ \bibnamefont
  {Kalthoff}}, \bibinfo {author} {\bibfnamefont {G\"otz~S.}\ \bibnamefont
  {Uhrig}}, \ and\ \bibinfo {author} {\bibfnamefont {J.~K.}\ \bibnamefont
  {Freericks}},\ }\bibfield  {title} {\enquote {\bibinfo {title} {{Emergence of
  Floquet behavior for lattice fermions driven by light pulses}},}\ }\href@noop
  {} {\bibfield  {journal} {\bibinfo  {journal} {Phys. Rev. B}\ }\textbf
  {\bibinfo {volume} {98}},\ \bibinfo {pages} {035138} (\bibinfo {year}
  {2018})}\BibitemShut {NoStop}%
\bibitem [{\citenamefont {Mahan}(2000)}]{MahanBook}%
  \BibitemOpen
  \bibfield  {author} {\bibinfo {author} {\bibfnamefont {Gerald~D.}\
  \bibnamefont {Mahan}},\ }\href@noop {} {\emph {\bibinfo {title}
  {Many-particle physics}}}\ (\bibinfo  {publisher} {Springer},\ \bibinfo
  {year} {2000})\BibitemShut {NoStop}%
\bibitem [{\citenamefont {Scalapino}\ \emph {et~al.}(1993)\citenamefont
  {Scalapino}, \citenamefont {White},\ and\ \citenamefont
  {Zhang}}]{Scalapino1993}%
  \BibitemOpen
  \bibfield  {author} {\bibinfo {author} {\bibfnamefont {Douglas~J.}\
  \bibnamefont {Scalapino}}, \bibinfo {author} {\bibfnamefont {Steven~R.}\
  \bibnamefont {White}}, \ and\ \bibinfo {author} {\bibfnamefont {Shoucheng}\
  \bibnamefont {Zhang}},\ }\bibfield  {title} {\enquote {\bibinfo {title}
  {Insulator, metal, or superconductor: The criteria},}\ }\href@noop {}
  {\bibfield  {journal} {\bibinfo  {journal} {Phys. Rev. B}\ }\textbf {\bibinfo
  {volume} {47}},\ \bibinfo {pages} {7995--8007} (\bibinfo {year}
  {1993})}\BibitemShut {NoStop}%
\end{thebibliography}%

%\newpage\

%\newpage

\renewcommand\thefigure{S\arabic{figure}}
\renewcommand\thefigure{S\arabic{figure}}

\renewcommand{\thesubsubsection}{\thesubsection.\arabic{subsubsection}}
\newcommand{\Avg}[1]{\langle #1 \rangle}
\newcommand{\Amp}[2]{\langle #1|#2\rangle}

\def \ekr{{\ve_{k_R+}}}
\def \Tef{{\tilde T}}
\def \mef{{\tilde \m}}
\def \lef{{\tilde \lm}}
\def \zef{{\tilde z}}
\def \bq {{\mathbf{q}}}
\def \bA {{\mathbf{A}}}
\def \bE {{\mathbf{E}}}
\def \bJ {{\mathbf{J}}}
\setcounter{section}{0}
\setcounter{figure}{0}
% \title{Supplementary material for Floquet topological insulators: from band structure engineering to novel non-equilibrium quantum phenomena} %in gapped Dirac materials} %and Berry curvature
% \author{Mark S. Rudner$^1$ and Netanel H. Lindner$^2$}
% \affiliation{$^{1}$Niels Bohr International Academy and the Center for Quantum Devices, Niels Bohr
% Institute, University of Copenhagen, 2100 Copenhagen, Denmark} \affiliation{$^{2}$Physics
% Department, Technion, 320003, Haifa, Israel}

% \maketitle

% Explicit gap opening

%\begin{widetext}
\onecolumngrid 
\vspace{0.3 in}
\begin{center}\large{\bf Supplementary material for ``Floquet topological insulators: from band structure engineering to novel non-equilibrium quantum phenomena''}\end{center}
%\end{widetext}
\vspace{0.2 in}
\twocolumngrid
\section{Gap opening due to circularly polarized driving fields}
In this section we provide an explicit calculation to demonstrate how a circularly polarized driving
 field can open a gap at the Dirac point in the massless Dirac model of Eq.~(1) in the main text~\cite{Oka2009,Kitagawa2010}.
%%%%%%%%%%%%%%%%%%%%%%%%%%%%%%%%%%%%%%%%%%%%%%%%%%%%%%%%%%%%%%%%%%%%%%%%%%%
\begin{figure}[b]
\includegraphics[width=\columnwidth]{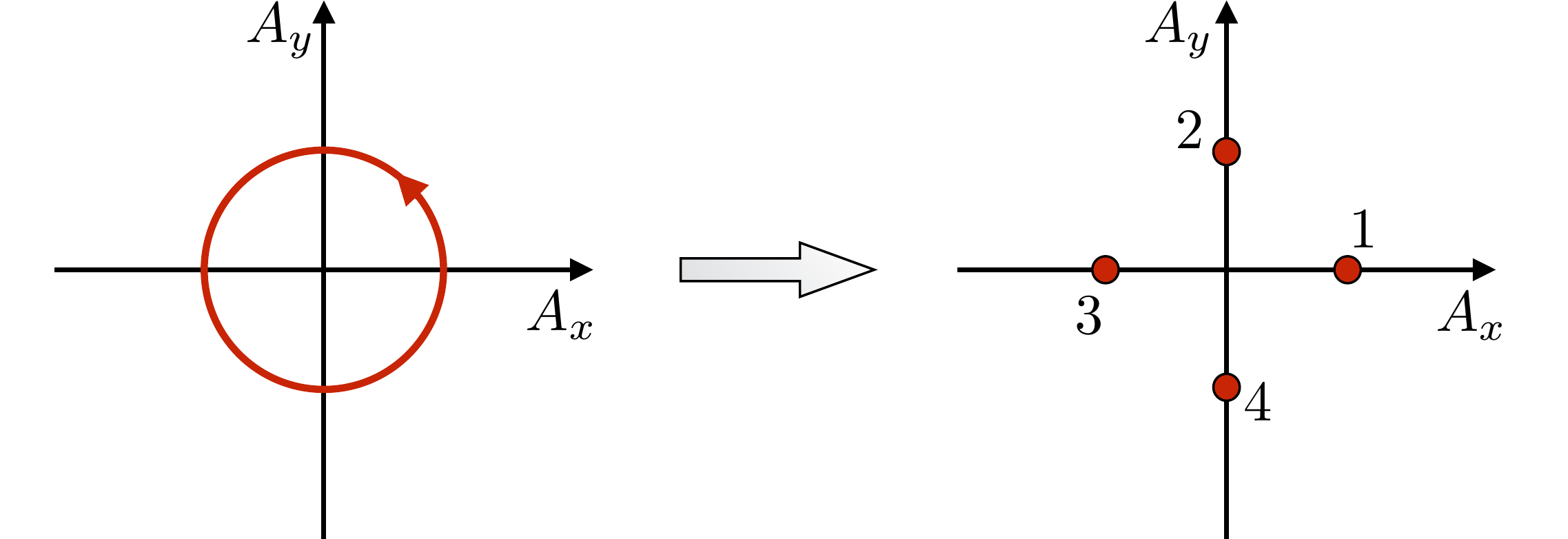}
\caption{To demonstrate Floquet gap opening at a Dirac point [Eq.~(1) of the main text], we consider a piecewise-constant driving protocol that mimics the effect of a circularly polarized field.
Rather than continuously rotating in the $xy$-plane, the vector potential with strength $A_0$ is taken to point in sequence along $x$, $y$, $-x$, and $-y$.
In each step, the field is applied for a duration $T/4$, such that the total driving period is $T \equiv 2\pi/\omega$.
The Floquet evolution operator $U(T)$ can be found exactly within this model.}
\label{fig:FourStepDrive}
%\vspace{-0.2 in}
\end{figure}
%%%%%%%%%%%%%%%%%%%%%%%%%%%%%%%%%%%%%%%%%%%%%%%%%%%%%%%%%%%%%%%%%%%%%%%%%%%
Rather than considering a continuously rotating circularly polarized harmonic driving field, we consider a four-step, piecewise-constant approximation to the circularly polarized drive as depicted in Fig.~\ref{fig:FourStepDrive}.
In each step $n$, the vector potential is held constant for a time $T/4$ with magnitude $A_0$ along the direction $\hat{\vec{e}}_n$ corresponding to point $n$ in the figure.
For each $\vec{k}$, the corresponding Bloch Hamiltonian for step $n$ is given by $H_n(\vec{k}) = v[\hbar\vec{k} - eA_0 \hat{\vec{e}}_n]\cdot\boldsymbol{\sigma}$.

The four-step drive maintains the chirality of the continuously rotating field, while allowing for a simple exact solution in the two-band model of Eq.~(1) in the main text.
To obtain the effective Hamiltonian, we calculate the Floquet operator $U(\vec{k}, T) = U_4(\vec{k})U_3(\vec{k})U_2(\vec{k})U_1(\vec{k})$, where $U_n(\vec{k}) = e^{-i H_n(\vec{k})T/(4\hbar)}$.
For each $n$ and $\vec{k}$, $U_n(\vec{k})$ can be obtained simply by exponentiating Pauli matrices.

To illustrate gap opening we focus on the Dirac point and evaluate $U(\vec{k}~=~0, T)~=~e^{-i \phi \sigma_y} e^{-i \phi \sigma_x} e^{i \phi \sigma_y} e^{i \phi \sigma_x}$, with $\phi = evA_0 T/(4\hbar)$.
In the high frequency (short period, $T$) limit $\phi \ll 1$, the exponentials can be expanded to obtain
%\begin{widetext}
\begin{eqnarray}
 U(\vec{k} = 0, T) %&=& e^{i \phi \sigma_y} e^{i \phi \sigma_x} e^{-i \phi \sigma_y} e^{-i \phi \sigma_x}\\
%&=& [(1 - \tfrac12\phi^2)\bm{1} + i\phi\sigma_y][(1 - \tfrac12\phi^2)\bm{1} + i\phi\sigma_x][(1 - \tfrac12\phi^2)\bm{1} - i\phi\sigma_y][(1 + \tfrac12\phi^2)\bm{1} - i\phi\sigma_y] + \mathcal{O}(\phi^3)\\
\label{eq:commutator} &=& \bm{1} + \phi^2 [\sigma_x, \sigma_y] + \mathcal{O}(\phi^3)\\
&=& \bm{1}+ 2i\phi^2 \sigma_z \approx e^{2i\phi^2\sigma_z}.
\end{eqnarray}
%\end{widetext}
Using $U(\vec{k}=0, T) = e^{-i H_{\rm eff}(\vec{k}=0) T/\hbar}$, and $T = 2\pi/\omega$, we find $H_{\rm eff}(\vec{k} = 0) = \tilde{\Delta} \sigma_z$ with $\tilde{\Delta} = -\pi (evA_0)^2/(4\hbar\omega)$.

As this explicit calculation shows, the chiral driving field induces a gap at the Dirac point [cf.~Eq.~(1) of the main text], with a magnitude that grows quadratically with the amplitude (linearly with the intensity) of the drive.
A very similar result is obtained for the case of the continuously rotating field, with a different prefactor.
The $\sigma_z$ term in the effective Hamiltonian arises due to the fact that $H(\vec{k}, t)$ does not commute with itself at different times.
Reversing the handedness of the drive would reverse the sign of the commutator in Eq.~(\ref{eq:commutator}), and hence also reverse the sign of the induced gap.

% vanishing $\nu_1
\section{Vanishing of the winding number $\nu_1$ for continuous time evolution}
As stated in Sec.~II of the main text, the winding number $\nu_1$ that counts the net winding of all Floquet bands around the quasienergy zone must vanish for any continuous evolution generated by a local, bounded Hamiltonian.
To see this, it is helpful to consider how the spectrum (band structure) of the evolution operator $U_{\rm 1D}(k, t) = \mathcal{T}e^{-(i/\hbar)\int_0^t dt' H_{\rm 1D}(k, t')}$ builds up as a function of time, $t$.
At $t = 0$, the evolution operator is the identity, $U_{\rm 1D}(k, t = 0) = {\bf 1}$.
At a small time $\delta t > 0$, the spectrum of $U_{\rm 1D}(k, \delta t)$ reflects the band structure of the instantaneous Hamiltonian, $H_{\rm 1D}(k, t = 0)$; crucially, the eigenvalues of $U_{\rm 1D}(k, \delta t)$ are periodic in $k$ and do not exhibit any nontrivial winding when $k$ traverses the Brillouin zone.
As time advances, the spectrum remains periodic in $k$ for all times, and by continuity cannot suddenly develop a net winding.
Therefore the spectrum of the Floquet operator $U_{\rm 1D}(k, T)$ cannot host a net winding of all of its bands; thus $\nu_1 = 0$.

\section{Definition of the time-averaged spectral function}
In Sec.~IV.A of the main text (see in particular Fig.~3) we introduced the ``time-averaged spectral function'' as a helpful tool for visualizing how mesoscopic transport occurs in Floquet systems.
In this section we give a formal definition of this quantity in terms of the system's retarded single-particle Floquet Green's function.
For completeness we review some basic features of Green's functions in Floquet systems, and discuss key similarities and differences between driven and non-driven systems (focusing on the non-interacting case).
Throughout this section we set $\hbar = 1$.

\subsection{Formal definitions}
We consider a Floquet system governed by a time-periodic Hamiltonian $H(t)$ satisfying $H(t + T) = H(t)$.
%Below we will work in a Heisenberg picture with respect to the evolution under Hamiltonian $H(t)$.
For a time-dependent Hamiltonian, we recall that the transformation to the Heisenberg picture is defined by the time evolution operator $U(t, t_0) = \mathcal{T}e^{-i\int_{t_0}^t dt' H(t')}$, which propagates the system forward in time from time $t_0$ to time $t$.
(Here $t_0$ is an arbitrary reference time, with respect to which the transformation is defined.)
The (time-independent) state of the system in the Heisenberg picture, $\ket{\psi}_H$, is obtained from the Schr\"{o}dinger picture state $\ket{\psi(t)}$ via $\ket{\psi}_H = U^\dagger(t, t_0)\ket{\psi(t)}$.
Importantly, a generic {\it time-dependent} operator $A(t)$ in the Schr\"{o}dinger picture transforms as
\begin{equation}
\label{eq:HeisenbergOp}A_H(t) = U^\dagger(t, t_0) A(t) U(t, t_0).
\end{equation}

Below we will work within the formalism of second quantization.
For brevity we will neglect spin throughout; the extension to include spin is straightforward.
We define the fermionic (``field'') creation operator $\Psi^\dagger(\vec{r})$, that creates a particle at position $\vec{r}$, and the annihilation operator $\Psi(\vec{r})$, that removes a particle from position $\vec{r}$.
These operators obey usual fermionic anticommutation relations, $\{\Psi(\vec{r}), \Psi^\dagger(\vec{r}')\} = \delta(\vec{r}-\vec{r}')$.

In the position representation, the single-particle retarded Green's function~\cite{Bruus2004} is defined %in terms of the Heisenberg picture creation and annihilation operators $\Psi_H(\vec{r}, t)$ and $\Psi^\dagger_H(\vec{r}', t')$, respectively,
as
\be
\label{eq:GR}G^R(\vec{r}t; \vec{r}'t') = -i\theta(t - t')\Avg{\{\Psi_H(\vec{r},t), \Psi^\dagger_H(\vec{r}',t') \}},
\ee
where the average is taken with respect to the many-body state of the system.
Here, the Heisenberg picture creation and annihilation operators $\Psi_H(\vec{r}, t)$ and $\Psi^\dagger_H(\vec{r}', t')$ are obtained from $\Psi(\vec{r})$ and $\Psi^\dagger(\vec{r}')$, respectively, using Eq.~(\ref{eq:HeisenbergOp}).
The retarded Green's function contains important information about the modes in which particles may propagate in the system, and will be the central object in the analysis below.

Similar to the case in equilibrium, for non- or weakly-interacting Floquet systems it is convenient to evaluate the Green's function in Eq.~(\ref{eq:GR}) in the basis of the single-particle Floquet states (``modes'') of the system in the absence of interactions, $\{\Ket{\psi_\nu(t)}\}$.
The Floquet modes are defined in the Schr\"{o}dinger picture as $\Ket{\psi_\nu(t)} = e^{-i\epsilon_\nu(t - t_0)}\Ket{\phi_\nu(t)}$, with $\Ket{\phi_\nu(t+T)} = \Ket{\phi_\nu(t)}$.
Here $\epsilon_\nu$ is the single-particle quasienergy associated with mode $\nu$.
In the position representation, mode $\nu$ is described by the time-periodic wave function $\phi_\nu(\vec{r}, t) = \Amp{\vec{r}}{\phi_\nu(t)}$.

In the {\it Schr\"{o}dinger picture}, we define the Floquet state creation operator at time $t$ as
\begin{equation}
c^\dagger_\nu(t) = \int d\vec{r}\, \phi_\nu(\vec{r},t) \Psi^\dagger(\vec{r}).
\label{eq: floquet creation}
\end{equation}
Thus, $c^\dagger_\nu(t)$ creates a particle in Floquet state $\nu$ at time $t$. [Note that we omit
the quasienergy phase factor from the definition of $c^\dagger_\nu(t)$.] The annihilation operator
$c_\nu(t)$ is defined analogously through Hermitian conjugation.

Crucially, for a {\it non-interacting system}, the Heisenberg operators $c^\dagger_{\nu, H}(t)$ and $c_{\nu,H}(t)$, defined via Eq.~(\ref{eq:HeisenbergOp}), take on a simple form in terms of the corresponding operators at the reference time $t_0$:
\begin{eqnarray}
\nonumber c^\dagger_{\nu, H}(t) &=& e^{i\epsilon_\nu(t-t_0)}c^\dagger_\nu(t_0)\\
\label{eq:Heisenberg_c} c_{\nu, H}(t) &=& e^{-i\epsilon_\nu(t-t_0)}c_\nu(t_0).
\end{eqnarray}
%We use the subscript $H_0$ to indicate relations that apply only for Heisenberg picture operators of non-interacting systems.
To see why Eq.~(\ref{eq:Heisenberg_c}) is true, note that the transformation to the Heisenberg picture corresponds to evolving backwards in time from $t$ to $t_0$: if a particle is created in Floquet mode $\nu$ at time $t$, then by its definition as a Floquet state it will evolve back to the corresponding mode $\nu$ at time $t_0$ under this transformation.
More formally, this property is reflected in the fact that in first quantized representation the evolution operator $U(t, t_0)$ for a single particle is diagonal with respect to the Floquet states, and can be written as $U(t, t_0) = \sum_\nu e^{-i\epsilon_\nu(t-t_0)}\Ket{\phi_\nu(t)}\Bra{\phi_\nu(t_0)}$.
%This property can be used to confirm the results in Eq.~(\ref{eq:Heisenberg_c}) above.

Using Eq.~(\ref{eq:Heisenberg_c}), we obtain a crucial relation that we will use below for the evaluation of the retarded Green's function for non-interacting Floquet systems:
\be
\label{eq:FloquetCommutation} \{c_{\nu,H}(t), c^\dagger_{\nu',H}(t')\} = e^{-i\epsilon_\nu (t-t')}\delta_{\nu\nu'}.
\ee
To obtain this result, we used the orthogonality of the single-particle Floquet states at equal times: $\Amp{\phi_\nu(t_0)}{\phi_{\nu'}(t_0)} = \delta_{\nu\nu'}$.
Note the similarities between the relations in Eqs.~(\ref{eq:Heisenberg_c}) and (\ref{eq:FloquetCommutation}) and the corresponding relations for the Heisenberg picture creation and annihilation operators in a non-driven, non-interacting system.

\subsection{Properties of the Floquet retarded\\ Green's function}
In equilibrium, with or without interactions, the single-particle retarded Green's function [Eq.~(\ref{eq:GR})] has many useful properties.
For example $G^R(t,t')$ (with position or orbital indices suppressed) depends on $t$ and $t'$ only through the time difference $t - t'$.
% after transforming to frequency space, $G^R(\omega)$ carries information about the energies of the characteristic modes of the system and.
After transforming to frequency space, the spectral function $A(\omega) = -\frac{1}{\pi} {\rm Im} [ G^R(\omega)]$ carries information about the energies of the characteristic single-particle modes of the system. %; here $G^R_\nu(\omega)$ is the single-particle retarded Green's function evaluated in the .
The trace (i.e., sum over all states) of the spectral function yields the density of states of the system.
Importantly, $A(\omega)$ is a positive semi-definite matrix (in the space of single particle orbitals), and for non-interacting systems does not depend on the state of the system.

In contrast, for a driven system, $G^R(t, t')$ is a function of both $t$ and $t'$, not only the
time difference. Nonetheless, analogous spectral information to that familiar from equilibrium
systems can be obtained for Floquet systems (see
Refs.~\cite{Tsuji2008,FoaTorresMultiTerminal,Usaj2014,Qin2017, Kalthoff2018,Uhrig2019}), provided
that the system is non-interacting and/or that it is in a time-periodic steady state. Under these
conditions, $G^R(t, t')$ is periodic in the average (or ``center of mass'') time $\bar{t} =
\frac12(t + t')$ with period $T$: shifting both $t$ and $t'$ by $T$ leaves the state of the system
invariant. As shown rigorously in Ref.~\cite{Uhrig2019}, a positive spectral density analogous to
that in equilibrium is then obtained by averaging $G^R(\bar{t} + \frac12 \tau,
\bar{t}-\frac12\tau)$, with $\tau = t - t'$, over one full period in the average time $\bar{t}$,
and then Fourier transforming with respect to the time-difference, $\tau$.

%\begin{widetext}
\onecolumngrid
\subsection{Evaluation of the Floquet retarded Green's function for a non-interacting system}
In the remainder of this section we illustrate the properties above with an explicit calculation of the retarded Green's function for a {\it non-interacting} Floquet system.
Returning to Eq.~(\ref{eq:GR}), we express the field operators $\Psi^\dagger_H(\vec{r}',t')$ and $\Psi_H(\vec{r},t)$ in terms of Floquet state creation and annihilation operators.
There is some freedom in how to do this; the most fruitful way is to write $\Psi^\dagger_H(\vec{r}',t') = U^\dagger(t', t_0) \Psi^\dagger(\vec{r}') U(t', t_0)$ and $\Psi(\vec{r}, t) = U^\dagger(t, t_0) \Psi(\vec{r}) U(t, t_0)$, and then to express $\Psi(\vec{r})$ and $\Psi^\dagger(\vec{r}')$ in terms of the (Schr\"{o}dinger picture) operators $\{c_\nu(t)\}$ and $\{c^\dagger_{\nu'}(t')\}$.
This yields $ \Psi_H(\vec{r},t) = \sum_\nu \phi^*_\nu(\vec{r}, t) c_{\nu, H}(t)$ and $\Psi^\dagger_H(\vec{r}',t') = \sum_\nu \phi_\nu(\vec{r}', t') c^\dagger_{\nu', H}(t)$.
%\bea
%\nonumber \Psi_H(\vec{r},t) &=& \sum_\nu \phi^*_\nu(\vec{r}, t) c_{\nu, H}(t)\\
% \Psi^\dagger_H(\vec{r}',t') &=& \sum_\nu \phi_\nu(\vec{r}', t') c^\dagger_{\nu', H}(t).
%\eea
Substituting into Eq.~(\ref{eq:GR}) and using Eq.~(\ref{eq:FloquetCommutation}), we obtain
\be
\label{eq:GR_non}G_0^R(\vec{r}t; \vec{r}'t') = -i\theta(t-t')\sum_\nu \phi_\nu^*(\vec{r},t)\phi_\nu(\vec{r}', t') e^{-i\epsilon_\nu (t-t')}.
\ee
Here we use the subscript 0 to emphasize that the result holds only for non-interacting systems.

Due to the periodicity of the wave functions $\{\phi_\nu(\vec{r},t)\}$, we express $\phi_\nu^*(\vec{r},t)\phi_\nu(\vec{r}', t')$ as a Fourier series in terms of harmonics $\{\phi_\nu^{(m)}(\vec{r})\}$ of the drive frequency, $\omega$: $\phi_\nu^*(\vec{r},t)\phi_\nu(\vec{r}', t')~=~\sum_{mm'} e^{im\omega (\bar{t} + \tau/2)}e^{-im'\omega (\bar{t}-\tau/2)} \big[\phi_\nu^{(m)}(\vec{r})\big]^*\phi_\nu^{(m')}(\vec{r}')$.
Inserting this expression into Eq.~(\ref{eq:GR_non}) and averaging with respect to $\bar{t}$ over one full period (for fixed $\tau$), we obtain the time-averaged single-particle retarded Green's function, $ \bar{G}_0^R(\vec{r}, \vec{r}'; \tau)$:
\bea
 \label{eq:GR_avg}\bar{G}_0^R(\vec{r}, \vec{r}'; \tau) &\equiv& \frac{1}{T}\int_{t_0}^{t_0 + T} d\bar{t}\,G_0^R(\vec{r, }\bar{t} + \tau/2; \vec{r}', \bar{t} - \tau/2)\\
\nonumber &=& -i\theta(\tau)\sum_{\nu,m} e^{-i(\epsilon_\nu - m\omega)\tau}\big[\phi_\nu^{(m)}(\vec{r})\big]^*\phi_\nu^{(m)}(\vec{r}').
\eea

Taking the Fourier transform  $\tilde{G}_0^R(\vec{r}, \vec{r}'; \Omega) = \lim_{\eta \to 0^+}\int_{-\infty}^\infty d\tau\, e^{i(\Omega + i\eta)\tau}\bar{G}_0^R(\vec{r}, \vec{r}'; \tau)$,  we obtain
\be
\label{eq:GR_avg2}\tilde{G}_0^R(\vec{r}, \vec{r}'; \Omega) = \sum_{\nu,m} \frac{\big[\phi_\nu^{(m)}(\vec{r})\big]^*\phi_\nu^{(m)}(\vec{r}')}{\epsilon_\nu - m\omega - \Omega + i\eta}.
\ee
Analogous to equilibrium, we define the ``time-averaged density of states'' via $\bar{\rho}_0(\Omega) = -\frac{1}{\pi} {\rm Tr}\, {\rm Im}[\tilde{G}_0^R(\Omega)]$:
\be
\label{eq:DOS} \bar{\rho}_0(\Omega) = \sum_{\nu,m} A^{(m)}_\nu \delta(\epsilon_\nu - m\hbar\omega - \hbar\Omega), \quad A^{(m)}_\nu = \Amp{\phi_\nu^{(m)}}{\phi_\nu^{(m)}},
\ee
where $A^{(m)}_\nu$ % = \int d\vec{r}\, |\phi^{(m)}_\nu(\vec{r})|^2$ %$A^{(m)}_\nu= \Amp{\phi_\nu^{(m)}}{\phi_\nu^{(m)}}$
captures the spectral weight of the $m$-th harmonic/sideband component $\ket{\phi_\nu^{(m)}}$ in the Floquet state $\ket{\psi_\nu(t)}$.
(Here $\ket{\phi_\nu^{(m)}}$ is the ket corresponding to the component $\phi^{(m)}_\nu(\vec{r})$ of the position-space Floquet state wave function defined above.)

As seen in Eq.~(\ref{eq:DOS}), the time-averaged density of states is comprised of delta-function peaks at frequencies corresponding to the quasienergies of the single-particle Floquet states of the system, plus or minus all integer multiples of the driving field photon energy, $\hbar \omega$.
The height of each peak is governed by the spectral weight $A^{(m)}_\nu$ of the corresponding sideband component of the Floquet state.

To obtain the time-averaged spectral function, we single out the contribution from a given Floquet
eigenstate $\nu$: $\mathcal{A}_\nu(\Omega) = \sum_m A^{(m)}_\nu \delta(\epsilon_\nu - m \omega -
\Omega)$. The time-averaged spectral function captures the fact that the spectral weight of each
Floquet state is spread in frequency (energy) over several discrete harmonics. In particular, as
discussed in the main text the time-averaged spectral function thus provides a helpful way to see
how states at specific {\it energies} in the leads couple to Floquet states with given
quasienergies in the system \cite{FoaTorresMultiTerminal,Farrell2015,Kundu2013,Kundu2017}.

\section{Floquet-Kubo formula}
In this section we discuss the linear response of a Floquet system to the application of an
additional weak perturbing field, oscillating at a frequency $\Omega$ that we assume to be much
smaller than the driving frequency $\omega$. The Hamiltonian, including the perturbation, is given
by $\hat H(t)=\hat H_0(t)+F(t)\hat B(t)$, where $F(t)$ is a general function of time, and $\hat
B(t)$ is an operator that may depend on time explicitly in the Schr\"{o}dinger picture, albeit in a
periodic manner with the same period as $\hat{H}_0$: $\hat B(t)=\hat B(t+T)$. The derivation given
in this section closely follows the derivation of the Kubo formula for equilibrium
systems~\cite{Bruus2004}. Throughout this section we will denote operators using hats to avoid
possible ambiguity about which objects are operators and which are simply real or complex numbers.

As a starting point, we assume that in the absence of the perturbing field the system has reached a time-periodic stationary state, described by a density matrix $\hat{\rho}_0(t)$, with $\hat{\rho}_0(t + T) = \hat{\rho}_0(t)$.
The change in the expectation value of a constant or time-periodic operator $\hat{A}(t)$ in response to the perturbing field $F(t)$ is given by
\begin{equation}
\delta \langle \hat A(t)\rangle = \int_{-\infty}^\infty dt'\, \tilde\chi(t,t') F(t'), \label{eq:linear response}
\end{equation}
where $\delta \langle \hat A(t)\rangle = \Tr \left[\hat\rho(t)\hat A(t)\right]-\Tr \left[\hat\rho_0(t)\hat A(t)\right]$.
Here,  $\hat\rho(t)$ is the state of the system in the presence of the perturbation. %, while $\rho_0(t)$ is the time-periodic sta.
Analogous to the case in equilibrium, the retarded response function $\tilde\chi(t,t')$ is defined as
\begin{equation}
\tilde\chi(t,t')=-i\theta(t-t')\Tr\left\{\hat \rho_0(t_0) \left[\hat A_I(t),\hat B_I(t')\right]\right\},
\label{eq: response function}
\end{equation}
where the interaction picture operator $\hat{A}_I(t)$ is defined by the transformation
\begin{equation}
\hat A_I(t)=\hat{U}_0^\dagger(t,t_0)\hat A(t) \hat{U}_0(t,t_0),
\end{equation}
with  $\hat{U}_0(t,t_0)=\mathcal{T}e^{-i \int_{t_0}^t \hat H_0(t')dt'}$, where $\mathcal{T}$ denotes time
ordering. The operator $\hat B_I(t')$ is defined in a similar manner. Note that although the arbitrary reference time
$t_0$ appears explicitly in Eq.~(\ref{eq: response function}), the response function $\tilde\chi(t,t')$ is in fact independent of $t_0$.
This can be checked by using the definitions of $\hat\rho_0(t_0)$ and of the operators $\hat{A}_I(t)$ and $\hat{B}_I(t')$.

Since the system is assumed to be in a \textit{steady state}, the response function $\tilde{\chi}(t,t')$ is
invariant under shifting both $t$ and $t'$ by $T$, the driving period.  To see why, we note the
steady state condition implies that $\hat{\rho}_0(t_0)=\hat{\rho}_0(t_0-T)$. The time-periodicity of the
Hamiltonian implies $\hat{U}_0(t+T,t_0)=\hat{U}_0(t,t_0-T)$, and therefore
\begin{equation}
\begin{split}
\Tr\left\{\hat\rho_0(t_0) \left[\hat A_I(t+T),\hat B_I(t'+T)\right]\right\}
&= \Tr\left\{\hat\rho_0(t_0-T) U_0^\dagger(t_0,t_0-T)\left[\hat A_I(t),\hat B_I(t')\right]U_0(t_0,t_0-T)\right\}\\
&=\Tr\left\{\hat\rho_0(t_0) \left[\hat A_I(t),\hat B_I(t')\right]\right\}.
\end{split}
\end{equation}
Transforming variables using $\tau \equiv t - t'$ we see that the function $ \chi(\tau,t') \equiv \tilde{\chi}(t' + \tau,t')$, given by
\begin{equation}
\chi(\tau,t')=-i\theta(\tau)\Tr\left\{\rho_0(t_0) \left[\hat A_I(t'+\tau),\hat B_I(t')\right]\right\},
\end{equation}
is periodic in  $t'$ with period $T$ (for fixed $\tau$). % (for fixed $t-t'$).
%Note that $ \chi(t-t',t')=\tilde \chi(t,t')$.
Thus $\chi(\tau, t')$ has a mixed Fourier representation in terms of one continuous frequency
variable, $\Omega$, conjugate to $\tau$, and a discrete set of harmonics, $m$, which capture its
periodicity in $t'$:
\begin{equation}
\label{eq:chi_transf}\chi(t-t',t')=\sum_m \int_{-\infty}^\infty d\Omega\, e^{-i \Omega (t-t')}e^{-i m\omega t'} \chi^{(m)}(\Omega).
\end{equation}
Inserting Eq.~(\ref{eq:chi_transf}) into Eq.~(\ref{eq:linear response}) and taking the Fourier
transform of both sides with respect to $t$, we obtain
\begin{equation}
\delta \langle \hat A\rangle(\Omega)=\sum_m  \chi^{(m)}(\Omega) F(\Omega-m\omega).
\label{eq: freq response}
\end{equation}

Suppose that i) we probe the observable $\hat A(t)$ at frequencies $\Omega$ that are much smaller than the
drive frequency, $\omega$, and ii) the Fourier transform of the perturbing field, ${F}(\Omega')$, only has support at frequencies $\Omega'\ll\omega$.
(The latter captures our original assumption that the system is perturbed at a frequency much less than that of the drive.)
Under these conditions, the only significant term in  the sum in  Eq.~(\ref{eq: freq response}) corresponds to $m=0$, giving
\begin{equation}
\delta\langle  \hat{A}\rangle(\Omega) =\chi^{(0)}(\Omega)F(\Omega).
\label{eq: low freq response}
\end{equation}
Note that $\chi^{(0)}(\Omega)$ is just the Fourier transform of the ``time-averaged response function,''  $\bar{\chi}(\tau) =\frac{1}{T}\int_0^T dt'
\chi(\tau,t')$, with respect to the variable $\tau$.

\subsection{Floquet-Kubo Formula for the electrical conductivity}

We now focus on the response of an electromagnetically-driven electronic Floquet system to a
uniform probing AC electric field oscillating at a frequency
$\Omega$, where as before $\Omega\ll \omega$~\cite{TorresFloquetKubo2005,Oka2009}. The extension to finite wave vector probe fields can
be performed using a similar approach. The electromagnetic gauge potential is then a sum of two
terms, $\bA(t)=\bA_0(t)+\delta \bA(t)$, where $\bA_0(t)=\bA_0(t+T)$ describes the time-periodic
driving field (which we aim to include exactly), and $\delta \bA(t)$ describes the weak probe
field. To second order in $\delta \bA(t)$, the Hamiltonian is given by
\begin{equation}
%\hat H(t)= \hat H_0(t) + \frac{\partial \hat{H}_0(t)}{\partial \delta  \bA(t)}\Bigg|_{\delta  \bA(t)=0} \delta  \bA(t)
%+ \frac{\partial^2 \hat{H}_0(t)}{\partial \delta  A_\alpha(t) \partial \delta A_\beta(t)}\Bigg|_{\delta  \bA(t)=0}\delta A_\alpha(t) \delta A_\beta(t)
\hat H(t)= \hat H(t)\Big|_{\delta  \bA(t)=0} +\ \frac{\partial \hat{H}(t)}{\partial \bA(t)}\Bigg|_{\delta  \bA(t)=0}\!\!\!\!\!\!\!\!\!\!\!\!\!\!\!\cdot\, \delta  \bA(t)
\ +\ \frac{\partial^2 \hat{H}(t)}{\partial A_\alpha(t) \partial A_\beta(t)}\Bigg|_{\delta  \bA(t)=0}\!\!\!\!\!\!\!\!\!\!\!\!\!\!\!\delta A_\alpha(t) \delta A_\beta(t),
\label{eq: ham 2nd order}
\end{equation}
where $\alpha, \beta = \{x, y, z\}$.
The current operator $\hat{\mathbf{J}}(t)=\left(\hat{J}_x(t),\hat{J}_y(t),\hat{J}_z(t)\right)$ is
given by
\begin{equation}
%\hat{\mathbf{J}}(t)=\frac{\partial \hat{H}_0(t)}{\partial \delta  \bA(t)}
\hat{\mathbf{J}}(t)=-\frac{\partial \hat{H}(t)}{\partial \bA(t)}. %\Bigg|_{\delta  \bA(t)=0}.
\end{equation}

Importantly, due to the time-dependence of $\bA_0(t)$ in $\hat H(t)$, the current operator
$\hat{\vec{J}}(t)$ is time-dependent (in the Schr\"{o}dinger picture). In analogy to common
practice in equilibrium systems, we now define a ``paramagnetic current operator''
$\hat{\bJ}^{(p)}(t)$ and the ``kinetic'' operator $\hat{K}_{\alpha\beta}(t)$ as
\cite{MahanBook,Scalapino1993}:
\begin{equation}
\hat{\bJ}^{(p)}(t)=-\frac{\partial \hat{H}(t)}{\partial  \bA(t)}\Bigg|_{\delta  \bA(t)=0},\quad \hat{K}_{\alpha\beta}(t)=-\frac{\partial^2 \hat{H}(t)}{\partial A_\alpha(t) \partial A_\beta(t)}\Bigg|_{\delta  \bA(t)=0}.
\label{eq: def para dia}
\end{equation}
%and the ``kinetic energy'' operator as
%\begin{equation}
%\hat{K}_{\alpha\beta}(t)=\frac{\partial^2 \hat{H}_0(t)}{\partial A_\alpha(t) \partial A_\beta(t)}\Bigg|_{\delta  \bA(t)=0} \qquad \alpha,\beta=x,y,z,
%\end{equation}

Using Eqs.~(\ref{eq: ham 2nd order})--(\ref{eq: def para dia}), the expectation value of the current, up to first order in $\delta \bA(t)$, can be written as a sum  of ``paramagnetic'' and ``diamagnetic'' contributions,
\begin{equation}
\langle\hat{J}_\alpha(t)\rangle=\langle \hat{J}_\alpha^{(p)}(t)\rangle + \sum_\beta\Tr \left[\hat{\rho}_0(t) \hat{K}_{\alpha\beta}(t)\right]\delta A_\beta(t).
\label{eq: para plus dia}
\end{equation}
For clarity we note that the expectation value $\langle\hat{J}_\alpha(t)\rangle \equiv
\Tr[\hat{\rho}(t)\hat{J}_\alpha(t)]$ of the time-dependent operator $\hat{J}_\alpha(t)$
is taken  with respect to the (full) state of the system $\hat\rho(t)$ at the same time, $t$. % and likewise for $\langle\hat{J}^{(p)}_\alpha(t)\rangle$ .
To obtain an equation valid within the
regime of linear response, in Eq.~(\ref{eq: para plus dia}) we expand the expectation value
$\langle\hat{J}^{(p)}_\alpha(t)\rangle$  up to linear order in the probe field $\delta\bA(t)$,
while we take the expectation value in the second term on the RHS %the expectation value is taken
with respect to the unperturbed state $\hat{\rho}_0(t)$.

 %the expectation value in the first term on the RHS of Eq.~(\ref{eq: para
%plus dia}) must be
%evaluated with respect to the state of the system taken to \addMR{zeroth and first} order in the probe field $\delta \bA(t)$, %in contrast to the second term where
%while in the second term the expectation value is taken with respect to the unperturbed state $\hat{\rho}_0(t)$.
The change in the paramagnetic current due to the probe field $\delta \bA(t)$ (relative to any time
periodic current that may flow in the steady state) is given by
\begin{equation}
\delta\langle \hat{J}_\alpha^{(p)}(t)\rangle =\int_{-\infty}^{\infty}dt'\chi_{\alpha\beta}(t-t',t') \delta A_\beta(t'),
\label{eq: para time}
\end{equation}
where
\begin{equation}
\chi_{\alpha\beta}(\tau,t')=-i\theta(\tau)\Tr\left\{\rho_0(t_0) \left[\hat{J}_{\alpha,I}^{(p)}(t'+\tau),\hat{J}_{\beta,I}^{(p)}(t')\right]\right\}.
\end{equation}
Taking the Fourier transform of Eq.~(\ref{eq: para time}) with respect to $t$, and using the
periodicity of $\chi_{\alpha\beta}(t-t',t')$ with respect to $t'$, we obtain
\begin{equation}
\delta \langle \hat{J}_\alpha^{(p)}\rangle(\Omega)=\int_{-\infty}^{\infty} dt' \sum_m e^{i(\Omega-m\omega)t'}\chi^{(m)}_{\alpha\beta}(\Omega)\delta A_\beta(t').
\label{eq: para omega_0}
\end{equation}
For the electromagnetic gauge potential, we choose a gauge for which $\bE(t)=-\frac{d}{dt}\bA(t)$.
Using integration by parts on Eq.~(\ref{eq: para omega_0}) to obtain a time derivative on $\delta
A_\beta(t)$ yields
\begin{equation}
\delta \langle \hat{J}_\alpha^{(p)}\rangle(\Omega)=\sum_m \frac{\chi^{(m)}_{\alpha\beta}(\Omega)}{i(\Omega-m\omega)}\delta E_\beta(\Omega-m\omega).
\label{eq: para omega}
\end{equation}
For electric fields that contain only low frequencies, $\Omega\ll\omega$, only the $m=0$
component contributes to Eq.~(\ref{eq: para omega}).

In the following we will assume that the gauge (probe) field $\delta\bA(t)$ appears only in the
quadratic part of the Hamiltonian, $\hat H^{(2)}(t)=\sum_{ij}  h_{ij}(t) \hat d^\dagger_i \hat
d_j$, where $\hat{d}^\dagger_i$ is an electronic creation operator with respect to a fixed,
time-independent basis. In terms of the basis of Floquet states, the quadratic part of the
Hamiltonian can be written as $\hat{H}^{(2)}(t)=\sum_{\bk} h_{\nu_1\nu_2}(\bk,t)
\hat{c}^\dagger_{\bk\nu_1}(t)\hat c_{\bk\nu_2}(t)$, where $\{\hat c^\dagger_{\bk\nu}(t)\}$ are
creation operators of electrons in single particle Floquet states, see Eq.~(\ref{eq: floquet
creation}), and $\nu$ is a Floquet band index. We then express the paramagnetic part of the current
operator as
\begin{equation}
\hat{J}^{(p)}_\alpha(t)=\sum_{\bk} j_{\alpha,\nu_1\nu_2}(\bk,t) \hat c^\dagger_{\bk\nu_1}(t)\hat c_{\bk\nu_2}(t),
\label{eq: para quad}
\end{equation}
where the matrix elements $j_{\alpha, \nu_1\nu_2}(\bk,t)$ are given by
\begin{equation}
\label{eq:j_def}j_{\alpha,\nu_1\nu_2}(\bk,t)=-\frac{\partial h_{\nu_1\nu_2}(\bk,t)}{\partial A_\alpha(t)}\Bigg|_{\delta  \bA(t)=0}=-\big\langle \phi_{\bk,\nu_1}(t)\big|\frac{\partial \check{h}(t)}{\partial  A_\alpha(t)}\big|\phi_{\bk,\nu_2}(t)\big\rangle\Big|_{\delta  \bA(t)=0},
\end{equation}
where $\check{h}(t)$ is the single-particle (first quantized) operator corresponding to the matrix
$h_{ij}(t)$.

We focus on the case of a {\it diagonal} steady state of the form
\begin{equation}
 \hat \rho_0(t)=\prod_{\bk\nu}[f_{\bk\nu}\hat c^\dagger_{\bk\nu}(t) \hat
c_{\bk \nu}(t)+(1-f_{\bk\nu})\hat c_{\bk\nu}(t) \hat c^\dagger_{\bk\nu}(t)],
\label{eq: steady rho}
\end{equation}
where  $f_{\bk\nu}$ is the population of the Floquet state created by $\hat c^\dagger_{\bk\nu}(t)$.
%To evaluate the response function, we use the analogue of Eq.~(\ref{eq:Heisenberg_c}) for the interaction picture.
As a useful preliminary for the evaluation of the response function, we note that for a steady
state of the form in Eq.~(\ref{eq: steady rho}), the analogue of Eq.~(\ref{eq:Heisenberg_c}) for
the interaction picture yields
\begin{equation}
\Tr\left[\hat\rho_0(t_0)\hat{c}^\dagger_{I, \bk\nu}(t_1) \hat{c}_{I, \bk'\nu'}(t_2)\right]=f_{\bk\nu}e^{i\varepsilon_{\bk\nu}(t_1-t_2)}\delta_{\bk\bk'}\delta_{\nu\nu'}.
\label{eq: chi prelim}
\end{equation}

We now use the results and definitions above to evaluate the linear response conductivity of the
Floquet system.
Using Eq.~(\ref{eq: para quad})--Eq.~(\ref{eq: chi prelim}), the response
function $\chi_{\alpha\beta}(\tau,t')$ is evaluated to be
\begin{equation}
%\begin{split}
\chi_{\alpha\beta}(\tau,t')=  -i\theta(\tau)\sum_{\bk}\sum_{\nu_1\nu_2}e^{i(\varepsilon_{\bk \nu_1}-\varepsilon_{\bk \nu_2})\tau}  \left(f_{\bk\nu_1}-f_{\bk\nu_2}\right)  j_{\alpha, \nu_1\nu_2}(\bk,t' + \tau)j_{\beta, \nu_2\nu_1}(\bk,t').
%\end{split}
\end{equation}\\

 Using the time-periodicity of $j_{\alpha,
\nu_1\nu_2}(\bk,t)$ [see definition in Eq.~(\ref{eq:j_def})], we expand $j_{\alpha,
\nu_1\nu_2}(\bk,t)=\sum_m e^{-i m \omega t}j^{(m)}_{\alpha, \nu_1\nu_2}(\bk)$. Averaging the time
variable $t'$ over one period and taking the Fourier transform with respect to $\tau$, we obtain
the time-averaged response function
\begin{equation}
\label{eq:chi0final}  \chi^{(0)}_{\alpha\beta}(\Omega)=\sum_{\bk}\sum_{\nu_1\nu_2}\sum_m\frac{\left(f_{\bk\nu_1}-f_{\bk\nu_2}\right)j^{(m)}_{\alpha, \nu_1\nu_2}(\bk)j^{(-m)}_{\beta,\nu_2\nu_1}(\bk)}{{\Omega}-m\omega+(\varepsilon_{\bk \nu_1}-\varepsilon_{\bk \nu_2})}.
\end{equation}
Using the gauge $\bE(t)=-\frac{d}{dt}\bA(t)$ as before, we obtain a contribution from the
diamagnetic (second) term in Eq.~(\ref{eq: para plus dia}) to the total current $\langle
\hat{J_\alpha}\rangle (\Omega)\equiv\int_{-\infty}^{\infty} dt\, e^{i\Omega t}\langle
\hat{J_\alpha}(t)\rangle$ that is given by:
% to $\langle \hat{J_\alpha}\rangle (\Omega)\equiv\int_{-\infty}^{\infty} dt e^{-i\Omega t}\langle \hat{J_\alpha}(t)\rangle_t$ which is given by
\begin{equation}
\int_{-\infty}^\infty dt\, e^{i\Omega t} \Tr\left[\hat\rho_0(t)\hat{K}_{\alpha\beta}(t)\right]\delta A_\alpha(t)=\sum_m\frac{ \mathcal{K}_{\alpha\beta}^{(m)}}{i(\Omega-m\omega)}\delta E_\alpha(\Omega+m\omega).
\label{eq: dia}
\end{equation}
In writing Eq.~(\ref{eq: dia}) we have used the time-periodicity of
$\mathcal{K}_{\alpha\beta}(t)=\Tr\left[\hat\rho_0(t)\hat{K}_{\alpha\beta}(t)\right]$ and expanded
$\mathcal{K}_{\alpha\beta}(t)=\sum_m \mathcal{K}^{(m)}_{\alpha\beta} e^{-im\omega t}$. Using this
expansion, Eq.~(\ref{eq: dia}) is obtained by integration by parts. Note that for electric fields
that contain only low frequency components $\Omega\ll\omega$, only the term $m=0$ in Eq.~(\ref{eq:
dia}) gives a contribution to the current. In this case, this second contribution to the current is
given by the time-average of $\mathcal{K}_{\alpha\beta}(t)$,
\begin{equation}
\mathcal{K}^{(0)}_{\alpha\beta}=\frac{1}{T}\int_0^T dt\, \mathcal {K}_{\alpha\beta}(t).
\end{equation}

Finally we obtain Ohm's law for the response to the probe field, $\delta \Avg{\hat{J}_\alpha}(\Omega) =
\sigma_{\alpha\beta}(\Omega) \delta E_\beta(\Omega)$: %, for low frequencies $\Omega\ll\omega$ we obtain:
\begin{equation}
\sigma_{\alpha\beta}(\Omega)=\frac{\chi^{(0)}_{\alpha\beta}(\Omega)+\mathcal{K}^{(0)}_{\alpha\beta}}{i\Omega},\quad \textrm{(for $\Omega\ll\omega$)}.
\end{equation}
Note the similarity between this result and the analogous expression for the conductivity in equilibrium systems, as well the form of the response function $\chi^{(0)}_{\alpha\beta}(\Omega)$ in Eq.~(\ref{eq:chi0final}).

%\begin{equation}
% \chi(\Omega)=\sum_{\substack{\nu\nu'l}}\frac{\mathcal{M}_{\nu\nu'}^{(l)}(f_{\nu'}-f_{\nu})}{\hbar\Omega+\varepsilon_{\nu'}-\varepsilon_{\nu}-l\hbar\omega+i0^+},
%\label{eq:Lindhard}
%\end{equation}
%\end{widetext}

%\bibliography{FloquetTopologicalInsulator_references}

\end{document}